\documentclass[aps,prd,longbibliography,notitlepage,superscriptaddress,nofootinbib]{revtex4-1}[11pt]
\usepackage{graphics}      
\usepackage{graphicx}      
\usepackage{longtable}     
\usepackage{url}           
\usepackage{bm,dsfont}            
\usepackage{amsmath}
\usepackage{amssymb}
\usepackage{float}
\usepackage{physics}
\usepackage{tensor}
\usepackage{slashed}
\usepackage{bbold,xcolor}
\usepackage{bm}
\usepackage[utf8]{inputenc}
\newcommand{\vect}[1]{\boldsymbol{#1}_{\perp}}
\usepackage{mathtools}

\usepackage{soul}

\usepackage{appendix}

\begin{document}

\title{Inclusive prompt photon-jet
correlations as a probe of gluon saturation
in electron-nucleus scattering at small $x$}

\author{Isobel Kolbé}
\email{ikolbe@uw.edu}
\affiliation{\small
Institute for Nuclear Theory, University of Washington, Seattle, Washington 98195-1550}

\author{Kaushik Roy}
\email{kaushik.roy.1@stonybrook.edu}
\affiliation{\small Physics Department, Brookhaven National Laboratory, Bldg. 510A, Upton, NY 11973, USA}
\affiliation{\small Department of Physics and Astronomy, Stony Brook University, Stony Brook, New York 11794, USA }

\author{Farid Salazar}
\email{farid.salazarwong@stonybrook.edu}
\affiliation{\small Physics Department, Brookhaven National Laboratory, Bldg. 510A, Upton, NY 11973, USA}
\affiliation{\small Department of Physics and Astronomy, Stony Brook University, Stony Brook, New York 11794, USA }

\author{Bj\"orn Schenke}
\email{bschenke@bnl.gov}
\affiliation{\small Physics Department, Brookhaven National Laboratory, Bldg. 510A, Upton, NY 11973, USA}

\author{Raju Venugopalan}
\email{raju@bnl.gov}
\affiliation{\small Physics Department, Brookhaven National Laboratory, Bldg. 510A, Upton, NY 11973, USA}

\date{\today}

\begin{abstract}
    We compute the differential cross-section for inclusive prompt photon$+$quark production in deeply inelastic scattering  of electrons off nuclei at small $x$ ($e+A$ DIS) in the framework of the Color Glass Condensate effective field theory. 
    The result is expressed as a convolution of the leading order (in the strong coupling $\alpha_{\mathrm{s}}$) impact factor for the process and universal dipole matrix elements, in the limit of hard photon transverse momentum relative to the nuclear saturation scale  $Q_{s,A}(x)$. 
    We perform a numerical study of this process for the kinematics of the Electron-Ion Collider (EIC), exploring in particular the azimuthal angle correlations between the final state photon and quark.
    We observe a systematic suppression and broadening pattern of the back-to-back peak in the relative azimuthal angle distribution, as the saturation scale is increased by replacing proton targets with gold nuclei.
    Our results suggest that photon+jet final states in inclusive $e+A$ DIS at high energies are in general a promising channel for exploring gluon saturation that is complementary to inclusive and diffractive dijet production.
    They also provide a sensitive empirical test of the universality of dipole matrix elements when compared to identical measurements in proton-nucleus collisions.
    However because photon+jet correlations at small $x$ in EIC kinematics require jet reconstruction at small $k_\perp$, it will be important to study their feasibility relative to  photon-hadron correlations.

\end{abstract}

\maketitle

\section{Introduction}\label{sec:intro}

It  is conjectured~\cite{Gribov:1984tu,Mueller:1985wy} that the  proliferation of soft (small $x$) gluons in hadron wavefunctions at high energies via bremsstrahlung is saturated by many-body screening and recombination effects when gluon occupancies are parametrically $O(1/\alpha_{\mathrm{s}})$, where $\alpha_{\mathrm{s}}$ is the QCD coupling. The discovery and characterization of this gluon saturation phenomenon is a major goal of  deeply inelastic scattering (DIS) experiments at the Electron-Ion Collider (EIC) and several signatures of gluon saturation have been discussed in this context~\cite{Accardi:2012qut,Aschenauer:2017jsk}.

We will discuss here the process $e+A\rightarrow e+\gamma+{\rm jet}+X$ corresponding to the computation of a photon+quark inclusive final state in DIS off a nucleus with mass number $A$, where the quark fragments to a jet and an arbitrary number $X$ of other final state hadrons are produced. This process (isolated   photon+jet) was previously measured at HERA in DIS off protons~\cite{Abramowicz:2012qt} and compared to perturbative QCD (pQCD) computations\footnote{Note that these authors with their collaborators have also computed inclusive photon production~\cite{Gehrmann-DeRidder:2006zbx,GehrmannDeRidder:2006vn} in the same framework and compared their results to HERA data.} to $O(\alpha^2_{\rm em} \alpha_{\mathrm{s}})$ in \cite{GehrmannDeRidder:2000ce,Aurenche:2014wla,Aurenche:2017zjk}. 

Our focus here is to explore the impact of gluon saturation on photon+jet correlations at the EIC. The distinguishing feature of this phenomenon is an emergent semi-hard saturation scale $Q_{s,A}(x)$ that increases with decreasing $x$ and increasing mass number $A$. We will study how the saturation scale scale modifies photon+jet correlations. Because all-twist many-body correlations become important in this regime, this requires one to go beyond leading twist perturbative computations at small $x$ and resum powers of $Q_{s,A}^2/Q^2$ to all orders at each order in the weak coupling expansion in $\alpha_{\mathrm{s}}$.

The physics of ``saturated" gluons in QCD  is described 
by the Color Glass Condensate (CGC) effective field theory (EFT)~\cite{McLerran:1993ni,McLerran:1993ka,McLerran:1994vd,Iancu:2000hn,Iancu:2001ad,Ferreiro:2001qy,Iancu:2003xm,Gelis:2010nm,Kovchegov:2012mbw,Blaizot:2016qgz} which employs a Born-Oppenheimer separation of the hadron wavefunction whereby the dynamics of small $x$ partons is sourced by static large $x$ parton color charges. The saturation scale $Q^2_s(x)$ appears as a dimensionful scale in this EFT and corresponds to the density of color sources that interact coherently with the small $x$ partons. The density of color sources grows with mass number, thus one obtains $Q_s^2\propto A^{1/3}$. Since the saturation scale is the only dimensionful scale at $x$ (and $A$) where $Q_{s}^2(x) \gg \Lambda_{\rm QCD}^2$, the typical momenta of partons is peaked at this scale. Since then $\alpha_{\mathrm{s}}(Q_s^2)\ll 1$, the EFT provides a systematic semi-classical weak coupling expansion in the strongly correlated, non-linear small $x$ regime of QCD~\cite{McLerran:1993ni,McLerran:1993ka,McLerran:1994vd} where the phase space occupancy of gluons $\sim 1/\alpha_{s}(Q_s(x)) \gg  1$ is large. 

Testing the quantitative reliability of perturbative calculations in the CGC EFT requires one to consider observables which can provide a clear picture of the underlying dynamics. Inclusive prompt photon production and prompt photon-jet angular correlations  are two  examples of such observables. These processes have been studied thus far in the CGC EFT only in the context of proton-proton/nucleus ($p+A$) collisions~\cite{Gelis:2002ki,Baier:2004tj,JalilianMarian:2005zw,JalilianMarian:2005gm,JalilianMarian:2008iz,Rezaeian:2009it,Rezaeian:2012wa,Rezaeian:2012ye,Klein-Bosing:2014uaa,Rezaeian:2016szi,Benic:2016yqt,Benic:2016uku,Benic:2017znu,Ducloue:2017kkq,Benic:2018hvb,Goncalves:2020tvh,Santos:2020jih}. In this paper, we will present the first computation of the leading order (LO) cross-section for the inclusive production of a prompt photon in association with a quark (jet) in electron-nucleus ($e+A$) scattering, in the Regge-Gribov kinematics of fixed $Q^{2}$, squared center-of-mass energy $s \rightarrow \infty$ and Bjorken $x_{\rm Bj} \rightarrow 0$. Our work employs the formalism for inclusive prompt photon+quark-antiquark ($\gamma+q\bar{q}$) dijet production to next-to-leading order (NLO) developed in~\cite{Roy:2018jxq,Roy:2019hwr,Roy:2019cux}. We will adapt this formalism to study the impact of gluon saturation on inclusive photon+quark
correlations. Due to the analytical and numerical complexity of the computations, we will only consider 
the LO formalism here; the extension to NLO will be left to future work. 

At LO in $\alpha_\mathrm{s}$, inclusive prompt photon+jet production in $e+A$ scattering at small $x$ proceeds through the fluctuation of the exchanged virtual photon into a long-lived quark-antiquark dipole plus a photon emitted from either parton\footnote{Note that this contribution is formally NLO in the pQCD power counting of Refs.~\cite{GehrmannDeRidder:2000ce,Aurenche:2014wla,Aurenche:2017zjk}.}. The quark and antiquark can multiple scatter coherently off the dense gluons in the nucleus either before or after emitting the photon. As a result of this multiple scattering, the quark and/or antiquark acquires a Wilson line color phase. Indeed, in the eikonal approximation valid at high energies, spatial correlators of these Wilson lines incorporate all the information about the nuclear target accessed in the DIS process at small $x$. Specifically, in the case of photon+dijet production, the quantum expectation value for the process depends on two-point ``dipole" and four-point ``quadrupole" correlators~\cite{Roy:2018jxq}.
The former is accessed in fully inclusive DIS while extraction of the latter requires more differential measurements such as inclusive dijet production~\cite{Dominguez:2011wm,Mantysaari:2019hkq} and the  photon+dijet process discussed in  \cite{Roy:2018jxq}.

An interesting question we will address is whether the photon+quark process at LO is sensitive to both dipole and quadrupole correlators. This is not evident from a brute force integration of the phase space of the antiquark in the expression for the cross-section for photon+quark+antiquark production. Towards this end, we will show that the results of  \cite{Roy:2018jxq} can be reexpressed in a form where, for a clearly identified photon+quark  hard process, the cross-section is sensitive only to dipole correlators. As a corollary, as the phase space for isolation cut on the photon is reduced, the cross-section will show increasing sensitivity to quadrupole correlators.
 
In addition to addressing the formal question on the relative sensitivity of dipole and quadrupole correlators, we will explore the discovery potential of this measurement in the kinematics of the EIC. The prospect of such measurements is exciting since they will  be performed in $e+A$ collisions at the EIC for the first time. Further, the EIC will  operate at unprecedented luminosities which opens up the possibility to measure rare events through comprehensive analyses of experimental data hitherto unavailable even in electron-proton  collisions. While  identical measurements are feasible in $p+A$ collisions at the Relativistic Heavy Ion Collider (RHIC) and the Large Hadron Collider (LHC),  final state interactions in such experiments complicate the unambiguous extraction of gluon saturation in the hadron wavefunctions. However because the computation of this process also exists in the CGC EFT, comparisons of the results of $e+A$ and $p+A$ collisions offer one the opportunity to systematically isolate initial and final state contributions and thereby extract universal features of gluon saturation. Specifically, we will be able to determine if the dipole and quadrupole operators extracted from different processes in DIS are the same as those extracted in $p+A$ collisions.

Turning now to empirical probes, we note that measurements of azimuthal angle correlations of final state particles are especially sensitive to gluon saturation. In conventional perturbative QCD, energy-momentum conservation dictates that a hard photon (or jet) is produced back-to-back with another jet because only transverse momentum $k_\perp \leq \Lambda_{\rm QCD}$ is assumed to be transferred from a hadron in the computation of the hard cross-section. However, as we noted, with the onset of gluon saturation, the momentum transfer from the hadron is given by $k_\perp \sim Q_s(x)$. This contributes to a systematic suppression and broadening of the back-to-back or ``away side" peak in the $\Delta \phi$ relative azimuthal distribution of the hard photon-jet cross-section or of dijets.

Much of the discussion on azimuthal angle correlations at small $x$ in the CGC and related frameworks has been restricted to dijet/dihadron production in $p(d)+A$ collisions~\cite{Kharzeev:2004bw,Marquet:2007vb,Albacete:2010pg,Dominguez:2010xd,Dominguez:2011wm,Stasto:2011ru,Iancu:2013dta,vanHameren:2014lna,Kotko:2015ura,vanHameren:2016ftb,Albacete:2018ruq,Stasto:2018rci,Giacalone:2018fbc,Fujii:2020bkl}. Similar studies have been performed for the case of $e+A$ scattering for the production of dijets/dihadrons~\cite{Zheng:2014vka,Dumitru:2015gaa,Dumitru:2018kuw,Mantysaari:2019hkq} and  trijets\footnote{See~\cite{Altinoluk:2018byz,Altinoluk:2020qet} for a discussion on angular correlations in dijet plus photon and three jet production respectively, for the case of $p+A$ collisions.}~\cite{Ayala:2016lhd}. We will explore here such correlations between the observed photon and quark in the context of $e+A$ collisions at small $x$. We will show that our results are similar to those for inclusive photon+hadron production in $p+A$ collisions in LHC kinematics~\cite{JalilianMarian:2012bd,Rezaeian:2012wa,Rezaeian:2012ye,Rezaeian:2016szi,Benic:2017znu} where  a  suppression and broadening of the ``back-to-back" peak in the azimuthal angle distribution is predicted.

The outline of the paper is as follows.  
In Section~\ref{sec:cross-section-setup}, we begin with general definitions for the various kinematic quantities involved in this DIS process of interest. 
We next formulate the computation of the amplitude and elucidate the different components necessary for obtaining the cross-section for inclusive prompt photon+quark production in $e+A$ DIS at small $x$. The key expression for the cross-section is written in  Eq.\,\eqref{eq:diff-cs-gammajet-general} in terms of contributions from different polarization states of the exchanged virtual photon. The most important ingredients in deriving this quantity are the amplitude and cross-section for $\gamma+q\bar{q}$ production in the hadronic subprocess (virtual photon-nucleus scattering), for a given polarization of the virtual photon.

We discuss the computation of the former in section ~\ref{sec:photon-dijet-amplitude}; the expression for the amplitude is provided in Eqs.~\eqref{eq:reduced-amplitude} and \eqref{eq:amplitude-total-LO}. Using this expression for the amplitude, we derive in Section~\ref{sec:photon-dijet-cross-section}, results for the cross-section for $\gamma+q\bar{q}$ production separately for the longitudinal and transverse polarizations of the virtual photon. The contribution from the longitudinally polarized photon to the cross-section is given by Eqs.~\eqref{eq:diff-cs-long-pol} and \eqref{eq:R-long-pol-gamma-dijet} and the net contribution from the two transverse polarization states is given by Eqs.~\eqref{eq:diff-cs-trans-pol} through \eqref{eq:R-T-ins-ins}. 

In Section~\ref{sec:antiquark-phase-space-integration}, divided into three subsections, we obtain results for the cross-section for $\gamma+q$ production using expressions for the  $\gamma+q\bar{q}$ cross-section obtained in the previous section and integrating over the phase space of the antiquark.  We consider the case of a longitudinally polarized virtual photon to demonstrate this procedure. The corresponding exercise for transverse polarizations is discussed in Appendix~\ref{sec:photon-jet-trans-pol}. In Section~\ref{sec:fragmentation-contribution-long-pol}, we discuss in detail the treatment of the singular contribution for the case when the emitted photon becomes collinear to the integrated antiquark. We use a simple cutoff procedure to regulate the collinear divergence and absorb it into the (anti)quark-to-photon fragmentation function; the cross-section proportional to this is given by Eq.\,\eqref{eq:fragmentation-cs-long-pol}. In Section~\ref{sec:direct-contribution-gamma-jet-long}, we combine the non-divergent terms from the integration over the antiquark phase space into what we call the ``direct" photon+quark contribution to the cross-section. The results are given in  Eqs.~\eqref{eq:direct-cs-long-pol}-\eqref{eq:direct-cs-long-pol-B}. We then consider in Section~\ref{sec:direct-photon-limit-long} a limiting form of the ``direct" photon+quark production cross-section in momentum space. We show that this contribution is  dominated by photons that are hard relative to $Q_{s}$. This momentum space expression for the direct photon+quark cross-section is given in  Eq.\,\eqref{eq:diff-cs-direct-limit-long-mspace} with its constituents defined in Eqs.~\eqref{eq:H-L-q-gamma}-\eqref{eq:K-tilde-L-4}.

One of the main advantages of the simplification in Section~\ref{sec:direct-photon-limit-long} is that it significantly aids in the numerical analyses for direct photon+quark production. In Section~\ref{sec:azimuthal-correlations} we employ these results to study for the first time the azimuthal angle correlations in direct photon-quark in $e+A$ scattering at small $x$. 
Since we are able to identify a kinematic region where only dipole Wilson line correlators contribute, we can study the energy evolution of the azimuthal angle correlations by computing the $x$ evolution of this correlator. This is to a good approximation described by the Balitsky-Kovchegov (BK) renormalization group (RG) equation~\cite{Balitsky:1995ub,Kovchegov:1999yj} that systematically resums leading logarithms containing all powers of $\alpha_{\mathrm{s}}\ln{x}$. In our work, we will employ  the running coupling (BK) equation that involves a specific scheme for running coupling effects that appear at next-to-leading log accuracy in the RG evolution~\cite{Albacete:2010sy}.

In section \ref{sec:azimuthal-correlations-setup} we review the setup of the calculation and in section \ref{sec:azimuthal-correlations-numerical-results} we present the numerical results. We focus our attention on the suppression and broadening of the back-to-back peak as we change the nuclear species from proton to gold (Au). We study the dependence of these observables on different kinematic variables, the transverse momenta and rapidity of of the final state photon and the $Q^2$ dependence of the virtual photon probe in the scattering process. We summarize our principal results in Section~\ref{sec:summary} and briefly discuss future work in the framework of this computation.

Appendices~\ref{sec:conventions}-\ref{sec:single-inclusive-parton-cs} supplement the material in the main text. The conventions for light cone coordinates are given in Appendix~\ref{sec:conventions}. Appendix~\ref{sec:Feynman rules} discusses the Feynman rules used in the CGC EFT calculation. In Appendix~\ref{sec:hadron-amplitude-squared}, we present  expressions for the squared amplitude for longitudinal and transverse polarizations of the virtual photon that are used to construct the perturbative components of the cross-section in Eqs.~\eqref{eq:diff-cs-long-pol} and \eqref{eq:diff-cs-trans-pol}. In Appendix~\ref{sec:gamma-matrix-identities}, we list the identities involving traces of gamma matrices and photon polarization vectors that are used to derive the results in the previous section. These are very general and may be useful for other small $x$ computations. Appendix~\ref{sec:coefficients-J-functions} contains explicit expressions for the various quantities that constitute the expression for the cross-section for $\gamma+q\bar{q}$ production from a transversely polarized virtual photon given by Eqs.~\eqref{eq:diff-cs-trans-pol}-\eqref{eq:R-T-ins-ins}.

Appendix~\ref{sec:photon-jet-trans-pol} details the computation of the cross-section for photon+quark production for the case of transversely polarized virtual photons. Here the discussion is 
organized in an identical manner to the discussion in Section~\ref{sec:antiquark-phase-space-integration} and  results for the fragmentation and direct photon+quark contributions are presented separately over three subsections. In Appendix~\ref{sec:fragmentation-contribution-trans-pol}, we present in Eq.\,\eqref{eq:fragmentation-cs-trans-pol} the combined contribution from the two transverse polarization states of $\gamma^{*}$ towards the fragmentation photon+quark cross-section. The corresponding results for the direct photon+quark production are provided in Appendix~\ref{sec:direct-photon-trans-pol} in Eqs.~\eqref{eq:direct-cs-trans-pol} through \eqref{eq:B-T-reg-ins}. In  Appendix~\ref{sec:direct-photon-limit-trans}, we obtain a simplified expression for this cross-section fully in momentum space; this is given by Eq.\,\eqref{eq:diff-cs-direct-photon-jet-limit-mspace-trans-pol}, with its constituents defined in Eqs.~\eqref{eq:H-T-q-gamma} through \eqref{eq:K-tilde-T-ins-4}. We will use the results of Appendix~\ref{sec:direct-photon-limit-trans} in our numerical analyses in Section~\ref{sec:azimuthal-correlations}. Finally, Appendix~\ref{sec:single-inclusive-parton-cs} contains expressions for inclusive quark production in DIS at small $x_{\rm Bj}$ for different polarizations of the virtual photon. These are used in the normalization of the observable that is numerically evaluated and depicted graphically in Section~\ref{sec:azimuthal-correlations}.

\section{General considerations and outline of the computation} \label{sec:cross-section-setup}

In this section, we will discuss features of the CGC EFT framework necessary to compute the cross-section for inclusive prompt photon+quark production in $e+A$ scattering at small $x$. These considerations are quite general and apply to any inclusive process in $e+A$ collisions.

We begin with a brief description of the relevant kinematics. A relativistic energetic electron with four momentum, $k_{e}$ is scattered off an ultrarelativistic nucleus with momentum $P_{N}$, per nucleon. The scattered electron has momentum $k'_{e}$ and the invariant mass of the recoiling system is given by $W$. We will work in a frame where the nucleus is right moving and has a large `$+$' component of nucleon light cone\footnote{See Appendix~\ref{sec:conventions} for the conventions for light cone coordinates used in this work.} (LC) momentum $P_{N}^{+}$ and both the nucleon and the virtual photon exchanged in the scattering have zero transverse momentum,
\begin{equation}
    q=(-\frac{Q^{2}}{2q^{-}},q^{-},\vect{0}) \, \, \, \,  , \, \, \, \, P_{N}=(P^{+}_{N},0,\vect{0}) \, .
\end{equation}
We assume in this work that $P_{N}^{-}=\frac{M^{2}_{N}}{2P^{+}_{N}} \simeq 0$, since $P_{N}^{+}\gg M_{N}$, where $M_{N}$ is the mass of the nucleon. The virtuality $Q^{2}$ of the virtual photon is completely determined by the four momenta of the incoming and outgoing electron, $Q^{2}=-(k_{e}-k'_{e})^{2}.$ We will consider the  deeply inelastic scattering (DIS) region of  $e+A$ collisions, where $Q^{2} \gg M^{2}_{N}$ and $W^{2} \gg M^{2}_{N}$, with $W^{2}=(q+P_{N})^2 \simeq 2q^{-} \, P^{+}_{N} $ (neglecting $q^2$ relative to $2P\cdot q\equiv 2 P_N^+ q^-$). For the kinematics of interest, the largest energy scale in the process is the center-of-mass energy, $\sqrt{s}$, where $s=(k_{e}+P_{N})^2=2k^{-}_{e} \, P^{+}_{N}$
since we will consider only $x_{\rm Bj} = \frac{Q^2}{s y} \leq 10^{-2}$, with fixed inelasticity $y = \frac{P\cdot q}{P\cdot k}$. We will neglect the mass of the electron throughout the calculation. 

As discussed in the introduction, the inclusive prompt photon+quark production in DIS at small $x$ proceeds through the hadronic subprocess in which the exchanged virtual photon ($\gamma^{*}$) fluctuates into a long-lived quark-antiquark ($q\bar{q}$) dipole, which multiple scatters off the classical background field of the nucleus, and the quark or antiquark emits a photon ($\gamma$) prior or subsequent to this scattering. In order to calculate the cross-section for the production of photon+quark, we need to first compute the cross-section for $\gamma+q\bar{q}$ production  and then integrate out the phase space of the antiquark. 

The 4-momentum assignments for this process are shown in Table~\ref{tab:example}.
\begin{table*}[!htbp]
\caption{\label{tab:example}4-momentum assignments used in the calculation}
\begin{ruledtabular}
\begin{tabular}{lll}
$q$: Exchanged virtual photon & $k_{e}$: Incoming electron &\, $k'_{e}$: Outgoing electron   \\
$k$: Quark, directed outward  & $p$: Antiquark, directed outward &\, $k_{\gamma}$:  Outgoing photon \\
$l$: Quark/antiquark internal momentum to be integrated over 
\end{tabular}
\end{ruledtabular}
\end{table*}

Following~\cite{Roy:2018jxq}, we can write the amplitude for the  electron-nucleus scattering process
\begin{equation}
  e(k_{e})+A(P_{N}) \rightarrow e'(k'_{e})+q(k)+\bar{q}(p)+\gamma(k_{\gamma})+X \, ,  
  \label{eq:eA-scattering-equation}
\end{equation}
as 
\begin{align}
    \mathcal{M}^{eA \rightarrow e'q\bar{q}\gamma X} =  M^{\mu}_{\rm{lep}} \,  \frac{i\, \Pi_{\mu \alpha}(q)}{-Q^2} \,  M^{\alpha}_{\rm{had}} \, ,
    \label{eq:DIS-amp-generic}
\end{align}
where $X$ represents anything the nucleus breaks into.
Here we have  represented respectively by $M_{\rm lep}$ and $M_{\rm had}$ the amplitudes for the leptonic and hadronic subprocesses. The former can be obtained as
\begin{equation}
 M^{\mu}_{\rm{lep}}=   (-i e) \, \big[ \bar{u}(k'_{e}) \gamma^\mu u(k_{e}) \big] \, .
 \label{eq:lepton-amplitude}
\end{equation}
$\Pi_{\mu \alpha}(q)$ denotes the gauge dependent expression for the  numerator in the propagator for the virtual photon. We will specify our choice of gauge in section~\ref{sec:photon-dijet-amplitude}.

We will now decompose the amplitude in Eq.\,\eqref{eq:DIS-amp-generic} in a basis of polarization vectors of the virtual photon. One of the advantages of decomposing the amplitude in this new basis is that it allows us to examine the behavior of observables as a function of the polarization of the initial state for the hadronic reaction, i.e., the virtual photon. Although completely equivalent, this is different from the computation in Ref.~\cite{Roy:2018jxq} where the decomposition was in terms of the basis vectors in Minkowski space\footnote{In the polarization basis we also have four basis vectors, three of them being the polarization vectors, $\epsilon^{\lambda}(q)$, $\lambda=0,\pm 1$. The fourth one, which is along the four momentum $q$, gives zero upon acting on both the lepton and hadronic amplitudes by virtue of the Dirac equation and Ward identity, respectively.}.

We begin with the well known completeness relation whereby we express the quantity $\Pi_{\mu \alpha}(q)$ as a sum over the three polarization states of the virtual photon,
\begin{align}
\Pi_{\mu \alpha}(q)=    \sum_{\lambda=0,\pm 1} (-1)^{\lambda +1} \epsilon^{\lambda\, *}_\mu(q) \epsilon^{\lambda}_\alpha(q)  \, .
    \label{eq:completeness-relation}
\end{align}
We will always use $\lambda$ (and variants $\lambda'$, $\bar{\lambda}$) to indicate the polarization states of the photon (real or virtual).

Taking the modulus squared of the amplitude in Eq.\,\eqref{eq:DIS-amp-generic}, and performing the necessary averaging and sum over electron spins and final state photon polarizations, we obtain 
\begin{align}
 \frac{1}{2} \left| \mathcal{M}^{eA \rightarrow e'q\bar{q}\gamma X} \right|^2 =\frac{1}{Q^4} L^{\mu\nu} \Pi_{\mu \alpha}(q) \Pi_{\nu \beta}(q) X^{\alpha \beta}\equiv  \frac{1}{Q^4} \sum_{\lambda, \lambda'=0,\pm 1} (-1)^{\lambda+\lambda'} \mathcal{L}^{\lambda \lambda'} \mathcal{X}_{\lambda' \lambda} \, ,
 \label{eq:DIS-amp-squared-general}
\end{align}
where $L^{\mu \nu}$ and $X^{\alpha \beta}$ are the lepton and hadron tensors, respectively and we have used the completeness relation in Eq.\,\eqref{eq:completeness-relation} to define 
\begin{align}
    \mathcal{L}^{\lambda \lambda'} = L^{\mu \nu} \epsilon^{\lambda \, *}_\mu(q) \,  \epsilon^{\lambda'}_\nu(q) \, ,  \label{eq:lepton-tensor-pol-basis} \\
    \mathcal{X}^{\lambda \lambda'} = X^{\alpha \beta} \epsilon^{\lambda \, *}_\alpha(q) \, \epsilon^{\lambda'}_\beta(q) \, , \label{eq:hadron-tensor-pol-basis}
\end{align}
in the basis of polarization vectors. One can interpret these as matrix elements of two $3\times 3$ matrices $\mathcal{L}$ and $\mathcal{X}$, whose diagonal and off-diagonal elements represent the cases of having identical and non-identical polarization states of the virtual photon in the amplitude and the complex conjugate amplitude respectively.

The lepton and hadron tensors can be expressed in terms of the amplitudes for their subprocesses as\footnote{Here the short form `pols.' 
refers to the sum over polarizations of the final state photon.}
\begin{align}
    L^{\mu \nu}&= \frac{1}{2} \sum_{\rm spins} M^{\mu \dagger}_{\rm lep} \, M^{\nu}_{\rm lep}= \frac{e^{2}}{2} \text{Tr}\big[ \slashed{k'_{e}} \gamma^{\mu} \slashed{k_{e}} \gamma^{\nu} \big] \, \, , \label{eq:lepton-tensor-general} \\
    X^{\alpha \beta}&= \sum_{\rm spins, pols.} M^{\alpha \dagger}_{\rm had} \, M^{\beta}_{\rm had} \, \, . \label{eq:hadron-tensor-general}
\end{align}
We can therefore rewrite the quantity in Eq.\,\eqref{eq:hadron-tensor-pol-basis} as
\begin{align}
    \mathcal{X}^{\lambda \lambda'}= \sum_{\rm spins, pols.} \mathcal{M}^{\lambda \,  \dagger}_{\rm had} \, \mathcal{M}^{\lambda'}_{\rm had} \,  ,
     \label{eq:hadron-tensor-pol-basis-2}
    \end{align}
    where
    \begin{align}
    \mathcal{M}^{\lambda}_{\rm had}=\epsilon^{\lambda}_{\alpha} (q) \, M^{\alpha}_{\rm had}  \, ,
     \label{eq:hadron-amplitude-pol-basis}
    \end{align}
is the hadron amplitude for a given polarization state of the virtual photon. 

In the CGC EFT, the expectation value of a generic operator $\mathcal{O}$ is expressed as 
\begin{equation}
   \langle \mathcal{O} \rangle_{Y}=  \int [\mathcal{D}\rho_A] W_{Y}[\rho_A]  \, \mathcal{O}[\rho_A] \,  ,
   \label{eq:CGC-average}
 \end{equation}
 where $\mathcal{O}[\rho_A]$ is the quantum expectation value for a particular configuration $\rho_A$ of the color sources\footnote{For a general discussion of field theories with strong time dependent sources, see for  example~\cite{Gelis:2006yv,Gelis:2006cr,Gelis:2007kn} and references therein.} and $W_{Y}[\rho_A]$ is a stochastic gauge invariant weight functional representing the 
 distribution of color charge configurations at a rapidity $Y=\ln(x_{0}/x)$
 in the target\footnote{$x_0 \simeq 0.01$ refers to the initial scale separating sources from fields in the target wavefunction. $x < x_0$ is the momentum fraction of gluons in the target probed by the projectile and is determined from the kinematics. This is the scale up to which the sources must be evolved to incorporate quantum effects. While it is usually different from the Bjorken variable $xps_{\mathrm{BJ}}=Q^2/2 P\cdot q$, the two can be equated to leading logarithmic accuracy.}.
 
We can therefore write the general expression for the differential cross-section for inclusive photon+quark production in the CGC EFT as
\begin{align}
    \frac{\mathrm{d} \sigma^{eA\rightarrow e'q \gamma X}}{\mathrm{d} \Omega_{e'} \, \mathrm{d} \Omega_{q} \, \mathrm{d} \Omega_{\gamma}}= \, \int \mathrm{d} \Omega_{\bar{q}} \, \left \langle \frac{1}{2}  \left| \mathcal{M}^{eA \rightarrow e'q\bar{q}\gamma X} \right|^2 \right \rangle_{Y} \, ,
\end{align}
where $\mathrm{d} \Omega$ represents the phase space of the final state particles. For the process of interest, we have chosen to integrate over the phase space of the antiquark ($\bar{q}$). 

For the particles produced from the hadronic subprocess ($q,\bar{q}, \gamma$) we can express their invariant phase space as
\begin{equation}
    \mathrm{d} \Omega_{q}=\frac{\mathrm{d}^{2}\bm{k}_{\perp} \, \mathrm{d} \eta_{q}}{2 (2\pi)^{3}} \quad , \quad   \mathrm{d} \Omega_{\bar{q}}=\frac{\mathrm{d}^{2}\bm{p}_{\perp} \, \mathrm{d} \eta_{\bar{q}}}{2 (2\pi)^{3}} \quad , \quad   \mathrm{d} \Omega_{\gamma}=\frac{\mathrm{d}^{2}\bm{k}_{\gamma \perp} \, \mathrm{d} \eta_{\gamma}}{2 (2\pi)^{3}} \, ,
    \label{eq:phase-space-q-qbar-gamma}
\end{equation}
 where in general $\mathrm{d}\eta=\mathrm{d}v^{-}/v^{-}$, with $\eta$ the rapidity of a particle of four momentum $v$ in the frame we are working in. For the quark, this can be written\footnote{Note that we are working in the massless limit, in which the rapidity and pseudorapidity are identical.} as $
    \eta_{q} = \frac{1}{2} \ln ( k^{-}/k^{+} ) $. 
 Similar relations follow for the antiquark and the emitted photon. In our convention, the rapidity is defined to be positive in the direction of  \emph{left moving} particles, namely, moving along the `$-$' LC direction\footnote{We remind the reader that we chose the target nucleus to be right moving 
 as opposed to a convention frequently adopted in the literature  
 where the nucleus is a left mover and the rapidity (of a particle with 4-momentum $v$) is defined as $
    \eta = \frac{1}{2} \ln ( v^{+}/v^{-} ) $, which is positive along the direction of right moving particles.}.

The CGC averaged (see Eq.\,\eqref{eq:CGC-average}) square of the amplitude in  Eq.\,\eqref{eq:DIS-amp-squared-general} can be written in the basis of polarization vectors as
\begin{equation}
  \left \langle \frac{1}{2}  \left| \mathcal{M}^{eA \rightarrow e'q\bar{q}\gamma X} \right|^2 \right \rangle_{Y}= \frac{1}{Q^4}  \sum_{\lambda, \lambda'=0,\pm 1} (-1)^{\lambda+\lambda'} \mathcal{L}^{\lambda \lambda'} \big \langle \mathcal{X}_{\lambda' \lambda} \big \rangle_{Y} \, .
  \label{eq:CGC-amp-squared-photon-dijet}
\end{equation}
 
We can express the phase space for the outgoing electron as
\begin{equation}
    \mathrm{d} \Omega_{e'}= \frac{\mathrm{d}W^{2} \,  \mathrm{d}Q^{2} \, \mathrm{d} \phi_{e'}}{4s(2\pi)^{3}} \, ,
    \label{eq:outgoing-electron-phase-space}
\end{equation}
where $\phi_{e'}$ is the azimuthal angle of the outgoing electron's 3-momentum, $\vec{k'_{e}}$. By evaluating the different matrix elements $\mathcal{L}^{\lambda \lambda'}$, using Eqs.~\eqref{eq:lepton-tensor-pol-basis} and \eqref{eq:lepton-tensor-general}, it is easy to show that integrating over $\phi_{e'}$ projects out only the products of the diagonal elements in the sum over polarizations of $\gamma^{*}$ appearing in the r.h.s of Eq.\,\eqref{eq:CGC-amp-squared-photon-dijet}. From the lepton side, this means that only the elements $\mathcal{L}^{0 \, 0} $ and $\mathcal{L}^{ \pm 1 \, \pm 1}$ enter the expression for the $\phi_{e'}$-integrated cross-section. 

We can now write the general expression for the differential cross-section for inclusive $\gamma+q$ production in $e+A$ DIS at small $x$ as
\begin{align}
    \frac{d \sigma ^{eA \rightarrow e'q\gamma X}}{\mathrm{d}W^{2} \mathrm{d}Q^{2} \mathrm{d}^{2} \vect{k} \mathrm{d}^{2} \vect{k_{\gamma}}    \mathrm{d} \eta_{q}  \mathrm{d} \eta_{\gamma}}&=f_{L}(Q^{2},W^{2}) \, \frac{d \sigma ^{\gamma^{*}_{L} A \rightarrow q \gamma X}}{\mathrm{d}^{2} \vect{k}  \mathrm{d}^{2} \vect{k_{\gamma}}    \mathrm{d} \eta_{q} \mathrm{d} \eta_{\gamma}}  +f_{T}(Q^{2},W^{2}) \sum_{\lambda=T=\pm 1} \,  \frac{d \sigma ^{\gamma^{*}_{\lambda} A \rightarrow q \gamma X}}{\mathrm{d}^{2} \vect{k}  \mathrm{d}^{2} \vect{k_{\gamma}}    \mathrm{d} \eta_{q} \mathrm{d} \eta_{\gamma}}   \, ,
    \label{eq:diff-cs-gammajet-general}
\end{align}
where the leptonic contribution is contained in the longitudinally and transversely polarized photon fluxes  defined respectively as
\begin{align}
    f_{L}(Q^{2},W^{2})&=\frac{y}{16\pi^2 Q^4 s} \, \mathcal{L}^{0 \, 0}= \frac{\alpha_{\rm em}}{\pi Q^{2}sy} (1-y) \, , \nonumber \\
    f_{T}(Q^{2},W^{2}) &=\frac{y}{16\pi^2 Q^4 s} \, \mathcal{L}^{\pm 1 \, \pm 1}= \frac{\alpha_{\rm em}}{4\pi Q^{2}sy} [1+(1-y)^{2}] \, .   \label{eq:photon-fluxes}
\end{align}
These are constituted, respectively,  of the components $\mathcal{L}^{0 \, 0}$ and $\mathcal{L}^{\pm 1 \, \pm 1}$ of the lepton tensor defined by Eqs.~\eqref{eq:lepton-tensor-pol-basis} and \eqref{eq:lepton-tensor-general}. The dependence on $W^2$ enters in the above quantities through the inelasticity variable $y$ we defined previously which can be written (neglecting the mass of the nucleon) as
\begin{equation}
    y=\frac{W^{2}+Q^{2}}{s} \, .
    \label{eq:inelasticity-DIS}
\end{equation}
Since  we are interested in the kinematics $s, W^{2} \gg Q^{2} $, we have $y \sim W^{2}/s$. 

In Eq.\,\eqref{eq:diff-cs-gammajet-general}, we have separated the leptonic contribution to the  $\gamma+q$ production cross-section from the hadronic part. The hadronic contribution to the cross-section is contained in the quantity,  $\mathrm{d}\sigma^{\gamma_{\lambda}^{*}A\rightarrow q\gamma X}$, which represents the $\gamma^{*}+A$ scattering for a given polarization $\lambda=L(0),T(\pm 1)$ of the virtual photon\footnote{In what follows we will always sum over the two transverse polarizations $T(\pm 1$), instead of averaging as typically done.}. In terms of the cross-section for $\gamma+q\bar{q}$ production in the virtual photon-nucleus scattering, this can be written as
\begin{equation}
    \frac{d \sigma ^{\gamma^{*}_{\lambda} A \rightarrow q \gamma X}}{\mathrm{d}^{2} \vect{k}  \mathrm{d}^{2} \vect{k_{\gamma}}    \mathrm{d} \eta_{q} \mathrm{d} \eta_{\gamma}} = \int \mathrm{d}^{2} \vect{p} \mathrm{d}\eta_{\bar{q}} \, \frac{d \sigma ^{\gamma^{*}_{\lambda} A \rightarrow q\bar{q} \gamma X}}{\mathrm{d}^{2} \vect{k} \mathrm{d}^{2} \vect{p}  \mathrm{d}^{2} \vect{k_{\gamma}}    \mathrm{d} \eta_{q} \, \mathrm{d} \eta_{\bar{q}} \, \mathrm{d} \eta_{\gamma}} \, ,
    \label{eq:virtual-photon-hadron-scatt-cs-general}
\end{equation}
where 
\begin{align}
    \frac{d \sigma^{\gamma_\lambda^* A \rightarrow q\bar{q}\gamma X}}{\mathrm{d}^{2} \vect{k} \mathrm{d}^{2} \vect{p}  \mathrm{d}^{2} \vect{k_{\gamma}}    \mathrm{d} \eta_{q} \, \mathrm{d} \eta_{\bar{q}} \, \mathrm{d} \eta_{\gamma}} = \frac{1}{8 (2\pi)^{9}}    \big  \langle  \mathcal{X}_{\lambda \lambda} \big  \rangle_{Y}  \, . \label{eq:diff-cs-gamma-dijet-definition}
\end{align}
The CGC averaged hadron tensor can be written in the basis of polarization vectors as
\begin{align}
   \langle \mathcal{X}_{\lambda \lambda} \rangle_{Y}= \left \langle \sum_{\rm spins, pols.} \mathcal{M}^{\lambda \,  \dagger}_{\rm had} \, \mathcal{M}^{\lambda}_{\rm had} \right \rangle_{Y} \,   , 
    \label{eq:hadron-tensor-pol-basis-averaged}
\end{align}
where $\mathcal{M}^{\lambda}_{\rm had}$ (see Eq.\,\eqref{eq:hadron-amplitude-pol-basis}) is the amplitude for the hadronic subprocess $\gamma^{*}A\rightarrow q\bar{q}\gamma X$ for a given polarization $\lambda$ of the virtual photon.

We summarize below the steps necessary to obtain the cross-section for $\gamma+q$ production in $e+A$ DIS at small $x$:
 \begin{enumerate}
     \item Compute the total amplitude, $\mathcal{M}_{\rm had}^{\lambda}$ for $\gamma+q\bar{q}$ production in virtual photon-nucleus scattering.
     \item  Compute the differential cross-section for $\gamma+q\bar{q}$ production as defined in Eq.\,\eqref{eq:diff-cs-gamma-dijet-definition} using the expression for the amplitude.
     \item Integrate over the phase space of the antiquark ($\bm{p}_{\perp}$ and $\eta_{\bar{q}}$) to get the differential cross-section for $\gamma+q$ production as defined in Eq.\,\eqref{eq:virtual-photon-hadron-scatt-cs-general}.
 \end{enumerate}
Over the next sections, we will sequentially demonstrate the realization of these steps.

\section{$\gamma+q\bar{q}$ production amplitude in $\gamma^*+A$ scattering at leading order} \label{sec:photon-dijet-amplitude}

We will work here in the basis of polarization vectors of the virtual photon which was discussed in the previous section. To further simplify the computation, and with a view to extending the current work to higher orders in $\alpha_{s}$, we choose to work in the light cone gauge $A^{-}=0$ using momentum space Feynman rules (see Appendix~\ref{sec:Feynman rules}). 

We choose for the polarization vectors of the virtual photon that we introduced in the previous section, the following convention:
\begin{align}
    \epsilon^{\lambda=0}(q) &= \left( \frac{Q}{q^-},0,\vect{0} \right) \, , \label{eq:pol-vect-long} \\
    \epsilon^{\lambda=\pm 1}(q) &= \left(0,0,\vect{\epsilon}^{\pm 1} \right),\ \ \ \ \text{where} \quad  \vect{\epsilon}^{\pm 1} = \frac{1}{\sqrt{2}} \left(1,\pm i \right) \, . \label{eq:pol-vect-trans}
\end{align}
It is easy to check that these vectors satisfy the completeness relation in Eq.\,\eqref{eq:completeness-relation} as
\begin{align}
 \sum_{\lambda=0,\pm 1} (-1)^{\lambda +1} \epsilon^{\lambda\, *}_\mu(q) \epsilon^{\lambda}_\alpha(q) =\Pi_{\mu \alpha}(q)= -g_{\mu \alpha}+\frac{q_{\mu} \, n_{\alpha}+q_{\alpha}n_{\mu}}{q \cdot n} \, , \qquad n_{\mu}=\delta_{\mu -} \, , 
\end{align}
where we have provided an explicit expression for  the numerator $\Pi_{\mu \alpha}$  of the photon propagator in $A^-=0$ gauge.
\begin{figure}[!htbp]
    \centering
    \includegraphics[scale=1.2]{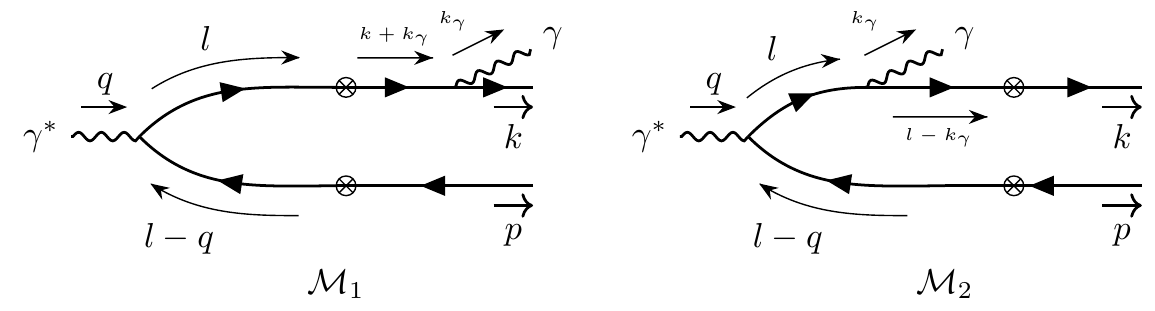}
    \includegraphics[scale=1.2]{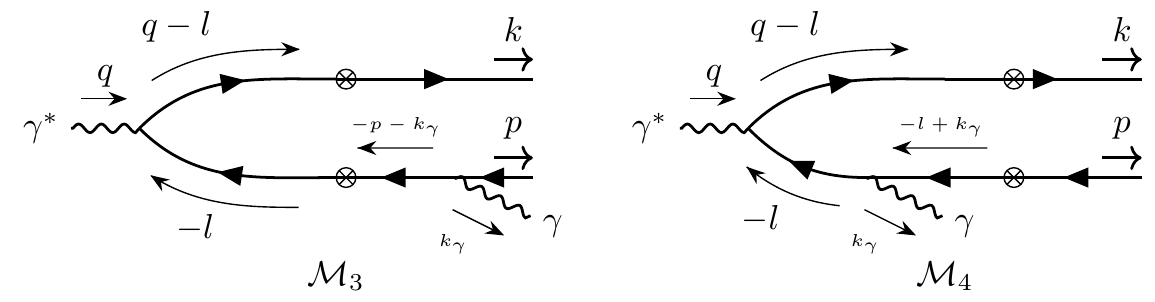}
    \label{fig:my_label}
    \caption{Diagrams contributing to the LO amplitude of $\gamma+q\bar{q}$ production at small $x$ in the CGC EFT. Top (bottom) diagrams correspond to the photon emission from the quark (antiquark). We explicitly show the choice of internal momenta. The crossed circles represent the effective vertices that contain all possible eikonal interactions with the background classical field of the nucleus.} \label{fig:photon_jet_LO_diagrams}
\end{figure}

 As shown in Fig. \ref{fig:photon_jet_LO_diagrams}, at leading order in the CGC power counting, there are four contributions to the amplitude~\cite{Roy:2018jxq}, 
\begin{equation}
    \mathcal{M}_{\rm had}^{\lambda} \, [\rho_A]=\sum_{r=1}^{4} \mathcal{M}_{r}^{\lambda} \, [\rho_A] \, .
    \label{eq:hadron-amplitude-total}
\end{equation}
The Feynman graphs along each row in Fig.~\ref{fig:photon_jet_LO_diagrams}  differ from each other in terms of the photon emission relative to the scattering off the background classical field by the quark-antiquark dipole. Also, the contributions from $\mathcal{M}_3$ and $\mathcal{M}_4$ are obtained from those of $\mathcal{M}_1$ and $\mathcal{M}_2$, respectively, by interchanging the quark and antiquark lines.

The momentum space expression for the amplitude of the hadronic subprocess labeled $\mathcal{M}_1$ can be written as\footnote{
The polarization vectors for the outgoing  photon are labeled by the index $\bar{\lambda}=\pm 1$. Since we will sum over $\bar{\lambda}$ in the amplitude squared, we do not put this label on the amplitude in the l.h.s of Eq.\,\eqref{eq:amp-LO-1-general}.}
\begin{align}
    \mathcal{M}^{\lambda}_{1;ij} [\rho_A]=\int_{l} \bar{u}(k) (-ieq_{f}) \slashed{\epsilon}^{\bar{\lambda} \, *} (k_{\gamma}) S^{0}_{ik}(k+k_{\gamma}) \mathcal{T}_{km}(k+k_{\gamma},l) S^{0}_{mn}(l) (-ieq_{f}) \slashed{\epsilon}^{\lambda}(q) S^{0}_{np}(l-q) \mathcal{T}_{pj}(l-q,-p) v(p) \, ,
    \label{eq:amp-LO-1-general}
\end{align}
where $i$,$j$, are color indices.

Further, $q_{f}$ represents the fractional charge of the quark/antiquark in units of the elementary charge $e$, the free (massless) quark Feynman propagator is given by
\begin{equation}
   S^{0}_{ij}(p)=\frac{i\slashed{p}}{p^{2}+i\epsilon} \, \delta_{ij} \,  ,
   \label{eq:quark-mom-propagator}
\end{equation}
and the all-twist resummed effective vertices that enter the momentum space dressed quark and antiquark propagators are represented by 
\begin{equation}
    \mathcal{T}_{ij}(p,p')=2\pi \, \delta(p^{-}-p'^{-}) \gamma^{-} \, \text{sign}(p^{-}) \int_{\bm{z}_{\perp}} e^{-i(\vect{p}-\bm{p'}_{\perp}) \cdot \bm{z}_{\perp}} \tilde{U}^{\text{sign}(p^{-})}_{ij}(\bm{z}_{\perp}) \, .
    \label{eq:quark-effective-vertex}
\end{equation}
Here $\int_{\vect{z}}$ is shorthand for $\int \mathrm{d}^2 \vect{z}$ and $\tilde{U}^{+1}$, $\tilde{U}^{-1}$ denote lightlike Wilson lines in the fundamental and antifundamental representation of $SU(N_{\mathrm{c}})$ respectively. Their labels $i$ and $j$ represent color indices in the fundamental representation of $SU(N_{c})$ and sign$(p^{-})$ is the sign function. The latter takes values of $\pm 1$, respectively for the case of quark and antiquark. From the explicit expression provided in Eq.\,\eqref{eq:Wilson-line-fund}, one can see that the functional dependence on the color charge density  $\rho_{A}$ enters the hadron amplitude in Eq.\,\eqref{eq:hadron-amplitude-total} through  $\tilde{U}$.   Note that the effective vertices on the quark and antiquark lines are proportional respectively to $\tilde{U}(\vect{x})$ and $\tilde{U}^{\dagger}(\vect{y})$, where $\bm{x}_{\perp}$ and $\bm{y}_{\perp}$ are the respective transverse coordinates.

The effective vertices given by Eq.\,\eqref{eq:quark-effective-vertex} represent all possible scatterings of the quark-antiquark dipole with the classical field of the nucleus \emph{including} the case of ``no scattering"~\cite{Roy:2018jxq} (see also Appendix~\ref{sec:Feynman rules}). Since we are interested in the physical amplitude which should necessarily involve interactions with the nucleus, we have to subtract from the expression for the amplitude in Eq.\,\eqref{eq:amp-LO-1-general} (and similarly for the amplitudes of the other processes), the possibility of ``no scattering" with the background field. This contribution is obtained by setting $\tilde{U}=\mathds{1}$ in the expressions for $\mathcal{T}$'s that enter the amplitudes for each process in Fig.~\ref{fig:photon_jet_LO_diagrams}.

Subtracting the ``no scattering" contribution from the amplitude in Eq.\,\eqref{eq:amp-LO-1-general}, and using the expressions in Eqs.~\eqref{eq:quark-mom-propagator} and \eqref{eq:quark-effective-vertex}, we can factor out the color structure and the phase of the quark and the antiquark to write the amplitude for $\mathcal{M}_1$ as
\begin{align}
    \mathcal{M}^{\lambda}_{1;ij} [\rho_A] =(e q_f)^2  \int_{\vect{x},\vect{y}} e^{-i \vect{k} \cdot \vect{x}-i\vect{p} \cdot \vect{y}} \left[\tilde{U}(\vect{x}) \tilde{U}^\dagger (\vect{y}) -\mathds{1} \right]_{ij} \, \mathcal{N}_{1}^{\lambda}(\vect{x},\vect{y}) \, , 
\end{align}
where
\begin{align}
     \mathcal{N}_1^{\lambda}(\vect{x},\vect{y})&= e^{-i \vect{k_\gamma} \cdot \vect{x}} (2\pi)^2 \int_{l} e^{i \vect{l}\cdot \vect{r}}  \delta(k^- + k_\gamma^- -l^-)  \delta(l^- - q^- + p^-) \frac{-i N_1^{\lambda}(l)}{\left[l^2+i \epsilon \right]\left[(l-q)^2 + i \epsilon \right]} \label{eq:N1full} \, .
\end{align}
Above we have introduced $\vect{r}=\vect{x}-\vect{y}$, and the numerator is given by,
\begin{align}
     N_1^{\lambda}(l) &= \frac{1}{2k \cdot k_\gamma}\bar{u}(k) \slashed{\epsilon}^{\bar{\lambda}*}(k_\gamma)(\slashed{k}+\slashed{k}_\gamma) \gamma^- \slashed{l} \slashed{\epsilon}^{\lambda}(q)(\slashed{l}-\slashed{q})\gamma^-v(p) \label{eq:N1def} \, .
\end{align}
The integration over $l^{-}$ can be performed trivially to yield an overall momentum conserving delta function: $\delta(q^{-}-k^{-}-p^{-}-k^{-}_{\gamma})$. This is a common feature of all CGC computations performed within the eikonal approximation\footnote{One can also think of this~\cite{Gelis:2002ki} as a manifestation of Fermi's golden rule; the invariance of the target nucleus with respect to the LC time $x^{+}$ implies that the projectile `$-$' momentum is conserved.}.

We next perform the integration over $l^{+}$ using  Cauchy's theorem of residues for complex contour integration. The numerator given by Eq.\,\eqref{eq:N1def} is independent of $l^{+}$ indicating that the value of the integral over $l^{+}$ will be entirely given by the residue theorem, in the limit of the contour radius going to infinity. The location of the $l^{+}$ poles, in this case, are determined by the signs of the external longitudinal momenta, $k^{-}$,  $p^{-}$ and $k_{\gamma}^{-}$. After these steps, we will finally express $\mathcal{N}_{1}^{\lambda}$ in terms of a transverse integration over $\vect{l}$. The amplitudes for the remaining three diagrams can be obtained following the same routine.

We can factor out the overall momentum conserving delta function to rewrite the total amplitude from the four processes in Fig.~\ref{fig:photon_jet_LO_diagrams} as
\begin{equation}
    \mathcal{M}^{\lambda}_{\rm had} \,  [\rho_A]= 2\pi \, \delta(q^{-}-k^{-}-p^{-}-k^{-}_{\gamma}) \, \mathcal{A}^{\lambda}_{\rm had} \,  [\rho_A] \, , 
    \label{eq:reduced-amplitude}
\end{equation}
where
\begin{equation}
  \mathcal{A}^{\lambda}_{\rm had} \,  [\rho_A]=  (e q_f)^2  \int_{\vect{x},\vect{y}} e^{-i \vect{k} \cdot \vect{x}-i\vect{p} \cdot \vect{y}} \left[\tilde{U}(\vect{x}) \tilde{U}^\dagger (\vect{y}) -\mathds{1} \right] \sum_{r=1}^{4} \mathcal{A}_{r}^{\lambda} (\vect{x},\vect{y})  \, .
  \label{eq:amplitude-total-LO}
\end{equation}
For the process $\mathcal{M}_{1}$, we can write $ \mathcal{A}_1(\vect{x},\vect{y})$ as 
\begin{equation}
      \mathcal{A}_1^{\lambda}(\vect{x},\vect{y})= e^{-i \vect{k_\gamma} \cdot \vect{x}} \int_{\vect{l}} \frac{A_1^{\lambda}(\vect{{l}})  \,  e^{i \vect{l}\cdot \vect{r}}}{\left( \vect{l}^2 + z_{\bar{q}} (1-z_{\bar{q}}) Q^2 \right)} \, ,  \label{eq:A1full} 
\end{equation}
where 
\begin{equation}
     A_1^{\lambda}(\vect{l}) = \frac{z_q}{q^-}  \left[ \bar{u}(k)  \left( \vect{\epsilon}^{\bar{\lambda}*,i} - \frac{z_\gamma}{2 z_q} \gamma^i \gamma^j \vect{\epsilon}^{\bar{\lambda}*,j} \right) \gamma^- \slashed{l} \slashed{\epsilon}^\lambda (q) (\slashed{l}-\slashed{q}) \gamma^- v(p) \right]\ \frac{z_q \vect{k_\gamma}^i -z_\gamma \vect{k}^i}{\left| z_q \vect{k_\gamma} -z_\gamma \vect{k} \right|^2} \label{eq:A1} \, .
\end{equation}
We introduced in this expression the ratios of the outgoing momenta to the component $q^{-}$ of the incoming virtual photon momentum\footnote{For the outward directed external momenta, we have $\{k^{-}, p^{-}, k_{\gamma}^{-} \} > 0$.},
\begin{equation}
z_{q}=\frac{k^{-}}{q^{-}}, \quad z_{\bar{q}}=\frac{p^{-}}{q^{-}}, \quad z_{\gamma}=\frac{k^{-}_{\gamma}}{q^{-}} \, .
\end{equation}
In deriving Eq.\,\eqref{eq:A1} from Eq.\,\eqref{eq:N1def}, we have made use of the Dirac equation: $\bar{u}(k) \slashed{k}=0$, the photon transversality condition, $k_{\gamma} \cdot \epsilon^{\bar{\lambda}}(k_{\gamma})=0$ and the Clifford algebra identity given by Eq.\,\eqref{eq:dirac-algebra}. 

There are a couple of advantages in writing the amplitude in the form given by Eq.\,\eqref{eq:A1}. The first one is apparent when one expands the Dirac structure in Eq.\,\eqref{eq:A1} as
\begin{equation}
  \gamma^- \slashed{l} \slashed{\epsilon}^\lambda (q) (\slashed{l}-\slashed{q}) \gamma^-= -4z_{\bar{q}} (1-z_{\bar{q}}) (q^{-})^{2} \, \gamma^{-} \, \epsilon^{\lambda , \, +}(q) +2q^{-} \big[ (1-z_{\bar{q}}) \, \gamma^{k} \,  \gamma^{r}-z_{\bar{q}} \, \gamma^{r} \, \gamma^{k} \big] \, \gamma^{-} \bm{l}_{\perp}^{r}    \, \bm{\epsilon}^{\lambda, \, k}_{\perp} \, . 
  \label{eq:gamma-structure-A1}
\end{equation}
For the choice of polarization vectors in Eqs.~\eqref{eq:pol-vect-long} and \eqref{eq:pol-vect-trans}, one can easily see that the amplitude naturally lends itself to a decomposition in the basis of  polarization vectors. For longitudinal polarization of the virtual photon, only the first term on the r.h.s of Eq.\,\eqref{eq:gamma-structure-A1} contributes to the amplitude, whereas only the second term survives for transverse polarization of $\gamma^{*}$. This feature aids in the computation of the squared amplitude in this basis.

The second advantage of Eq.\,\eqref{eq:A1} becomes manifest when we square this amplitude and sum over the polarizations of the final state photon. In this process, one generates from the numerator of the square of Eq.\,\eqref{eq:A1} a term $\vert z_{q} \bm{k}_{\gamma \perp}-z_{\gamma} \bm{k}_{\perp} \vert^{2}$ that cancels an identical  factor in the denominator. This makes the (collinear) singularity for $ z_{q} \bm{k}_{\gamma \perp}=z_{\gamma} \bm{k}_{\perp}$ transparent in the resulting expression.

Through a similar exercise used to derive  the amplitude $\mathcal{A}_1^{\lambda}$ in Eq.\,\eqref{eq:A1}, we can write the amplitude for the process labeled $\mathcal{M}_{2}$ in Fig.~\ref{fig:photon_jet_LO_diagrams} as
\begin{align}
    \mathcal{A}_2^{\lambda}(\vect{x},\vect{y})=&\  e^{-i \vect{k_\gamma} \cdot \vect{x}} \int_{\vect{l}} \frac{-A_{2,\rm{reg}}^{\lambda}(\vect{l}) \,  e^{i \vect{l}\cdot \vect{r}}}{\left(\vect{l}^2 + z_{\bar{q}} (1-z_{\bar{q}}) Q^2 \right) \left(Q^2 + \frac{\vect{k_\gamma}^2}{z_\gamma}  + \frac{(\vect{l}-\vect{k_\gamma})^2}{z_q} + \frac{\vect{l}^2}{z_{\bar{q}}} \right)} \nonumber \\
    +&\ e^{-i \vect{k_\gamma} \cdot \vect{x}} \int_{\vect{l}} \frac{-A_{2,\rm{ins}}^{\lambda}(\vect{l}) \,  e^{i \vect{l}\cdot \vect{r}}}{\left(Q^2 + \frac{\vect{k_\gamma}^2}{z_\gamma}  + \frac{(\vect{l}-\vect{k_\gamma})^2}{z_q} + \frac{\vect{l}^2}{z_{\bar{q}}} \right)} \, ,
    \label{eq:A2full}
\end{align}
where
\begin{align}
    A_{2,\rm{reg}}^{\lambda}(\vect{l}) &=\frac{1}{(q^-)z_q z_\gamma } \left[  \bar{u}(k)  \left( \vect{\epsilon}^{\bar{\lambda}*,i} + \frac{z_\gamma}{2(1- z_{\bar{q}})} \gamma^j \gamma^i \vect{\epsilon}^{\bar{\lambda}*,j} \right) \gamma^- \slashed{l} \slashed{\epsilon}^{\lambda}(q) (\slashed{l}-\slashed{q})\gamma^- v(p) \right]  \left[(1-z_{\bar{q}}) \vect{k_\gamma}^i -z_\gamma \vect{l}^i \right] \, , \label{eq:A2reg}\\
    A_{2,\rm{ins}}^{\lambda}(\vect{l})&=\frac{1}{(1-z_{\bar{q}})}  \left[\bar{u}(k) \gamma^i \slashed{\epsilon}^{\lambda}(q) \gamma^- v(p)\right] \vect{\epsilon}^{\bar{\lambda} *,i} \label{eq:A2ins} \, .
\end{align}
In writing Eq.\,\eqref{eq:A2full}, we have split the amplitude into two contributions labeled `regular' (reg) and `instantaneous' (ins).  The Dirac structure for the `regular' term in Eq.\,\eqref{eq:A2reg} is similar to that of $A_1^{\lambda}(\vect{l})$ in Eq.\,\eqref{eq:A1}. The nomenclature for the second contribution follows from light-cone perturbation theory, in which the `instantaneous' term corresponds to the emission of the photon from the $q\bar{q}$ splitting vertex. This can be seen from the Dirac structure of Eq.\,\eqref{eq:A2ins} where there is no propagator between the two photon vertices.

Because of their similar Dirac structure, the decomposition in Eq.\,\eqref{eq:gamma-structure-A1} also applies to the `regular' term in Eq.\,\eqref{eq:A2reg}. Using the identity
\begin{equation}
    \gamma^{i} \slashed{\epsilon}^{\lambda}(q) \gamma^{-} = -\gamma^{i} \gamma^{k} \gamma^{-}  \, \bm{\epsilon}_{\perp}^{\lambda,\, k}  \, ,
    \label{eq:gamma-structure-inst}
\end{equation}
one may verify that the `instantaneous' contribution is non-zero only for transverse polarizations of the virtual photon. Finally, there is no collinearly  singular structure in $\mathcal{A}_2$, because the photon is emitted by the (off-shell) quark before it scatters off the shock wave.

In this section, we have presented the expressions for the amplitudes of diagrams $\mathcal{M}_{1}$ and $\mathcal{M}_{2}$. For the momentum assigments shown in Fig.~\ref{fig:photon_jet_LO_diagrams}, the expressions for $\mathcal{A}_{3}^{\lambda} (\vect{x},\vect{y})$ and $\mathcal{A}_{4}^{\lambda} (\vect{x},\vect{y})$ can be obtained from the corresponding expressions for  $\mathcal{A}_{1}^{\lambda} (\vect{x},\vect{y})$ and $\mathcal{A}_{2}^{\lambda} (\vect{x},\vect{y})$, respectively, by implementing the following replacements in Eqs.~\eqref{eq:A1full} and \eqref{eq:A2full}: $  \vect{x}\leftrightarrow\vect{y} \, \, , p \leftrightarrow k $
(and $v(p) \leftrightarrow \bar{u}(k)$), an overall `$-$' sign, and reversing the order of the spinor structures.

 \section{Differential cross-section for $\gamma+q\bar{q}$ production in $\gamma^{*}+A$ scattering} \label{sec:photon-dijet-cross-section}
 
 In order to obtain the differential cross-section for $\gamma+q\bar{q}$ production, we have to multiply the net amplitude from the four processes in Fig.~\ref{fig:photon_jet_LO_diagrams}, as written in Eq.\,\eqref{eq:reduced-amplitude}, with the complex conjugate amplitude, and then perform the CGC averaging over the color sources. The seemingly problematic $(2\pi) \delta(0^{-})$ infinite  factor that arises from the square of $ (2\pi) \, \delta(q^{-}-k^{-}-p^{-}-k^{-}_{\gamma})$ (note that $q^{-}=k^{-}_{e}-k'^{-}_{e})$ in Eq.\,\eqref{eq:reduced-amplitude} can be fixed by using a properly normalized wavepacket description~\cite{Gelis:2002ki,Roy:2018jxq} instead of a plane wave for the incoming electron. This brings in a factor of $1/2k_{e}^-$. We can represent this schematically as 
 \begin{equation}
\big[  (2\pi) \, \delta(q^{-}-k^{-}-p^{-}-k^{-}_{\gamma})  \big]^2 \rightarrow \frac{1}{2k^{-}_{e}} \,  (2\pi) \, \delta(q^{-}-k^{-}-p^{-}-k^{-}_{\gamma})= \frac{q^{-}}{k^{-}_{e}}  \, \frac{1}{2q^{-}} \,  (2\pi) \, \delta(q^{-}-k^{-}-p^{-}-k^{-}_{\gamma}) \, .
 \end{equation}
In writing the cross-section for $\gamma+q$ production in $e+A$ DIS in Eq.\,\eqref{eq:diff-cs-gammajet-general}, we have absorbed the factor of $q^{-}/k^{-}_{e}=y$ (see Eq.\,\eqref{eq:inelasticity-DIS}) in the definition of the longitudinally and transversely polarized photon fluxes defined in Eqs.~\eqref{eq:photon-fluxes}. The unpolarized cross-section for inclusive $\gamma+q\bar{q}$ production in virtual photon-nucleus collisions, defined by Eq.\,\eqref{eq:hadron-tensor-pol-basis-averaged}, can therefore be written as
\begin{align}
    \frac{d \sigma^{\gamma_\lambda^* A \rightarrow q\bar{q}\gamma X}}{\mathrm{d}^{2} \vect{k} \mathrm{d}^{2} \vect{p}  \mathrm{d}^{2} \vect{k_{\gamma}}    \mathrm{d} \eta_{q} \, \mathrm{d} \eta_{\bar{q}} \, \mathrm{d} \eta_{\gamma}} = \frac{1}{8 (2\pi)^{9}} \, \frac{1}{2q^{-}} (2\pi) \delta(q^- - k^- -p^- - k_\gamma^-)    \big  \langle   \tilde{\mathcal{X}}_{\lambda \lambda} \big  \rangle_{Y}  \, , \label{eq:diff-cs-gamma-dijet-general}
\end{align}
where $\big  \langle   \tilde{\mathcal{X}}_{\lambda \lambda} \big  \rangle_{Y} $ can be written in terms of the reduced amplitudes (see Eq.\,\eqref{eq:amplitude-total-LO}) for the four diagrams in Fig.~\ref{fig:photon_jet_LO_diagrams} as
\begin{align}
  \big  \langle   \tilde{\mathcal{X}}_{\lambda \lambda} \big  \rangle_{Y}&= \left \langle \sum_{\rm spins, pols.} \mathcal{A}^{\lambda \, \dagger }_{\rm had} [\rho_{A}] \, \mathcal{A}^{\lambda}_{\rm had} [\rho_{A}] \right \rangle_{Y} \, \nonumber \\
 &= N_{\mathrm{c}} (e q_f)^4 \int_{\vect{x},\vect{y},\vect{x'},\vect{y'}} \!\!\!\!\!\!\!\!\!\!\!\!\!\!\!\!\!\!\!\!\!\!\!\!\!\!\! e^{-i \vect{k} \cdot (\vect{x} - \vect{x}')-i \vect{p} \cdot (\vect{y} - \vect{y}')} \,  \Xi(\vect{x},\vect{y};\vect{y}',\vect{x}' \vert Y) \sum_{\text{spins}, \text{pols.}}  \sum_{m,n=1}^{4}  \mathcal{A}^{\lambda \, \dagger}_m(\vect{x}',\vect{y}') \mathcal{A}_n^{\lambda} (\vect{x},\vect{y}) \, .
  \label{eq:modified-hadron-tensor-gamma-dijet}
\end{align}
Here and henceforth, counterparts of the coordinates and internal momenta in the Hermitian conjugates of the amplitudes carry prime labels. In the second line of Eq.\,\eqref{eq:modified-hadron-tensor-gamma-dijet}, the first sum is over the helicities of the Dirac spinors and polarizations of the final state photon, while the second sum runs over the four diagrams in Fig.~\ref{fig:photon_jet_LO_diagrams} giving a total of 16 contributions at the level of the cross-section.  These sums generate traces over the gamma matrices contained in the  expressions given by Eqs.~\eqref{eq:A1}, \eqref{eq:A2reg} \eqref{eq:A2ins} and their $q\leftrightarrow \bar{q}$ counterparts. Only 6 of these contributions are independent; the remaining 10 can be obtained simply by complex conjugation and $q \leftrightarrow \bar{q}$ interchange. These perturbatively computable pieces constitute the process dependent LO coefficient function or in the language of Regge theory, the LO `impact factor" for $\gamma+q\bar{q}$ production. The impact factor for photon+dijet production in $e+A$ DIS at small $x$ was recently computed to NLO~\cite{Roy:2019cux,Roy:2019hwr}.

Similarly, the color traces over the Wilson lines contained in the squared amplitude  and the subsequent CGC averaging give rise to the following combination of dipole and quadrupole correlators:
\begin{align}
    \Xi(\vect{x},\vect{y};\vect{y'},\vect{x'} \vert Y) = 1-S^{(2)}_{Y}(\vect{x},\vect{y})-S^{(2)}_{Y}(\vect{y'},\vect{x'}) + S^{(4)}_{Y}(\vect{x},\vect{y};\vect{y'},\vect{x'}) \, , \label{eq:color-structure-LO}
    \end{align}
   where the dipole and quadrupole Wilson line correlators are defined at a rapidity $Y$ as
    \begin{align}
    S^{(2)}_{Y}(\vect{x},\vect{y}) &= \frac{1}{N_{\mathrm{c}}} \left \langle \Tr(\tilde{U}(\vect{x}) \tilde{U}^\dagger(\vect{y})) \right \rangle_{Y} \, , \label{eq:dipole} \\
    S^{(4)}_{Y}(\vect{x},\vect{y};\vect{y}',\vect{x}') &= \frac{1}{N_{\mathrm{c}}} \left \langle \Tr(\tilde{U}(\vect{x}) \tilde{U}^\dagger(\vect{y}) \tilde{U}(\vect{y}') \tilde{U}^\dagger(\vect{x}')) \right \rangle_{Y} \, . \label{eq:quadrupole}
\end{align}
The CGC averaging inherent in these gauge invariant objects encodes non-trivial correlations between the gauge fields contained in the Wilson lines, thereby representing the coherent many-body gluodynamics characteristic of the saturation regime of QCD. 

These dipole and quadrupole Wilson line correlators appear in a variety of observables for both $e+A$ and $p+A$ collisions and can be thought of as building blocks of high energy QCD. Although intrinsically non-perturbative, the energy or rapidity evolution of these correlators can be derived using  perturbative techniques in the CGC EFT, resulting in the well-known JIMWLK\footnote{JIMWLK is an acronym for Jalilian-Marian, Iancu, McLerran, Weigert, Leonidov, Kovner.}~\cite{JalilianMarian:1997jx,JalilianMarian:1997dw,Weigert:2000gi,Iancu:2001ad,Iancu:2000hn,Ferreiro:2001qy}  or BK renormalization group equations. However, one needs to adopt a model to describe the non-perturbative initial distribution of the sources that enters the CGC averaging procedure. A simple and well-known model, which is valid for a large nucleus and moderate $x\sim 0.01$, is the McLerran-Venugopalan (MV)  model~\cite{McLerran:1993ni,McLerran:1993ka,McLerran:1994vd}. In this model, the large $x$ “valence”
charges are described as static  sources on the light cone, which are Gaussian distributed with 
\begin{equation}
    \langle \rho_{A}^{a} (x^{-},\bm{x}_{\perp}) \,  \rho_{A}^{b} (y^{-},\bm{y}_{\perp}) \rangle=\delta^{ab} \, \delta(x^{-}-y^{-}) \, \lambda_{A}(x^{-}) \, ,
\end{equation}
where
\begin{equation}
    \int dx^{-} \, \lambda_{A}(x^{-})=\mu_A^2 \, ,
\end{equation}
and $\mu^2_A$ is the average color charge squared of the sources per color and per unit transverse area of a nucleus with mass number A. Explicit expressions for the dipole and quadrupole many-body correlators are available~\cite{Kovchegov:2012mbw,Blaizot:2004wv,Dominguez:2011wm,Dominguez:2012ad} in the MV model. (See also~\cite{Dusling:2017aot,Shi:2017gcq,Fukushima:2017mko} for computations of higher point correlators.)

We will now provide expressions for the differential cross-section in terms of these ``soft" (non-perturbative) and ``hard" (perturbative) components explained in the previous paragraphs. For the case in which the virtual photon is longitudinally polarized, we can write the differential cross-section for $\gamma+q\bar{q}$ production in $\gamma^{*}_{L} + A$ scattering very compactly as
\begin{align}
    \frac{d \sigma^{\gamma^*_{L} A \rightarrow q\bar{q}\gamma X}}{d ( \rm{PS}_3)}=& \frac{16 N_{\mathrm{c}} \alpha^2_{\rm em} q^4_f }{(2\pi)^6} z^3_q z^3_{\bar{q}} \, \delta(1-z_{\rm tot})  \int_{\vect{x},\vect{y},\vect{x'},\vect{y'}}\!\! \!\!\!\!\!\!\!\!\!\!\!\!\!\!\!\!\!\!\!\!\!\!\!\!\!\!\!
    e^{-i \vect{k} \cdot (\vect{x}-\vect{x}')-i\vect{p} \cdot (\vect{y}-\vect{y}') }\ \  \Xi (\vect{x},\vect{y};\vect{y}',\vect{x}'\vert Y) \,  \mathcal{R}^{q\bar{q}\gamma}_L  \, ,
    \label{eq:diff-cs-long-pol}
\end{align}
where $d (\rm{PS}_3)$ = $d \eta_q d \eta_{\bar{q}} d \eta_\gamma d^2 \vect{k} d^2 \vect{p} d^2 \vect{k_\gamma}$ represents the three particle phase space measure and $z_{\rm tot} = z_q+z_{\bar{q}}+z_\gamma$. 
We can express the momentum fractions of the final state particles in terms of their  rapidities as
\begin{equation}
    z_{q}=\frac{k_{\perp}}{\sqrt{2} q^{-} } \, e^{\eta_{q}} \, \, , \, \,  z_{\bar{q}}=\frac{p_{\perp}}{\sqrt{2} q^{-}} \, e^{\eta_{\bar{q}}} \, \, ,\, \,   z_{\gamma}=\frac{k_{\gamma \perp}}{\sqrt{2} q^{-}} \, e^{\eta_{\gamma}} \, .
    \label{eq:mom-frac-in-terms-of-rapidity}
\end{equation}
The coefficient function $\mathcal{R}^{q\bar{q}\gamma}_L$ can be written as
\begin{align}
    \mathcal{R}^{q\bar{q}\gamma}_L (k,p,k_{\gamma};\vect{x},\vect{y},\vect{x'},\vect{y'}) &=   \xi_{qq} \left[ \mathcal{J}^{i}_{L,1}(\vect{r}) +  \mathcal{J}^{i}_{L,2}(\vect{r}) \right]\left[ \mathcal{J}^{i*}_{L,1}(\vect{r}') +  \mathcal{J}^{i*}_{L,2}(\vect{r}') \right] e^{-i \vect{k_\gamma} \cdot (\vect{x} - \vect{x}')}
    \nonumber \\
    &- \xi_{q\bar{q}} \left[ \mathcal{J}^{i}_{L,1}(\vect{r})  + \mathcal{J}^{i}_{L,2}(\vect{r}) \right] \left[ \mathcal{J}^{i*}_{L,3}(\vect{r})  + \mathcal{J}^{i*}_{L,4}(\vect{r}) \right] e^{-i\vect{k_\gamma} \cdot (\vect{x} - \vect{y}')} \nonumber \\
    &- \xi_{\bar{q}q}  \left[ \mathcal{J}^{i}_{L,3}(\vect{r})  + \mathcal{J}^{i}_{L,4}(\vect{r})  \right] \left[ \mathcal{J}^{i*}_{L,1}(\vect{r}')  + \mathcal{J}^{i*}_{L,2}(\vect{r}')  \right] e^{-i\vect{k_\gamma} \cdot (\vect{y} - \vect{x}')}
    \nonumber \\
    & + \xi_{\bar{q}\bar{q}} \left[ \mathcal{J}^{i}_{L,3}(\vect{r}) +  \mathcal{J}^{i}_{L,4}(\vect{r}) \right]\left[ \mathcal{J}^{i*}_{L,3}(\vect{r}') +  \mathcal{J}^{i*}_{L,4}(\vect{r}') \right] e^{-i \vect{k_\gamma} \cdot (\vect{y} - \vect{y}')} \, .
    \label{eq:R-long-pol-gamma-dijet}
\end{align}
We have explicitly shown here the momentum and coordinate dependence of $\mathcal{R}_{L}^{q\bar{q}\gamma}$. The terms in each line are organized according to their phases and the labels $1,\ldots,4$ denote the respective contributions from the four allowed processes in Fig.~\ref{fig:photon_jet_LO_diagrams}. The topology  of the cut-graphs that represent the different contributions along each line in Eq.\,\eqref{eq:R-long-pol-gamma-dijet} is similar. The first (fourth) line, for example, corresponds to the cases in which the final state photon is emitted by the quark (antiquark) in the direct and  conjugate amplitudes. Similarly, the second and third lines represent the cross-terms in which the photon is emitted by the quark (antiquark) in the direct amplitude and by  the antiquark (quark) in the conjugate amplitude. 

The vector functions $\mathcal{J}^{i}_{L,p}$,  ($p=1,\ldots,4$) are proportional to the light cone wavefunctions for the splitting of $\gamma_{L}^{*}$ into $ q \bar{q} \gamma$; these are distinct because of the different locations of the emission of the photon. We have not shown the momentum dependence in these functions for the sake of brevity. This will be clear from their expressions which are provided below.  Finally, the $\xi$'s are  coefficients that depend on the momentum fractions; these are defined to be
\begin{align}
   \xi_{qq} &=1+ \left(\frac{1-z_{\bar{q}} }{z_q} \right)^2  \quad , \quad   \xi_{\bar{q}\bar{q}} =1+ \left(\frac{1-z_{q} }{z_{\bar{q}}} \right)^2  \label{eq:xi-qq-xi-qbar-xi-bar}  \, , \\
  \xi_{q\bar{q}} &= \xi_{\bar{q}q} = \left[ \frac{1-z_q}{z_{\bar{q}}} + \frac{1-z_{\bar{q}}}{z_{q}} \right] \, . \label{eq:xi-q-qbar}
\end{align}
We now provide the expressions for $\mathcal{J}^{i}_{L,1}$ and $\mathcal{J}^{i}_{L,2}$. The expressions for $\mathcal{J}^i_{L,3}$ and  $\mathcal{J}^i_{L,4}$ are respectively obtained simply by interchange of quark and antiquark: $k \leftrightarrow p$ and $\vect{r} \leftrightarrow -\vect{r}$.
\begin{align}
    \mathcal{J}^{i}_{L,1}(\vect{r}) &= Q \,  (1-z_{\bar{q}}) \frac{\left(z_q \vect{k_\gamma}^i -z_\gamma \vect{k}^i \right) }{ \left(z_q \vect{k_\gamma} -z_\gamma \vect{k} \right)^2} \int \frac{d^2 \vect{l}}{(2 \pi)^2}  \frac{e^{i \vect{l} \cdot \vect{r}}}{\vect{l}^2 + \Delta_1}, \ \ \ \ \ \Delta_1 = z_{\bar{q}} (1 - z_{\bar{q}}) Q^2 \, ,  \label{eq:J-L-1} \\
    \mathcal{J}^{i}_{L,2}(\vect{r}) &=- \frac{Q}{z_\gamma z_q} \int\frac{d^2 \vect{l}}{(2\pi)^2} \frac{\left[(1-z_{\bar{q}})\vect{k_\gamma}^i -z_\gamma \vect{l}^i \right] e^{i \vect{l}\cdot \vect{r} }}{ \left( \vect{l}^2 + \Delta_1  \right) \left( Q^2 +\frac{\vect{l}^2}{z_{\bar{q}}}+\frac{\vect{k_\gamma}^2}{z_\gamma} + \frac{(\vect{l} - \vect{k_\gamma})^2}{z_q}  \right)} \label{eq:J-L-2} \, . 
\end{align}
In Appendix~\ref{sec:hadron-amplitude-squared}, we discuss in detail the steps leading to the expression for $\mathcal{R}^{q\bar{q}\gamma}_{L}$ in terms of the $\xi$ coefficients and the vector functions $\mathcal{J}_{L}^{i}$. We also provide all the necessary results needed to do this exercise for the transversely polarized virtual photon. This involves explicitly computing the squared amplitudes appearing in the modified hadron tensor in Eq.\,\eqref{eq:modified-hadron-tensor-gamma-dijet} for the different polarization states of $\gamma^{*}$ and then organizing the resulting expressions in the symmetric form displayed in Eq.\,\eqref{eq:R-long-pol-gamma-dijet}. (See also in this regard, Eqs.~\eqref{R-T-reg-reg}-\eqref{eq:R-T-ins-ins} in the upcoming discussion.)

In a similar fashion, we can obtain the differential cross-section for $q\bar{q}\gamma$ production in the case of transversely polarized virtual photons, $\gamma^{*}_{T}$. Summing over the two polarization states, we get
\begin{align}
  \sum_{T=\pm 1}  \frac{d \sigma^{\gamma^*_{T} A \rightarrow q\bar{q}\gamma X}}{d (\rm{PS}_3)}=& \frac{8 N_{\mathrm{c}} \alpha^2_{\rm em} q^4_f   }{(2\pi)^6} z_{q} z_{\bar{q}}  \,  \delta(1-z_{\rm tot}) \int_{\vect{x},\vect{y},\vect{x'},\vect{y'}} \!\!\!\!\!\!\!\!\!\!\!\!\!\!\!\!\!\!\!\!\!\!\!\!\!\!\!\!\!
    e^{-i \vect{k} \cdot (\vect{x}-\vect{x}')-i\vect{p} \cdot (\vect{y}-\vect{y}') }\ \  \Xi (\vect{x},\vect{y};\vect{y}',\vect{x}' \vert Y) \,  \mathcal{R}^{q\bar{q}\gamma}_T \, .
    \label{eq:diff-cs-trans-pol}
\end{align}
 We separate the perturbative component into three parts as
\begin{align}
    \mathcal{R}^{q\bar{q}\gamma}_{T}= \mathcal{R}^{q\bar{q}\gamma}_{T;\rm{reg-reg}} + \mathcal{R}^{q\bar{q}\gamma}_{T;\rm{reg-ins}}+ \mathcal{R}^{q\bar{q}\gamma}_{T;\rm{ins-ins}} \, ,
     \label{eq:R-trans-pol-gamma-dijet}
\end{align}
which follows from the decomposition of the amplitude in Eq.\,\eqref{eq:A2full} into a `regular' and an `instantaneous' term. Comparing Eqs.~\eqref{eq:A1} and \eqref{eq:A2reg}, we immediately see that the `regular' term has a gamma matrix structure identical to the contribution from the graph $\mathcal{M}_1$ in Fig.~\ref{fig:photon_jet_LO_diagrams}. We demonstrated using the identity in  Eq.\,\eqref{eq:gamma-structure-A1} that there is indeed a piece in this `regular' structure that yields non-zero contributions for transversely polarized $\gamma^{*}$. We also showed there that the `instantaneous' term contributes only for transverse polarizations of the virtual photon (see Eqs.~\eqref{eq:A2ins} and \eqref{eq:gamma-structure-inst}). At the amplitude squared level, we will therefore have direct (reg--reg, ins--ins) and cross (reg--ins or ins--reg) terms. It is therefore evident that there are going to be significantly more contributions to the cross-section in the case of $\gamma^*_T$ as compared to $\gamma^*_L$.

However, we can carefully organize these terms and express them compactly as shown in the following equations. The first term is given by
\begin{align}
    \mathcal{R}^{q\bar{q}\gamma}_{T;\rm{reg-reg}}&=\left[\zeta_{qq} \delta^{im} \delta^{jn} + \chi_{qq} \epsilon^{im} \epsilon^{jn}  \right]\left[ \mathcal{J}^{ij}_{T,1}(\vect{r})  + \mathcal{J}^{ij}_{T,2}(\vect{r})  \right] \left[ \mathcal{J}^{mn*}_{T,1}(\vect{r}')  + \mathcal{J}^{mn*}_{T,2}(\vect{r}')  \right] e^{-i \vect{k_\gamma} \cdot (\vect{x} - \vect{x}')} \nonumber \\
    &+\left[\zeta_{q\bar{q}} \delta^{im} \delta^{jn} - \chi_{q\bar{q}} \epsilon^{im} \epsilon^{jn}  \right]\left[ \mathcal{J}^{ij}_{T,1}(\vect{r})  + \mathcal{J}^{ij}_{T,2}(\vect{r})  \right] \left[ \mathcal{J}^{mn*}_{T,3}(\vect{r}')  + \mathcal{J}^{mn*}_{T,4}(\vect{r}')  \right] e^{-i \vect{k_\gamma} \cdot (\vect{x} - \vect{y}')} \nonumber \\
    & + \left[\zeta_{\bar{q}q} \delta^{im} \delta^{jn} - \chi_{\bar{q}q} \epsilon^{im} \epsilon^{jn}  \right]\left[ \mathcal{J}^{ij}_{T,3}(\vect{r})  + \mathcal{J}^{ij}_{T,4}(\vect{r})  \right] \left[ \mathcal{J}^{mn*}_{T,1}(\vect{r}')  + \mathcal{J}^{mn*}_{T,2}(\vect{r}')  \right] e^{-i \vect{k_\gamma} \cdot (\vect{y} - \vect{x}')} \nonumber \\
    &+\left[\zeta_{\bar{q}\bar{q}} \delta^{im} \delta^{jn} + \chi_{\bar{q}\bar{q}} \epsilon^{im} \epsilon^{jn}  \right]\left[ \mathcal{J}^{ij}_{T,3}(\vect{r})  + \mathcal{J}^{ij}_{T,4}(\vect{r})  \right] \left[ \mathcal{J}^{mn*}_{T,3}(\vect{r}')  + \mathcal{J}^{mn*}_{T,4}(\vect{r}')  \right] e^{-i \vect{k_\gamma} \cdot (\vect{y} - \vect{y}')}  \,.
    \label{R-T-reg-reg}
\end{align}
Comparing the above equation with Eq.\,\eqref{eq:R-long-pol-gamma-dijet}, we immediately notice a similarity in their structures, albeit with tensor functions, $\mathcal{J}_{T}^{ij}$ and different coefficients, $\zeta$ and $\xi$. The explicit expressions for these are provided in Appendix~\ref{sec:coefficients-J-functions}. Similarly, the second and third terms on the r.h.s of Eq.\,\eqref{eq:R-trans-pol-gamma-dijet} can be written respectively as
\begin{align}
    \mathcal{R}^{q\bar{q}\gamma}_{T;\rm{reg-ins}} &= \kappa_{qq} \Big[  \left(\mathcal{J}^{ii}_{T,1} (\vect{r})  + \mathcal{J}^{ii}_{T,2} (\vect{r}) \right) \mathcal{J}^{*}_{T\rm{ins},2}(\vect{r}') + \mathcal{J}_{T\rm{ins},2}(\vect{r}) \left(\mathcal{J}^{ii*}_{T,1} (\vect{r}')  + \mathcal{J}^{ii*}_{T,2} (\vect{r}') \right) \Big]e^{-i \vect{k_\gamma}\cdot (\vect{x}-\vect{x}') } \nonumber \\
    &+ \kappa_{q\bar{q}} \Big[ \left(\mathcal{J}^{ii}_{T,1} (\vect{r})  + \mathcal{J}^{ii}_{T,2} (\vect{r}) \right) \mathcal{J}^{*}_{T\rm{ins},4}(\vect{r}') + \mathcal{J}_{T\rm{ins},2}(\vect{r}) \left(\mathcal{J}^{ii*}_{T,3} (\vect{r}')  + \mathcal{J}^{ii*}_{T,4} (\vect{r}') \right)  \Big] \, e^{-i \vect{k_\gamma}\cdot (\vect{x}-\vect{y}') } \nonumber \\
    &+ \kappa_{\bar{q}q} \Big[ \left(\mathcal{J}^{ii}_{T,3} (\vect{r})  + \mathcal{J}^{ii}_{T,4} (\vect{r}) \right) \mathcal{J}^{*}_{T\rm{ins},2}(\vect{r}') + \mathcal{J}_{T\rm{ins},4}(\vect{r}) \left(\mathcal{J}^{ii*}_{T,1} (\vect{r}')  + \mathcal{J}^{ii*}_{T,2} (\vect{r}') \right)   \Big] \, e^{-i \vect{k_\gamma}\cdot (\vect{y}-\vect{x}') }  \nonumber
    \\
    &+ \kappa_{\bar{q}\bar{q}}  \Big[  \left(\mathcal{J}^{ii}_{T,3} (\vect{r})  + \mathcal{J}^{ii}_{T4} (\vect{r}) \right) \mathcal{J}^{*}_{T\rm{ins},4}(\vect{r}') + \mathcal{J}_{T\rm{ins},4}(\vect{r}) \left(\mathcal{J}^{ii*}_{T,3} (\vect{r}')  + \mathcal{J}^{ii*}_{T,4} (\vect{r}') \right) \Big] e^{-i \vect{k_\gamma}\cdot (\vect{y}-\vect{y}') } \, ,
    \label{eq:R-T-reg-ins}
\end{align}
and
\begin{align}
    \mathcal{R}^{q\bar{q}\gamma}_{T;\rm{ins-ins}} &= \sigma_{qq} \,  \mathcal{J}_{T\rm{ins},2}(\vect{r})\mathcal{J}^{*}_{T\rm{ins},2}(\vect{r'}) \,  e^{-i \vect{k_\gamma} \cdot (\vect{x}-\vect{x}')} + \sigma_{\bar{q}\bar{q}} \, \mathcal{J}_{T\rm{ins},4}(\vect{r})\mathcal{J}^{*}_{T\rm{ins},4}(\vect{r'})  \, e^{-i \vect{k_\gamma} \cdot (\vect{y}-\vect{y}')}
    \label{eq:R-T-ins-ins} \, .
\end{align}
The structures of these terms differ slightly because only the amplitudes of processes $\mathcal{M}_2$ and $\mathcal{M}_4$ in Fig.~\ref{fig:photon_jet_LO_diagrams} contain `instantaneous' contributions. Expressions for all the coefficients, $\kappa$ and functions, $\mathcal{J}_{T\rm ins}$ are provided in Appendix~\ref{sec:coefficients-J-functions}.

\section{Integrating the antiquark phase space} \label{sec:antiquark-phase-space-integration}

Now that we have the results for the cross-section for $\gamma+q\bar{q}$ production separately for longitudinal and transverse polarizations of the virtual photon, we need to integrate over the phase space of the antiquark to obtain the desired cross-section for $\gamma+q$ production as given by Eq.\,\eqref{eq:virtual-photon-hadron-scatt-cs-general}. We will show here the steps leading to the explicit expressions for this quantity by considering the case of a longitudinally polarized virtual photon. The corresponding computation for the case of transverse polarization is considerably more tedious and expressions for the $\gamma^{*}_{T}A \rightarrow q\gamma X$  cross-section are provided in Appendix~\ref{sec:photon-jet-trans-pol}.

\subsection{Collinear singularity and the fragmentation contribution} \label{sec:fragmentation-contribution-long-pol}

To begin with, the integration over the antiquark rapidity ($\mathrm{d}\eta_{\bar{q}}=\mathrm{d}z_{\bar{q}}/z_{\bar{q}}$) in Eq.\,\eqref{eq:virtual-photon-hadron-scatt-cs-general} can be done trivially using the delta function that enters the expression for the cross-section, in Eq.\,\eqref{eq:diff-cs-long-pol}. This sets $z_{\bar{q}}=1-z_{q}-z_{\gamma}$.  For the transverse momentum integration, we notice that all such terms in the $\gamma^{*}_{L}A \rightarrow q\bar{q}\gamma  X$ cross-section (see Eqs.~\eqref{eq:diff-cs-long-pol} and \eqref{eq:R-long-pol-gamma-dijet}) that do not have an  explicit $\vect{p}$-dependence will yield a two dimensional Dirac delta upon integration over the transverse phase space of the antiquark. In terms of the vector functions constituting the expression in Eq.\,\eqref{eq:R-long-pol-gamma-dijet}, these would be the ones \emph{not} proportional to $\mathcal{J}^{i}_{L,3}$ (or its complex conjugate) which we write below for reference,
\begin{equation}
    \mathcal{J}^{i}_{L,3}(\vect{r}) = Q \,  (1-z_q) \frac{\left(z_{\bar{q}} \vect{k_\gamma}^i -z_\gamma \vect{p}^i \right) }{ \left|z_{\bar{q}} \vect{k_\gamma} -z_\gamma \vect{p} \right|^2}  \int_{\vect{l}} \frac{e^{-i \vect{l} \cdot \vect{r}}}{\vect{l}^2 + \Delta_3} , \ \ \ \ \ \ \Delta_3 = z_{q} (1 - z_{q}) Q^2  \, .
    \label{eq:J-L-3}
\end{equation}
For the remaining terms in Eq.\,\eqref{eq:R-long-pol-gamma-dijet}, we have those proportional to $\mathcal{J}_{L,3}^{i}$ (or its complex conjugate) and \emph{one} term proportional to $\mathcal{J}^{i}_{L,3} \, \mathcal{J}_{L,3}^{i \, *} $.  We get finite results for the former after performing  the integration over $\vect{p}$. In this subsection, we will mainly be concerned with the latter which contains a singular contribution after the $\vect{p}$-integration. Physically, the term proportional to $\mathcal{J}^{i}_{L,3} \, \mathcal{J}_{L,3}^{i \, *} $ represents the case in which the photon is emitted from the antiquark after the scattering off the nucleus in the direct amplitude, and by the scattered antiquark in the conjugate amplitude. Using the expression for $\mathcal{J}^{i}_{L,3}$ given in Eq.\,\eqref{eq:J-L-3}, we see that the square of this quantity gives
\begin{equation}
    \mathcal{J}^{i}_{L,3}(\vect{r}) \, \mathcal{J}^{i \, *}_{L,3}(\vect{r}')= \frac{Q^2}{(2\pi)^{2}} \Big( \frac{1-z_{q}}{z_{\gamma}} \Big)^{2} \, \frac{1}{(\vect{p}-z_{\bar{q}}/z_{\gamma} \, \vect{k_{\gamma}} )^{2}} \, K_{0}(\Delta_{3}^{1/2} r_{\perp}) \,  K_{0}(\Delta_{3}^{1/2} r'_{\perp}) \, ,
    \label{eq:J-L-3-squared}
\end{equation}
where we have used the identity
\begin{equation}
    \int_{\vect{l}} \frac{e^{-i \vect{l} \cdot \vect{r}}}{\vect{l}^2 + \Delta}= \frac{1}{2\pi} K_{0}(\Delta^{1/2} r_{\perp}) \, .
    \label{eq:Bessel-K-0}
\end{equation}
When we are integrating over $\vect{p}$, we will encounter a region where the photon is collinear to the outgoing antiquark. This is apparent from the  integral
$$ \int \mathrm{d}^2 \vect{p} \frac{e^{-i\vect{p}.(\vect{y}-\vect{y'})}}{(\vect{p}-z_{\bar{q}}/z_{\gamma} \, \vect{k_{\gamma}} )^{2}}\,, $$ 
 for 
\begin{equation}
    \frac{\vect{p}}{z_{\bar{q}}}= \frac{\vect{k_{\gamma}}}{z_{\gamma}} \, .
\end{equation}
Above we have a $\vect{p}$-dependent phase that enters the cross-section as given by Eq.\,\eqref{eq:diff-cs-long-pol} and the denominator is taken from the squared expression in Eq.\,\eqref{eq:J-L-3-squared}.

There are many ways to regulate the collinear singularity. The simplest way is to introduce an infrared (IR) regulator $\Lambda$ in order to write the integral over the transverse momentum of the antiquark as\footnote{Strictly speaking, we are imposing a cutoff on the redefined variable $\vert \vect{p}-\frac{z_{\bar{q}}}{z_{\gamma} } \, \vect{k_{\gamma}} \vert$ which is proportional to the invariant mass of the photon-antiquark system. }
\begin{equation}
    \int_{\Lambda} \mathrm{d}^{2} \vect{p} \,  \frac{e^{-i\vect{p}.(\vect{y}-\vect{y'})}}{(\vect{p}-z_{\bar{q}}/z_{\gamma} \, \vect{k_{\gamma}} )^{2} } = 2\pi  \, \ln \Big(\frac{1}{\vert \vect{y}-\vect{y}' \vert \Lambda} \Big) \, e^{-i \frac{z_{\bar{q}}}{z_{\gamma}} \vect{k_{\gamma}}.(\vect{y}-\vect{y'})}  \, .
     \label{eq:collinear-integral}
\end{equation}
Note that $1/\vert \bm{y}_{\perp}-\bm{y'}_{\perp} \vert$ acts as a UV regulator for the $\vect{p}$ integration; these transverse coordinates are also present in the fundamental Wilson lines constituting the LO color structure in Eq.\,\eqref{eq:color-structure-LO}. The logarithm can be broken into two parts as \begin{equation}
 \ln \Big(\frac{1}{\vert \vect{y}-\vect{y}' \vert \Lambda} \Big) = \ln \Big( \frac{1}{\vert \vect{y}-\vect{y}' \vert \mu_{F}} \Big) + \overbrace{\ln \Big( \frac{\mu_{F}}{\Lambda} \Big)}^{\rm frag.} \, ,   
 \label{eq:collinear-log-break-up}
\end{equation}
by introducing a factorization scale $\mu_{F}$. The contributions from momenta below $\mu_{F}$, contained in the second logarithm on the r.h.s of Eq.\,\eqref{eq:collinear-log-break-up}, is absorbed into the (anti)quark-to-photon fragmentation function $D_{\bar{q}\rightarrow \gamma}(\lambda_{\gamma/\bar{q}},\mu_{F})$. Here  $\lambda_{\gamma/\bar{q}} \, (0 \leq \lambda_{\gamma/\bar{q}} \leq 1) $ represents the fraction of momentum of the antiquark carried by a photon. In terms of the partonic variables used in our computation, this is given by
\begin{equation}
  \lambda_{\gamma/\bar{q}}  = \frac{z_{\gamma}}{1-z_{q}}  \, .
  \label{eq:photon-mom-frac-q-bar}
\end{equation}
We will call this contribution proportional to $D_{\bar{q} \rightarrow \gamma}(\lambda_{\gamma/\bar{q}},\mu_{F})$ the ``fragmentation" (F) contribution. The fragmentation function $D_{\bar{q} \rightarrow \gamma}(\lambda_{\gamma/\bar{q}},\mu_{F})$ accompanies the cross-section for the corresponding $\gamma^{*}A \rightarrow q\bar{q} \gamma X$ hadronic subprocess in which the antiquark hadronizes and then emits a ``decay" or fragmentation photon.
We define the (anti)quark-to-photon fragmentation function\footnote{There is a theoretical uncertainty related to the choice of the factorization scale $\mu_{F}$. For example, one can choose $\mu_{F}$ to be the invariant mass of the photon-antiquark pair or the transverse momentum of the photon. Although this choice is arbitrary, a knowledge of the fragmentation function at a certain scale can be used to evaluate its value at another scale, thanks to the physical cross-section being independent of $\mu_{F}$. This leads to the evolution equation
\begin{equation}
    \frac{\partial (D_{\bar{q}\rightarrow \gamma}(\lambda_{\gamma/\bar{q}},\mu_{F}))}{\partial(\ln(\mu_{F}^{2}))}=\frac{\alpha_{\rm em}q^{2}_{f}}{2\pi} P_{\gamma \bar{q}}(\lambda_{\gamma/\bar{q}}) \, ,
\end{equation}
which also has higher order corrections in the coupling, that are not written above.} at $\mu_{F}$ as~\cite{GehrmannDeRidder:2006vn}
\begin{equation}
    D_{\bar{q} \rightarrow \gamma} (\lambda_{\gamma/\bar{q}},\mu_{F})= \frac{\alpha_{\rm em}q_{f}^{2}}{2\pi} P_{\gamma \bar{q}} (\lambda_{\gamma/\bar{q}}) \, \ln \Big( \frac{\mu_{F}^{2}}{\mu_{0}^{2}} \Big) + D_{\bar{q} \rightarrow \gamma} (\lambda_{\gamma/\bar{q}},\mu_{0}) \,  ,
    \label{eq:photon-fragmentation function}
\end{equation}
where $\mu_{0} \sim \Lambda \sim \Lambda_{\rm QCD}$ and the (anti)quark-to-photon fragmentation function at the initial scale, $D_{\bar{q} \rightarrow \gamma} (\lambda_{\gamma/\bar{q}},\mu_{0})$  is obtained from experimental data~\cite{Buskulic:1995au}. The familiar leading order photon splitting function is
\begin{equation}
  P_{\gamma \bar{q}} (\lambda_{\gamma/\bar{q}})=\frac{1+(1-\lambda_{\gamma/\bar{q}})^{2}}{\lambda_{\gamma/\bar{q}}}   \, .
  \label{eq:gamma-splitting-function}
\end{equation}
One can think of this as an RG procedure in which the (anti)quark-to-photon fragmentation function at some initial scale, $\mu_{0}$ is matched on to that at an evolved scale $\mu_{F}$. The quantum corrections are logarithmic in nature and the hard coefficients are given by the splitting functions.

The contribution in $\int \mathrm{d}^2 \vect{p} \text{exp}(-i \vect{p}.(\vect{y}-\bm{y'}_{\perp})) \,  \mathcal{J}^{i}_{L,3} \, \mathcal{J}_{L,3}^{i \, *} $ proportional to the first logarithm on the r.h.s of Eq.\,\eqref{eq:collinear-log-break-up} will be  absorbed into the remaining terms that are finite after the $\vect{p}$ integration and together they constitute the ``direct" (D) contribution\footnote{They therefore have a logarithmic sensitivity to $\mu_F$ that goes away when this contribution is combined with the fragmentation contribution to obtain the physical cross-sectiion.}. The inclusive prompt photon+quark production cross-section given by Eq.\,\eqref{eq:virtual-photon-hadron-scatt-cs-general} can therefore be expressed as
\begin{equation}
    \frac{\mathrm{d} \sigma^{\gamma^{*}_{L,T}A\rightarrow q\gamma X}}{\mathrm{d}^2 \vect{k}\mathrm{d}^2 \vect{k_{\gamma}} \mathrm{d}\eta_{q}\mathrm{d}\eta_{\gamma}  } = \frac{\mathrm{d} \sigma^{\gamma^{*}_{L,T}A \rightarrow q\gamma X}_{\rm D}}{\mathrm{d}^2 \vect{k}\mathrm{d}^2 \vect{k_{\gamma}} \mathrm{d}\eta_{q}\mathrm{d}\eta_{\gamma}}+ \frac{\mathrm{d} \sigma^{\gamma^{*}_{L,T}A \rightarrow q \gamma X}_{F} }{\mathrm{d}^2 \vect{k}\mathrm{d}^2 \vect{k_{\gamma}} \mathrm{d}\eta_{q}\mathrm{d}\eta_{\gamma}}\, 
.
\label{eq:cs-in-terms-of-d-and-f}
\end{equation}

We will discuss the ``direct'' contribution in the next section. Here we obtain results for the ``fragmentation" contribution to the $\gamma+q$ production cross-section.
 For this we will use the expression for the $\gamma+q\bar{q}$ production  cross-section for longitudinally polarized virtual photon given by Eqs.~\eqref{eq:diff-cs-long-pol} and \eqref{eq:R-long-pol-gamma-dijet} and the results in Eqs.~\eqref{eq:collinear-integral} and \eqref{eq:collinear-log-break-up}. It is easy to check that the prefactor $\xi_{\bar{q}\bar{q}}$ in the term proportional to $\mathcal{J}^{i}_{L,3} \, \mathcal{J}_{L,3}^{i \, *} $, defined in Eq.\,\eqref{eq:xi-qq-xi-qbar-xi-bar} is  proportional to the photon splitting function defined in Eq.\,\eqref{eq:gamma-splitting-function},
\begin{equation}
  \xi_{\bar{q}\bar{q}}= \frac{\lambda_{\gamma/\bar{q}}}{(1-\lambda_{\gamma/\bar{q}})^{2}}   \, P_{\gamma \bar{q}} (\lambda_{\gamma/\bar{q}}) \, .
  \label{eq:xi-squared-qq}
\end{equation}

For the longitudinally polarized virtual photon, the fragmentation contribution to the inclusive $\gamma+q$ production can therefore be written as
\begin{align}
    \frac{d \sigma^{\gamma^{*}_{L} A \rightarrow q \gamma X}_{F}}{\mathrm{d}^{2} \vect{k}  \mathrm{d}^{2} \vect{k_{\gamma}}    \mathrm{d} \eta_{q} \mathrm{d} \eta_{\gamma}} & = \frac{8N_{c}\alpha_{\rm em}q^{2}_{f}Q^{2}z_{q}^{3}(1-z_{q})^{2}}{(2\pi)^{6}} \int_{\vect{x},\vect{y},\vect{x'},\vect{y'}} \!\!\!\!\!\!\!\!\!\!\!\!\!\!\!\!\!\!\!\!\!\!\!\!\!\!\! e^{-i\vect{k}\cdot(\vect{x}-\vect{x'})} e^{ \frac{-i}{\lambda_{\gamma/\bar{q}}}\vect{k_{\gamma}}\cdot(\vect{y}-\vect{y'})} \, \Xi(\vect{x},\vect{y};\vect{y'},\vect{x'} \vert Y) \nonumber \\
    & \times \frac{D_{\bar{q}\rightarrow \gamma}(\lambda_{\gamma/\bar{q}},\mu_{F})}{\lambda_{\gamma/\bar{q}}} \, K_{0}(\Delta_{3}^{1/2} r_{\perp}) \,  K_{0}(\Delta_{3}^{1/2} r'_{\perp}) \, .
    \label{eq:fragmentation-cs-long-pol}
\end{align}
This can be further expressed in terms of the lightcone wavefunction for the longitudinally polarized $\gamma^{*}$ as
\begin{align}
     \frac{d \sigma^{\gamma^{*}_{L} A \rightarrow q \gamma X}_{F}}{\mathrm{d}^{2} \vect{k}  \mathrm{d}^{2} \vect{k_{\gamma}}    \mathrm{d} \eta_{q} \mathrm{d} \eta_{\gamma}} & =\frac{N_{c}\alpha_{\rm em}q^{2}_{f} \, k_{\perp} e^{\eta_{q}}}{\sqrt{2} \, (2\pi)^8}  \int_{\vect{x},\vect{y},\vect{x'},\vect{y'}} \!\!\!\!\!\!\!\!\!\!\!\!\!\!\!\!\!\!\!\!\!\!\!\!\!\!\! e^{-i\vect{k}\cdot(\vect{x}-\vect{x'})} e^{ \frac{-i}{\lambda_{\gamma/\bar{q}}}\vect{k_{\gamma}}\cdot(\vect{y}-\vect{y'})} \, \Xi(\vect{x},\vect{y};\vect{y'},\vect{x'} \vert Y) \nonumber \\
     &\times \frac{D_{\bar{q}\rightarrow \gamma}(\lambda_{\gamma/\bar{q}},\mu_{F})}{\lambda_{\gamma/\bar{q}}} \sum_{\alpha,\beta} \Psi^{L \, *}_{\alpha \, \beta} (q^{-},z_{q},r'_{\perp}) \, \Psi^{L }_{\alpha \, \beta} (q^{-},z_{q},r_{\perp}) \, , 
     \label{eq:fragmentation-cs-long-pol-wavefunctions}
\end{align}
where we have used the relations in Eq.\,\eqref{eq:mom-frac-in-terms-of-rapidity} to express $z_{q}$ in terms of momentum and rapidity of the quark, and the lightcone wavefunction is defined to be~\cite{Dominguez:2011wm}
\begin{equation}
    \Psi^{L }_{\alpha \, \beta} (q^{-},z_{q},r_{\perp})= 2\pi \sqrt{\frac{4}{q^-}} \, z\, (1-z) \, Q \, K_{0}(\Delta^{1/2} r_{\perp}) \, \delta_{\alpha \beta} \, ,
    \label{eq:LC-wavefunction-long}
\end{equation}
where $z$ is the momentum fraction carried by the quark from a longitudinally polarized virtual photon with momentum $q$ that splits into a $q\bar{q}$ dipole with size $r_{\perp}.$ Here we have $\Delta=Q^{2}z(1-z)$ for massless quarks and $\alpha,\beta$ are the helicities of the quark and antiquark. We obtain similar expressions for the case of transverse polarization of the virtual photon; the computation is described in Appendix~\ref{sec:fragmentation-contribution-trans-pol}.

We immediately notice a similarity between  Eq.\,\eqref{eq:fragmentation-cs-long-pol-wavefunctions} and the LO cross-section for inclusive dijet production in  $\gamma^{*}_{L}+A$  scattering~\cite{Dominguez:2011wm}. Interestingly, both these cross-sections contain the quadrupole Wilson line correlator, which is sensitive to the Weizs\"{a}cker-Williams (WW) unintegrated gluon distribution in the so-called ``correlation" limit\footnote{See \cite{Dumitru:2018kuw} for numerical studies of inclusive dijet production in DIS at small $x$ in the correlation limit at an EIC and \cite{Mantysaari:2019hkq} which goes beyond this limit for both inclusive and diffractive dijets in small $x$ DIS.} of back-to-back hard dijets with small momentum imbalance~\cite{Dominguez:2011wm,Dominguez:2011br}. This makes the study of the fragmentation photon contribution for inclusive prompt photon plus jet production in DIS at small $x$ appealing in its own right. 

The numerical analysis, however, is more challenging when one encounters the quadrupole. In the context of DIS, numerical results that incorporate the quadrupole (in addition to the dipole) have been recently obtained for inclusive dijet production in Ref.~\cite{Mantysaari:2019hkq}. This means that it is indeed feasible to provide numerical estimates of the fragmentation contribution for inclusive photon plus jet production, albeit with scale and scheme  uncertainties associated with the fragmentation function $D_{\bar{q}\rightarrow \gamma}(\lambda_{\gamma/\bar{q}},\mu_{F}) $. We will explore this calculation in the future; in this present work, we will limit ourselves to computing the direct contribution in Eq.\,\eqref{eq:cs-in-terms-of-d-and-f}.

To quantify the full impact of the direct contribution,  the numerical results that we will present in this paper will be for  kinematics in which the fragmentation contribution is suppressed, which as evident from Eq.\,\eqref{eq:fragmentation-cs-long-pol-wavefunctions} is the case for large $k_{\gamma \perp}/\lambda_{\gamma/\bar{q}}$. 

\subsection{Obtaining the direct photon+quark contribution} \label{sec:direct-contribution-gamma-jet-long}

In this section, we will derive an expression for the direct contribution to the differential cross-section in Eq.\,\eqref{eq:cs-in-terms-of-d-and-f}. In order to obtain this for the case of the longitudinally polarized virtual photon, we will compute the antiquark phase space integrated (finite) result for the terms in Eq.\,\eqref{eq:diff-cs-long-pol} modulo the piece of $\int \mathrm{d}^2 \vect{p} \text{exp}(-i \vect{p}.(\vect{y}-\bm{y'}_{\perp})) \,  \mathcal{J}^{i}_{L,3} \, \mathcal{J}_{L,3}^{i \, *} $ that has been absorbed into the fragmentation function. As discussed before, most of the terms are independent of $\vect{p}$ and hence will lead to $\vect{y}=\vect{y}'$ after integration over $\vect{p}$ and $\vect{y'}$. For such terms, the color structure of the Wilson line correlators simplifies to
 \begin{equation}
   \Theta(\vect{x},\bm{x'}_{\perp},\bm{y}_{\perp} \vert Y)=  1-S^{(2)}_{Y}(\vect{x},\vect{y})-S^{(2)}_{Y}(\vect{y},\vect{x'})+S^{(2)}_{Y}(\vect{x},\vect{x'}) \,  ,
    \label{eq:coincidence-color-structure}
\end{equation}
 which only contains  dipole correlators. We have chosen to denote the coincidence limit $\bm{y}_{\perp} = \bm{y'}_{\perp}$ of the LO color structure in Eq.\,\eqref{eq:color-structure-LO} by $\Theta$ to avoid confusion with the color structure for the remaining terms, which contain at least one $\mathcal{J}_{L,3}^{i}$ (or its complex conjugate) and therefore possess the most general color structure with both dipole and quadrupole correlators.

 The direct photon+quark contribution to the cross-section can therefore be split into the sum of two terms as 
\begin{align}
    \frac{d \sigma^{\gamma^{*}_{L} A \rightarrow q \gamma X}_{D}}{\mathrm{d}^{2} \vect{k}  \mathrm{d}^{2} \vect{k_{\gamma}}    \mathrm{d} \eta_{q} \mathrm{d} \eta_{\gamma}} &= \mathcal{A}_{L}(\eta_{q},\eta_{\gamma},\vect{k},\vect{k_{\gamma}})+\mathcal{B}_{L} (\eta_{q},\eta_{\gamma},\vect{k},\vect{k_{\gamma}};\mu_{F}) \, ,
    \label{eq:direct-cs-long-pol}
 \end{align}
 where $\mathcal{A}_{L}$ has the color structure given by Eq.\,\eqref{eq:coincidence-color-structure}, representative of terms in the amplitude independent of $\vect{p}$, and $\mathcal{B}_{L}$, which has the general color structure given by Eq.\,\eqref{eq:color-structure-LO}. The first piece can be expressed as 
 \begin{align}
    \mathcal{A}_{L}(\eta_{q},\eta_{\gamma},\vect{k},\vect{k_{\gamma}})&= \frac{16 N_{c}\alpha_{\rm em}^{2}q_{f}^{4} z_{q}^{3}(1-z_{q}-z_{\gamma})^{2}}{(2\pi)^{4}} \int_{\vect{x},\bm{x}_{\perp}^{\prime},\vect{y}} \!\!\!\!\!\!\!\!\!\!\!\!\!\!\!\!\!\! e^{-i\vect{k}\cdot(\vect{x}-\bm{x}_{\perp}^{\prime})} \,  \Theta(\vect{x},\vect{x}',\vect{y} \vert Y) \nonumber \\
     \times & \Big \{ \xi_{qq}\,
     \big[\mathcal{J}_{L,1}^{i}(\vect{r})+\mathcal{J}^{i}_{L,2}(\vect{r}) \big] \big[\mathcal{J}_{L,1}^{i\, *}(\vect{r'}) +\mathcal{J}_{L,2}^{i\, *}(\bm{r}_{\perp}^{\prime}) \big] \, e^{-i \vect{k_{\gamma}}\cdot(\vect{r}-\vect{r}')} +\xi_{\bar{q} \bar{q}} \, \mathcal{J}_{L,4}^{i}(\vect{r}) \mathcal{J}_{L,4}^{i\, *}(\bm{r}_{\perp}^{\prime}) \nonumber \\
     -& \ \ \xi_{q\bar{q}} \, \big[\mathcal{J}_{L,1}^{i}(\vect{r})+\mathcal{J}^{i}_{L,2}(\vect{r}) \big] \, \mathcal{J}_{L,4}^{i\, *}(\bm{r}_{\perp}^{\prime}) e^{-i \vect{k_{\gamma}}\cdot\vect{r}} - \xi_{\bar{q}q} \mathcal{J}_{L,4}^{i}(\vect{r}) \,  \big[\mathcal{J}_{L,1}^{i\, *}(\vect{r'}) +\mathcal{J}_{L,2}^{i\, *}(\bm{r}_{\perp}^{\prime}) \big] e^{i \vect{k_{\gamma}}\cdot\bm{r}_{\perp}^{\prime}}  \Big \} \, .
     \label{eq:direct-cs-long-pol-A}
 \end{align}
The expressions for the various coefficients $\xi_{qq}$, $\xi_{\bar{q}\bar{q}}$ and $\xi_{q\bar{q}} $ are given respectively in Eqs.~\eqref{eq:xi-qq-xi-qbar-xi-bar} and \eqref{eq:xi-q-qbar}. We should equate $z_{\bar{q}}=1-z_{q}-z_{\gamma}$ in these expressions because we are integrating over the antiquark phase space. The same argument also holds for the functions $\mathcal{J}^{i}_{L,1}$, $\mathcal{J}^{i}_{L,2}$ and $\mathcal{J}^{i}_{L,4}$, given respectively by Eq.\,\eqref{eq:J-L-1}, Eq.\,\eqref{eq:J-L-2} and the $q\leftrightarrow\bar{q}$ interchanged counterpart of Eq.\,\eqref{eq:J-L-2}.  

It should be noted that if we assume translational invariance in the transverse direction, then for $\bm{y}_{\perp}=\bm{y'}_{\perp}$, the dipoles appearing in the color structure given by Eq.\,\eqref{eq:coincidence-color-structure} are dependent only on the relative separations, $\vect{r}=\vect{x}-\vect{y}$,  $\bm{r}_{\perp}^{\prime}=\bm{x}_{\perp}^{\prime}-\bm{y'}_{\perp}$, and $\bm{r}_{\perp}-\bm{r'}_{\perp}=\vect{x}-\bm{x'}_{\perp}$. We will use this in the next section where we obtain limiting expressions for the direct photon+quark contribution and express them  entirely in momentum space.

To obtain a compact expression for the second contribution $\mathcal{B}_{L}$ in Eq.\,\eqref{eq:direct-cs-long-pol}, we will first isolate the piece in the r.h.s of Eq.\,\eqref{eq:collinear-integral} that contains $\ln (1 / |\vect{y}-\vect{y'}| \mu_{F} )$. As discussed previously, this comes from the integration in transverse momenta over $\bm{p}_{\perp}$ of the quantity $ \mathcal{J}^{i}_{L,3} \, \mathcal{J}^{i \, *}_{L,3}$, for momenta above the fragmentation scale $\mu_{F}$.  We will also use the following result\footnote{To avoid confusion, we note that  $\mathcal{J}^{i}_{L,3}(\vect{r})$ contains an implicit dependence on $\vect{p}$.} for terms in the amplitude squared that contain a single $\mathcal{J}^{i}_{L,3}$ or its complex conjugate,
\begin{equation}
    \int \mathrm{d}^{2}\vect{p} e^{-i\vect{p}\cdot\bm{r}_{yy' \perp}} \mathcal{J}^{i}_{L,3}(\vect{r})=2\pi i \, \frac{\bm{r}^{i}_{yy' \perp}}{\bm{r}^{2}_{yy' \perp}} e^{-i  \frac{(1-z_{q}) }{z_{\gamma}} \,  \vect{k_{\gamma}}\cdot\bm{r}_{yy' \perp}} \, e^{i\vect{k_{\gamma}}\cdot\bm{r}_{yy' \perp}} \,  \hat{\mathcal{J}}_{L,3} (r_{\perp}) \, ,
    \label{eq:integral-over-J3-L}
\end{equation}
where $\bm{r}_{yy' \perp}=\vect{y}-\vect{y'}$ and we have introduced the shorthand notation
\begin{equation}
  \hat{\mathcal{J}}_{L,3}  (r_{\perp})   = Q\, \frac{1-z_{q}}{z_{\gamma}} \int_{\vect{l}} \frac{e^{-i\vect{l}\cdot\vect{r}}}{\vect{l}^{2}+\Delta_{3}}=\frac{Q}{2\pi} \, \frac{1-z_{q}}{z_{\gamma}} \, K_{0}(\Delta^{1/2}_{3} r_{\perp}) \, .
  \label{eq:J-hat-L-3}
\end{equation}
The corresponding expression for  $\mathcal{J}_{L,3}^{i \, *}$ is obtained by replacing $\hat{\mathcal{J}}_{L,3} (r_{\perp})$ on the r.h.s.~of  Eq.\,\eqref{eq:integral-over-J3-L} by $\hat{\mathcal{J}}_{L,3}^{*}  (r'_{\perp})$, where
\begin{equation}
  \hat{\mathcal{J}}_{L,3}^{*}  (r'_{\perp})   = Q \, \frac{1-z_{q}}{z_{\gamma}} \int_{\vect{l'}} \frac{e^{i\vect{l'}\cdot\vect{r'}}}{\vect{l'}^{2}+\Delta_{3}}=\frac{Q}{2\pi} \, \frac{1-z_{q}}{z_{\gamma}} \, K_{0}(\Delta^{1/2}_{3} r'_{\perp}) \, .
  \label{eq:J-hat-star-L-3}
\end{equation}
As before, $\vect{r}=\vect{x}-\vect{y}$ and $\vect{r}'=\vect{x}'-\vect{y}'$. 

We obtain finally,
\begin{align}
    \mathcal{B}_{L}&(\eta_{q}, \eta_{\gamma}, \vect{k},\vect{k_{\gamma}};\mu_{F})= \frac{16 N_{c}\alpha_{\rm em}^{2}q_{f}^{4} z_{q}^{3}(1-z_{q}-z_{\gamma})^{2}}{(2\pi)^{6}} \int_{\vect{x},\bm{x}_{\perp}^{\prime},\vect{y}, \bm{y'}_{\perp} } \!\!\!\!\!\!\!\!\!\!\!\!\!\!\!\!\!\! e^{-i\vect{k}\cdot(\vect{x}-\bm{x}_{\perp}^{\prime})} e^{-i \frac{1-z_{q}}{z_{\gamma}} \vect{k_{\gamma}}\cdot\bm{r}_{yy' \perp}} \,  \Xi(\vect{x},\vect{y};\vect{y'},\vect{x'} \vert Y) \nonumber \\
    \times & \Big \{ \xi_{\bar{q} \bar{q}} \Big[ \pi \, \ln \Big( \frac{1}{\bm{r}_{yy' \perp}^{2} \, \mu_{F}^{2}} \Big) \, \hat{\mathcal{J}}_{L,3}(r_{\perp}) \, \hat{\mathcal{J}}^{*}_{L,3}(r'_{\perp}) + 2\pi i \frac{\bm{r}^{i}_{yy' \perp}}{\bm{r}^{2}_{yy' \perp}} \, \big(\hat{\mathcal{J}}_{L,3}(r_{\perp}) \, \mathcal{J}^{i \, *}_{L,4} (\bm{r'}_{\perp}) + \mathcal{J}^{i }_{L,4} (\bm{r}_{\perp}) \, \hat{\mathcal{J}}^{*}_{L,3}(r'_{\perp}) \big) \Big] \nonumber \\
    - & \xi_{q \bar{q}}    2\pi i \frac{\bm{r}^{i}_{yy' \perp}}{\bm{r}^{2}_{yy' \perp}}  \, \big[\mathcal{J}_{L,1}^{i}(\vect{r})+\mathcal{J}^{i}_{L,2}(\vect{r}) \big] \, \hat{\mathcal{J}}^{*}_{L,3}(r'_{\perp}) e^{-i \vect{k_{\gamma}}\cdot\vect{r}}  -\xi_{\bar{q} q}    2\pi i \frac{\bm{r}^{i}_{yy' \perp}}{\bm{r}^{2}_{yy' \perp}}  \, \hat{\mathcal{J}}_{L,3}(r_{\perp}) \,  \big[\mathcal{J}_{L,1}^{i\, *}(\vect{r'}) +\mathcal{J}_{L,2}^{i\, *}(\bm{r}_{\perp}^{\prime}) \big] e^{i \vect{k_{\gamma}}\cdot\bm{r}_{\perp}^{\prime}}   \Big \} \, .
    \label{eq:direct-cs-long-pol-B}
\end{align}
The $\hat{\mathcal{J}}_{L,3}$'s are defined in Eqs.~\eqref{eq:J-hat-L-3} and \eqref{eq:J-hat-star-L-3}. It is assumed that the replacement $z_{\bar{q}}=1-z_{q}-z_{\gamma}$ has been made in the expressions for the different $\xi$'s and $\mathcal{J}$'s appearing in the above equation.

The direct photon+quark production cross-section for the case of transverse polarization of the virtual photon can be similarly broken up into two pieces depending on the color structure. Explicit expressions for these are provided in Appendix~\ref{sec:direct-photon-trans-pol}.

\subsection{Simplifying the direct photon+quark contribution} \label{sec:direct-photon-limit-long}

We will now consider the direct photon+quark production cross-section obtained in the previous section in a limit where the scale $1/(\bm{y}_{\perp}-\bm{y'}_{\perp})^{2}$ is much larger than  the fragmentation scale $\mu_{F}^{2}$. In this limit, we are in a region entirely dominated by direct photons. A strong motivation to study this limit is that the color structure of the $\mathcal{B}_{L}$ term in the direct photon+quark cross-section (see Eqs.~\eqref{eq:direct-cs-long-pol} and \eqref{eq:direct-cs-long-pol-B}) simplifies greatly to include only dipole correlators when $\bm{y}_{\perp} \rightarrow \bm{y'}_{\perp}$. This will allow us to express the direct photon+quark cross-section entirely in momentum space, which makes the result much more amenable to numerical analysis compared to the general result, which contains the  quadrupole correlator. We will again restrict our discussion here to longitudinally polarized virtual photons. The same techniques can be applied to obtain results for transverse polarizations; these are provided in Appendix~\ref{sec:direct-photon-limit-trans}.

As is evident from the constituent expressions for the direct photon+quark production cross-section, we only need to examine the limiting form for the second contribution $\mathcal{B}_{L}$, since $\mathcal{A}_{L}$ (in Eqs.~\eqref{eq:direct-cs-long-pol} and \eqref{eq:direct-cs-long-pol-A}) already contains \emph{only} dipole correlators. Taking the naive limit $\bm{y}_{\perp}\rightarrow \bm{y'}_{\perp}$ in Eq.\,\eqref{eq:direct-cs-long-pol-B} is problematic because the logarithm in $\bm{r}_{yy' \perp}$ will diverge. We will therefore perform an approximation in which we Taylor expand the color structure around $\bm{r}_{yy' \perp}=0$ and perform the integration over $\bm{r}_{yy' \perp}$ using known identities.

From the expression for $\mathcal{B}_{L}$ given by Eq.\,\eqref{eq:direct-cs-long-pol-B}, we see that the integrals of interest can be expressed schematically in the following form 
\begin{align}
    I_{1}(\bm{v}_{\perp})&=\int_{\bm{u}_{\perp}} \ln \Big( \frac{1}{\mu_{F} u_{\perp}} \Big) \, e^{-i\bm{v}_{\perp}\cdot\bm{u}_{\perp}}f(\bm{u}_{\perp}) \, , \label{eq:integral-type-1-B}\\
     I_{2}^{i}(\bm{v}_{\perp})&=i\int_{\bm{u}_{\perp}}  \frac{\bm{u}_{\perp}^{i}}{\bm{u}_{\perp}^{2}}  e^{-i\bm{v}_{\perp}\cdot\bm{u}_{\perp}}f(\bm{u}_{\perp}) \, , \label{eq:integral-type-2-B}
\end{align}
where, for our computation, we have  $\bm{u}_{\perp}=\bm{r}_{yy' \perp}$, $\bm{v}_{\perp}=\frac{(1-z_{q})}{z_{\gamma}} \, \bm{k}_{\gamma \perp}$ and $f(\bm{u}_{\perp})$ represents the color structure denoted by $\Xi(\bm{x}_{\perp},\vect{y};\bm{y'}_{\perp},\bm{x'}_{\perp}\vert Y)$ in Eq.\,\ref{eq:color-structure-LO}. We can now do a Taylor expansion for the function $f$ around $\bm{u}_{\perp}=\bm{0}_{\perp}$ and rewrite the integral in Eq.\,\eqref{eq:integral-type-1-B} as
\begin{align}
    I_{1}(\bm{v}_{\perp}) \approx f(\bm{0}_{\perp}) \int_{\bm{u}_{\perp}} \ln \Big( \frac{1}{\mu_{F} u_{\perp}} \Big)\, e^{-i\bm{v}_{\perp}\cdot\bm{u}_{\perp}}+i \frac{\partial}{\partial v^{k}} \Big( \int_{\bm{u}_{\perp}} \ln \Big( \frac{1}{\mu_{F} u_{\perp}} \Big)\, e^{-i\bm{v}_{\perp}\cdot\bm{u}_{\perp}} \Big) \, (\partial_{k} f(\bm{0}_{\perp}))+\ldots \, .
\end{align}
We will now use the well-known identity (see for example Appendix A of \cite{Kovchegov:2012mbw})
\begin{equation}
    \int_{\bm{u}_{\perp}} \ln \Big( \frac{1}{a u_{\perp}} \Big) \, e^{-i\bm{v}_{\perp}\cdot\bm{u}_{\perp}}=2\pi \int u_{\perp} \mathrm{d} u_{\perp} \,  \ln \Big( \frac{1}{a u_{\perp}} \Big) J_{0}(u_{\perp}v_{\perp})=\frac{2\pi}{v_{\perp}^{2}} \, , 
    \label{eq:identity-leading-term-first-integral}
\end{equation}
where $J_{0}$ is a Bessel function of the first kind. In addition, we make the argument that the physical scale that sets the scale for the gradients of the Wilson line correlators contained in the function $f$, is the saturation momentum scale $Q_{s}$, i.e. $\partial_{k}f(\bm{0}_{\perp})\sim Q_{s}$. 

We can then write the integrals in Eqs.~\eqref{eq:integral-type-1-B} and \eqref{eq:integral-type-2-B} as 
\begin{align}
    I_{1}(\bm{v}_{\perp}) & \approx \frac{2\pi}{v_{\perp}^{2}} \, f( \bm{0}_{\perp}) \left( 1+O\left(\frac{Q_{s}}{v_{\perp}} \right) +\ldots \right) \, , \label{eq:limiting-form-int-type-1-B} \\
    I_{2}^{i} (\bm{v}_{\perp}) & \approx \frac{2\pi}{v_{\perp}^{2}} \bm{v}_{\perp}^{i} \, f(\bm{0}_{\perp}) \left( 1+O\left(\frac{Q_{s}}{v_{\perp}} \right) +\ldots \right) \, , \label{eq:limiting-form-int-type-2-B} 
\end{align}
where the relative magnitude between consecutive terms is $O(Q_{s}/v_{\perp})$.

In the limit, 
\begin{equation}
    \frac{1}{\vert \bm{y}_{\perp}-\bm{y'}_{\perp} \vert } \sim v_{\perp}=\frac{1-z_{q}}{z_{\gamma}} \, k_{\gamma \perp} \gg Q_{s} \, , 
    \label{eq:limit}
\end{equation} 
we can therefore consider only the leading term in the expansion and study qualitative features of the cross-section which is proportional to $f(\bm{0}_{\perp})$ or equivalently only contains dipoles as in Eq.\,\eqref{eq:coincidence-color-structure}.

We will now use the above arguments to obtain the leading term in the limit given by Eq.\,\eqref{eq:limit} for the term $\mathcal{B}_{L}$ in  Eq.\,\eqref{eq:direct-cs-long-pol-B}. We consider the dipoles to be translationally invariant so that
\begin{equation}
    S^{(2)}_{Y}(\vect{x},\vect{y}) \rightarrow S^{(2)}_{Y}(\vect{r}) \, \, , \, \, S^{(2)}_{Y}(\bm{x'}_{\perp},\bm{y'}_{\perp}) \rightarrow S^{(2)}_{Y}(\bm{r'}_{\perp}) \, \, , \, \, 
     S^{(2)}_{Y}(\vect{x},\vect{x}') \rightarrow S^{(2)}_{Y}(\vect{r}-\bm{r}_{\perp}^{\prime}) \, ,
\end{equation}
for $\bm{y}_{\perp}-\bm{y'}_{\perp}=0$, and we can write
\begin{equation}
    \Theta(\bm{r}_{\perp},\bm{r'}_{\perp} \vert Y)=1-S^{(2)}_{Y}(\vect{r})-S^{(2)}_{Y}(\bm{r'}_{\perp})+S^{(2)}_{Y}(\vect{r}-\bm{r}_{\perp}^{\prime})  \, .
    \label{eq:coincidence-color-structure-trans-invariance}
\end{equation}
Finally, we make the following simple redefinitions in order to obtain a compact expression for the direct photon+quark cross-section that may be expressed entirely in momentum space:
\begin{align}
   \mathcal{J}_{L,1}^{i} (\bm{r}_{\perp}) \, e^{-i\bm{k}_{\gamma \perp}\cdot\bm{r}_{\perp}}&= \mathcal{K}_{L,1}^{i} (\bm{r}_{\perp})=\int_{\vect{l}} e^{i\vect{l}\cdot\vect{r}} \widetilde{\mathcal{K}}^{i}_{L,1}(\vect{l}+\vect{k_{\gamma}})  \, , \label{eq:K-L-1} \\
    \mathcal{J}_{L,2}^{i} (\bm{r}_{\perp}) \, e^{-i\bm{k}_{\gamma \perp}\cdot\bm{r}_{\perp}}&= \mathcal{K}_{L,2}^{i} (\bm{r}_{\perp}) =\int_{\vect{l}} e^{i\vect{l}\cdot\vect{r}} \widetilde{\mathcal{K}}^{i}_{L,2}(\vect{l}+\vect{k_{\gamma}}) \, , \label{eq:K-L-2} \\
    \frac{z_{\gamma}}{1-z_{q}} \, \frac{\bm{k}_{\gamma \perp}^{i}}{\bm{k}_{\gamma \perp}^{2}} \, \hat{\mathcal{J}}_{L,3}(r_{\perp})&= \mathcal{K}_{L,3}^{i} (\bm{r}_{\perp}) =  \int_{\vect{l}} e^{i\vect{l}\cdot\vect{r}} \widetilde{\mathcal{K}}^{i}_{L,3}(\vect{l}) \, , \label{eq:K-L-3} \\
    \mathcal{J}^{i}_{L,4}(\bm{r}_{\perp})&=\mathcal{K}_{L,4}^{i} (\bm{r}_{\perp}) =  \int_{\vect{l}} e^{i\vect{l}\cdot\vect{r}} \widetilde{\mathcal{K}}^{i}_{L,4}(\vect{l}) \, . \label{eq:K-L-4} 
\end{align}
The functions $\mathcal{J}^{i}_{L,1}$ and  $\mathcal{J}^{i}_{L,2}$ are given by  Eqs.~\eqref{eq:J-L-1} and \eqref{eq:J-L-2}.  $\mathcal{J}^{i}_{L,4}$ is obtained from Eq.\,\eqref{eq:J-L-2} by $q\leftrightarrow \bar{q}$ interchange while $\hat{J}_{L,3}$ is given in Eq.\,\eqref{eq:J-hat-L-3}. The expressions for the complex conjugates can be obtained straightforwardly from the above equations. In Eqs.~\eqref{eq:K-L-1}-\eqref{eq:K-L-4}  appearing above, $\widetilde{\mathcal{K}}_{L}$'s represent the momentum space versions of their coordinate space counterparts.

In terms of these newly defined functions $\mathcal{K}_{L,p}^{i}$ ($p=1,\ldots,4$), we can combine the terms $\mathcal{A}_{L}$ and $\mathcal{B}_{L}$ in Eqs.~\eqref{eq:direct-cs-long-pol-A} and \eqref{eq:direct-cs-long-pol-B} to obtain the following expression for the \emph{leading} contribution to the direct photon+quark cross-section-in the limit described by Eq.\,\eqref{eq:limit}:
\begin{align}
     \frac{d \sigma^{\gamma^{*}_{L} A \rightarrow q \gamma X}_{D}}{\mathrm{d}^{2} \vect{k}  \mathrm{d}^{2} \vect{k_{\gamma}}    \mathrm{d} \eta_{q} \mathrm{d} \eta_{\gamma}} \Bigg \vert_{\rm{lead.}}&= \frac{16N_{c} \alpha_{\rm em}^{2}q^{4}_{f} z_{q}^{3}(1-z_{q}-z_{\gamma})^{2}}{(2\pi)^{4}} \, S_{\perp}  \int_{\bm{r}_{\perp}, \bm{r'}_{\perp}} \!\!\!\!\!\!\!\!\! e^{-i\bm{k}_{\perp}.(\bm{r}_{\perp}-\bm{r'}_{\perp})} \Theta(\bm{r}_{\perp},\bm{r'}_{\perp} \vert Y) \, \mathcal{R}^{q\gamma}_{L}(\bm{r}_{\perp},\bm{r'}_{\perp};k,k_{\gamma}) \, ,
     \label{eq:diff-cs-direct-photon-jet-limit-cspace}
\end{align}
where $S_{\perp}$ represents the transverse area of the target nucleus and
\begin{align}
  \mathcal{R}^{q\gamma}_{L} &=\xi_{qq}  \Big[ \mathcal{K}_{L,1}^{i}(\bm{r}_{\perp})+\mathcal{K}_{L,2}^{i}(\bm{r}_{\perp}) \big] \big[ \mathcal{K}_{L,1}^{i \, *}(\bm{r'}_{\perp})+\mathcal{K}_{L,2}^{i \, * }(\bm{r'}_{\perp}) \Big]-\xi_{q\bar{q}} \Big[ \mathcal{K}_{L,1}^{i}(\bm{r}_{\perp})+\mathcal{K}_{L,2}^{i}(\bm{r}_{\perp}) \Big] \Big[ \mathcal{K}_{L,3}^{i \, *}(\bm{r'}_{\perp})+\mathcal{K}_{L,4}^{i \, * }(\bm{r'}_{\perp}) \Big] \nonumber \\
  & -\xi_{\bar{q}q} \Big[ \mathcal{K}_{L,3}^{i}(\bm{r}_{\perp})+\mathcal{K}_{L,4}^{i}(\bm{r}_{\perp}) \Big] \Big[ \mathcal{K}_{L,1}^{i \, *}(\bm{r'}_{\perp})+\mathcal{K}_{L,2}^{i \, * }(\bm{r'}_{\perp}) \Big] + \xi_{\bar{q}\bar{q}} \Big[ \mathcal{K}_{L,3}^{i}(\bm{r}_{\perp})+\mathcal{K}_{L,4}^{i}(\bm{r}_{\perp}) \big] \big[ \mathcal{K}_{L,3}^{i \, *}(\bm{r'}_{\perp})+\mathcal{K}_{L,4}^{i \, * }(\bm{r'}_{\perp}) \Big] \, .
  \label{eq:R-direct-qgamma-limit-long-cspace}
\end{align}
We can now define the CGC averaged dipole correlators as
\begin{equation}
    C_{Y}(\vect{l}) = \int_{\vect{r}} e^{-i \vect{l}\cdot \vect{r}} \, S^{(2)}_{Y}(\vect{r}) \, , 
\end{equation}
and use Eqs.~\eqref{eq:K-L-1}-\eqref{eq:K-L-4} to express Eq.\,\eqref{eq:diff-cs-direct-photon-jet-limit-cspace} in terms of momentum space quantities as
\begin{align}
     \frac{d \sigma^{\gamma^{*}_{L} A \rightarrow q \gamma X}_{D}}{\mathrm{d}^{2} \vect{k}  \mathrm{d}^{2} \vect{k_{\gamma}}    \mathrm{d} \eta_{q} \mathrm{d} \eta_{\gamma}} \Bigg \vert_{\rm{lead.}}&= \frac{16N_{c} \alpha_{\rm em}^{2}q^{4}_{f} z_{q}^{3}(1-z_{q}-z_{\gamma})^{2}}{(2\pi)^{4}} \, S_{\perp} \int_{\bm{l}_{\perp}} 
    C_{Y}(\vect{l}) \mathcal{H}_L^{q\gamma}(\vect{l};k,k_\gamma)  \, ,
    \label{eq:diff-cs-direct-limit-long-mspace}
\end{align}
where
\begin{align}
    \mathcal{H}_L^{q\gamma}(\vect{l};k,k_\gamma)  &= \xi_{qq} \sum_{n,m=1}^2 \left[\widetilde{\mathcal{K}}^{i}_{L,n}(\vect{k}+\vect{k_\gamma}) -\widetilde{\mathcal{K}}^{i}_{L,n}(\vect{k}+\vect{k_\gamma}-\vect{l})\right] \left[\widetilde{\mathcal{K}}^{i}_{L,m}(\vect{k}+\vect{k_\gamma}) -\widetilde{\mathcal{K}}^{i}_{L,m}(\vect{k}+\vect{k_\gamma}-\vect{l})\right]  \nonumber \\
    &-\xi_{q\bar{q}}\sum_{n,m=1}^2 \left[\widetilde{\mathcal{K}}^{i}_{L,n}(\vect{k}+\vect{k_\gamma}) -\widetilde{\mathcal{K}}^{i}_{L,n}(\vect{k}+\vect{k_\gamma}-\vect{l})\right] \left[\widetilde{\mathcal{K}}^{i}_{L,m+2}(\vect{k}) -\widetilde{\mathcal{K}}^{i}_{L,m+2}(\vect{k}-\vect{l})\right] \nonumber \\
    &-\xi_{\bar{q}q}\sum_{n,m=1}^2 \left[\widetilde{\mathcal{K}}^{i}_{L,n+2}(\vect{k}) -\widetilde{\mathcal{K}}^{i}_{L,n+2}(\vect{k}-\vect{l})\right] \left[\widetilde{\mathcal{K}}^{i}_{L,m}(\vect{k}+\vect{k_\gamma}) -\widetilde{\mathcal{K}}^{i}_{L,m}(\vect{k}+\vect{k_\gamma}-\vect{l})\right] \nonumber \\
    & +\xi_{\bar{q}\bar{q}}\sum_{n,m=1}^2 \left[\widetilde{\mathcal{K}}^{i}_{L,n+2}(\vect{k}) -\widetilde{\mathcal{K}}^{i}_{L,n+2}(\vect{k}-\vect{l})\right] \left[\widetilde{\mathcal{K}}^{i}_{L,m+2}(\vect{k}) -\widetilde{\mathcal{K}}^{i}_{L,m+2}(\vect{k}-\vect{l})\right] \, .
    \label{eq:H-L-q-gamma}
\end{align}
In Eq.\,\eqref{eq:diff-cs-direct-limit-long-mspace}, $C_{Y}(\vect{l})$ contains the non-perturbative input that has to be described by a model such as the MV or Golec-Biernat Wusthoff (GBW)~\cite{GolecBiernat:1998js} models. The MV model provides initial conditions for the nucleus in the CGC approach whereas the GBW model describes high energy DIS data using a simple parametrization for the dipole amplitude.

For the longitudinally polarized virtual photon, the functions $\widetilde{\mathcal{K}}_{L}$ constituting the momentum space expression for the cross-section in Eq.\,\eqref{eq:diff-cs-direct-limit-long-mspace} are obtained as
\begin{align}
    \widetilde{\mathcal{K}}_{L,1}^{i}(\vect{l})&=(z_{q}+z_{\gamma}) \frac{z_{q}\bm{k}_{\gamma \perp}^{i} -z_{\gamma} \bm{k}_{\perp}^{i}}{(z_{q} \bm{k}_{\gamma \perp}-z_{\gamma} \bm{k}_{\perp})^{2}} \, \frac{Q}{\bm{l}_{\perp}^{2}+\Delta_{1}} \,  , \label{eq:K-tilde-L-1} \\
   \widetilde{\mathcal{K}}_{L,2}^{i}(\vect{l})&=  -\frac{Q}{z_{q}z_{\gamma}} \, \frac{(z_{q}+z_{\gamma}) \bm{k}_{\gamma \perp}^{i}-z_{\gamma}\bm{l}_{\perp}^{i}}{(\bm{l}_{\perp}^{2}+\Delta_{1}) \, \Big( Q^{2}+\frac{\bm{l}^{2}_{\perp}}{1-z_{q}-z_{\gamma}} +\frac{\bm{k}_{\gamma \perp}^{2}}{z_{\gamma}}+\frac{(\bm{l}_{\perp}-\bm{k}_{\gamma \perp})^{2}}{z_{q}} \Big)} \, , \label{eq:K-tilde-L-2} \\
   \widetilde{\mathcal{K}}_{L,3}^{i}(\vect{l})&= \frac{\bm{k}_{\gamma \perp}^{i}}{\bm{k}_{\gamma \perp}^{2}} \, \frac{Q}{\bm{l}_{\perp}^{2}+\Delta_{3}}   \, , \label{eq:K-tilde-L-3} \\
  \widetilde{\mathcal{K}}_{L,4}^{i}(\vect{l})&= -\frac{Q}{z_{\gamma}(1-z_{q}-z_{\gamma})} \frac{(1-z_{q}) \bm{k}_{\gamma \perp}^{i}+z_{\gamma} \bm{l}_{\perp}^{i}}{(\bm{l}_{\perp}^{2}+\Delta_{3}) \, \Big( Q^{2}+\frac{\bm{l}^{2}_{\perp}}{z_{q} }+\frac{\bm{k}_{\gamma \perp}^{2}}{z_{\gamma}}+\frac{(\bm{l}_{\perp}+\bm{k}_{\gamma \perp})^{2}}{1-z_{q}-z_{\gamma}} \Big)} \, . \label{eq:K-tilde-L-4}
\end{align}
The coefficients $\xi$'s are given in Eqs.~\eqref{eq:xi-qq-xi-qbar-xi-bar}-\eqref{eq:xi-q-qbar} and $$\Delta_{1}=(z_{q}+z_{\gamma}) (1-z_{q}-z_{\gamma}) Q^{2} \quad \, , \quad \Delta_{3}=z_{q}(1-z_{q}) Q^{2}\, .$$

Similar expressions for transversely polarized virtual photons are provided in Appendix~\ref{sec:direct-photon-limit-trans}. Substituting these results into  Eq.\,\eqref{eq:diff-cs-gammajet-general}, we obtain a simple momentum space expression for the differential cross-section for prompt ``direct" photon+quark production in $e+A$ DIS at small $x$.
Using suitable models for the non-perturbative content, we will now evaluate these expressions numerically and study observables that provide novel ways to study gluon saturation.

\section{Azimuthal angle correlations in direct photon+quark production}
\label{sec:azimuthal-correlations}

Azimuthal angle correlations in two-particle production have been proposed previously to be a sensitive probe of gluon saturation inside nuclear matter \cite{Kharzeev:2004bw}. Gluon saturation leads to a suppression of the back-to-back peak in the correlation of dihadrons/dijets in $e+A$~\cite{Zheng:2014vka} and $p+A$~\cite{Lappi:2012nh,Albacete:2010pg,Albacete:2018ruq,Stasto:2018rci} collisions, and in the correlation of photons and hadrons in $p+A$ collisions~\cite{JalilianMarian:2012bd, Goncalves:2020tvh}. To go beyond these studies, we compute azimuthal angle correlations of photons and quarks in $e+A$ DIS for the first time. We focus our attention on the production of direct photons, and leave the analysis of fragmentation photons for future work. Towards this end, we will study the behavior of the leading contribution to the differential cross-section (normalized by the transverse area of the target) as a function of the azimuthal angle between the direct photon and the quark $\Delta \phi_{q\gamma}$, and integrated over transverse momenta and rapidities:
\begin{align}
    \frac{\mathrm{d} N_{q\gamma}(\Delta \phi_{q\gamma})}{\mathrm{d} Q^2 \mathrm{d} W^2} &= \frac{1}{S_\perp} \sum_{f=u,d,s} \int \frac{\mathrm{d} \sigma_{D} ^{eA \rightarrow e'q\gamma X}}{\mathrm{d}W^{2} \mathrm{d}Q^{2} d k^2_\perp \mathrm{d} \eta_{q} \mathrm{d} k^2_{\gamma,\perp} \mathrm{d} \eta_{\gamma} \mathrm{d} (\Delta \phi_{q\gamma}) } \Bigg \vert_{\rm{lead.}} \mathrm{d} k^2_\perp \mathrm{d} \eta_q \mathrm{d} k^2_{\gamma,\perp} \mathrm{d} \eta_\gamma \, . \label{eq-direct-qgamma}
\end{align}
The differential cross-section for the leading contribution to  direct photon+quark production can be found by computing Eq.\,\eqref{eq:diff-cs-gammajet-general} together with Eq.\,\eqref{eq:diff-cs-direct-limit-long-mspace} and Eq.\,\eqref{eq:diff-cs-direct-photon-jet-limit-mspace-trans-pol}. 
We further normalize our results by the single inclusive quark production cross-section (see Appendix\,\ref{sec:single-inclusive-parton-cs})
\begin{align}
    \frac{\mathrm{d} N_{q}}{\mathrm{d} Q^2 \mathrm{d} W^2} &= \frac{1}{S_\perp} \sum_{f=u,d,s}  \int \frac{d \sigma ^{eA \rightarrow e'q X}}{\mathrm{d}W^{2} \mathrm{d}Q^{2} \mathrm{d} k^2_\perp   \mathrm{d} \eta_{q}} \mathrm{d} k^2_\perp \mathrm{d} \eta_q \, , \label{eq-semi-inclusive-q}
\end{align}
to define the correlation coefficient (associated yield) of direct photons for a quark/jet trigger\footnote{This observable has been studied in the production of dihadrons in $p+A$ collisions in \cite{Albacete:2010pg} and \cite{Albacete:2018ruq} .}:
\begin{align}
    \mathcal{C}(\Delta \phi_{q\gamma}) = \frac{ \mathrm{d} N_{q\gamma}(\Delta \phi_{q\gamma}) }{\mathrm{d} Q^2 \mathrm{d} W^2}\Big / \frac{\mathrm{d} N_{q}}{\mathrm{d} Q^2 \mathrm{d} W^2} \, , \label{eq:coincidence-probability.}
\end{align}
where we have summed over the contributions of the three light-quark flavors in defining Eqs. \eqref{eq-direct-qgamma} and \eqref{eq-semi-inclusive-q}. 

\subsection{Set-up: proton and nuclear dipole correlators from running coupling Balitsky-Kovchegov evolution}
\label{sec:azimuthal-correlations-setup}

Azimuthal angle correlations from the direct photon+quark alone depend on $C_Y(\vect{l})$, the two-point dipole correlator in momentum space. We obtain $C_Y(\vect{l})$ from the Fourier transform of the coordinate space solution to the running coupling Balitsky-Kovchegov (rcBK) equation \cite{Balitsky:1995ub,Kovchegov:1999yj,Kovchegov:1999ua,Balitsky:2008zza,Albacete:2007yr}:
\begin{align}
    \frac{\mathrm{d} S^{(2)}_Y(\vect{r})}{\mathrm{d} Y} = \int_{\vect{r_1}} \mathcal{K}^{rc}(\vect{r},\vect{r_1},\vect{r_2}) \left[  S^{(2)}_Y(\vect{r_1}) S^{(2)}_Y(\vect{r_2}) - S^{(2)}_Y(\vect{r}) \right]\,,
\end{align}
where $\vect{r_2}=\vect{r}-\vect{r_1}$, and the running coupling kernel $\mathcal{K}^{rc}$ is given by \cite{Balitsky:2006wa},
\begin{align}
    \mathcal{K}^{rc}(\vect{r},\vect{r_1},\vect{r_2}) = \frac{N_{\mathrm{c}} \alpha_{\mathrm{s}} (\vect{r})}{2 \pi^2} \left[ \frac{\vect{r}^2}{\vect{r_1}^2\vect{r_2}^2} + \frac{1}{\vect{r_1}^2}\left( \frac{\alpha_{\mathrm{s}}(\vect{r_1})}{\alpha_{\mathrm{s}}(\vect{r_2})}-1 \right) + \frac{1}{\vect{r_2}^2} \left( \frac{\alpha_{\mathrm{s}}(\vect{r_2})}{\alpha_{\mathrm{s}}(\vect{r_1})}-1 \right) \right]\,.
\end{align}
The coordinate space coupling in the above is given by 
\begin{align}
    \alpha_{\mathrm{s}}(\vect{r}) = \frac{12 \pi}{(33 - 2 N_f) \log(\frac{4 C^2}{\vect{r}^2\Lambda^2_{\rm QCD}})}\,,
\end{align}
with $\Lambda_{\rm QCD}=0.241 \ \rm{GeV}$, $N_f=3$, and $C$ a dimensionless parameter controlling the running of the coupling.

The initial conditions for the evolution are given by the modified MV model with a regulator $e_c'$ (MVe) used in prior phenomenological studies of data from HERA and from proton-nucleus collisions at RHIC aand the LHC~\cite{Lappi:2013zma}: 
\begin{align}
    S_{Y=0}^{(2)}(\vect{r}) = \exp[-\frac{1}{4}Q^2_{sp,0} r^2_\perp \log(\frac{1}{r \Lambda_{\rm QCD}}+e_c')] \, . \label{eq:proton_dipole}
\end{align}

\begin{figure}[!htbp]
    \centering
    \includegraphics[scale=0.45]{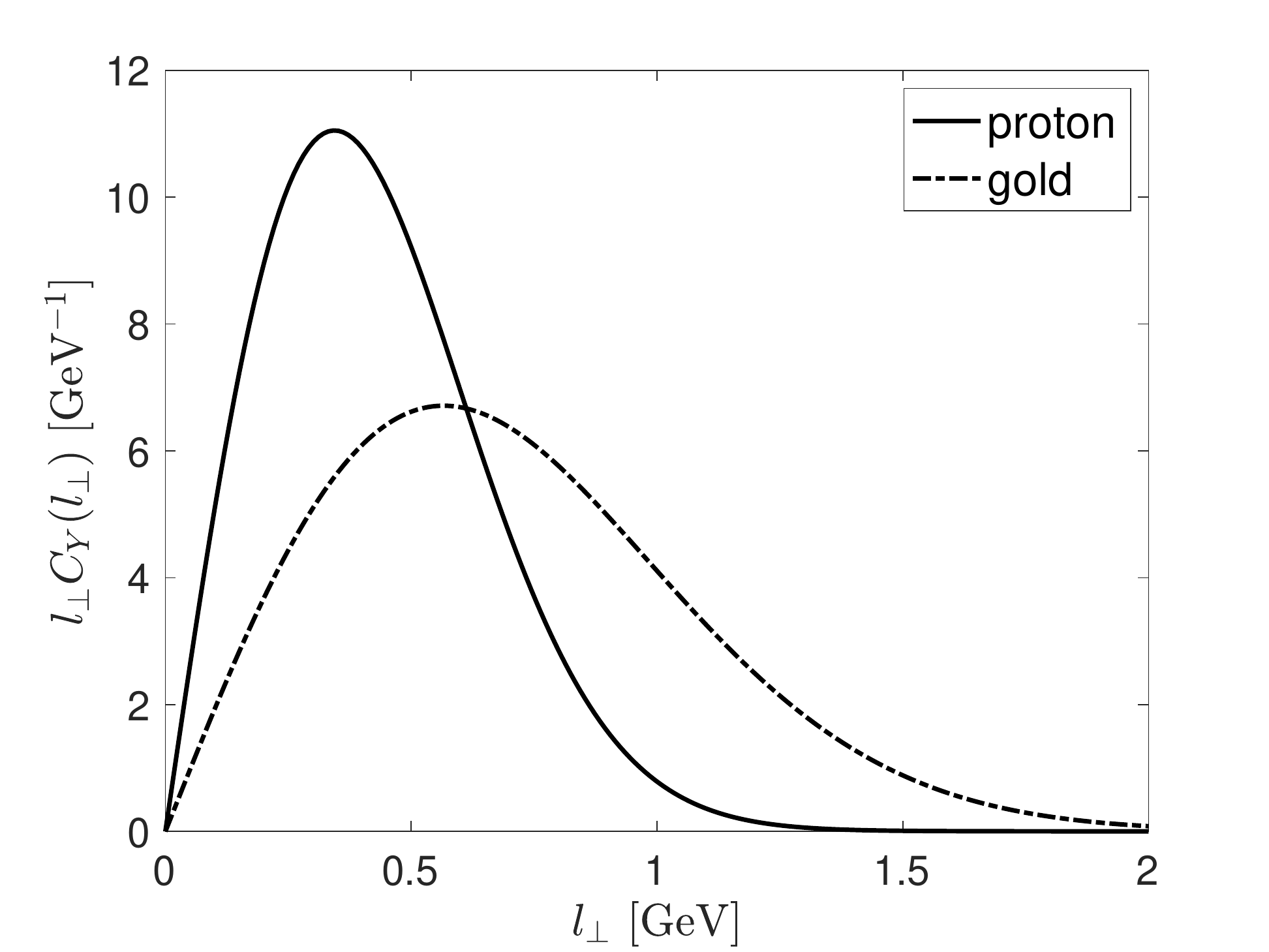}
    \caption{Proton and gold dipole correlators in momentum space at the initial rapidity $Y=0$ prior to rcBK evolution with $Y$. The nuclear dipole correlator is broader and peaked at a larger value of momentum relative to the proton dipole correlator. Note that in our convention for the Fourier transform we have $\int_{\vect{l}} C_Y(\vect{l}) =1 $.}
    \label{fig:dipole_correlator}
\end{figure}

The parameters are $C = 7.2$, $e_c' = 51.4$ and $Q^2_{sp,0} = 0.060\ \mathrm{GeV^2}$. Note that the latter two correspond to a saturation scale $Q^2_s = 0.238 \ \mathrm{GeV^2}$, defined by $S^{2}_{Y=0}(r^2 = 2 / Q^2_s) = \exp(-1/2)$. They have been choosen to fit inclusive DIS HERA $e+p$ data, and provide a good phenomenological description of the proton dipole correlator \cite{Lappi:2013zma}. (See also~\cite{Albacete:2010sy}.) The nuclear dipole correlator is computed via the optical Glauber model, generalizing Eq.\,\eqref{eq:proton_dipole} by the replacement,
\begin{align}
    Q^2_{sA,0}(\vect{b}) \rightarrow A S_{p,\perp} T_A(\vect{b}) Q^2_{sp,0} \, ,
\end{align}
where $A$ is the total number of nucleons in the nuclear target, $S_{p,\perp} = 16.4 \ \rm{mb}$ is the effective area of the proton\footnote{This is also a parameter of the MVe fit in~\cite{Lappi:2013zma}.}, and 
\begin{align}
    T_A(\vect{b}) = \int_{-\infty}^{\infty} \frac{n \ dz}{1+\exp[\frac{\sqrt{\vect{b}^2+z^2}-R_A}{a}]} \,,
\end{align}
is the nuclear thickness function obtained by integrating the Woods-Saxon distribution along the $z-$direction, with  $n$ chosen such that $\int_{\vect{b}}T_A(\vect{b}) = 1$.
For the gold nucleus ($A=197$), we take $R_A = 6.37$ fm, and $a = 0.54$ fm. 

Determining  the impact parameter in electron-nucleus collisions is difficult due to the lack of centrality classes. We will here, in this exploratory study, evaluate the saturation scale at the mean impact parameter ($\langle \vect{b} \rangle = 0.59 \times R_A$) as a proxy for minimum bias. This results in the  relation $Q^2_{sA,0}(\langle \vect{b} \rangle) =2.7 \times  Q^2_{sp,0} $ between the nuclear and proton saturation scales at the initial scale $x_0$ corresponding to the initial rapidity $Y=0$. The dipole correlator in momentum space can be computed with the given input parameters for both the proton and the gold nucleus. The result is plotted in  Fig.\,\ref{fig:dipole_correlator}.

Given these initial conditions, the rcBK equation is evolved up to $Y=\log(x_0/x_g)$, where $x_g$ is determined by the kinematics of the final state particles. For $q+\gamma$ production we have,
\begin{align}
    x_g= \frac{Q^2}{W^2} + \frac{k_\perp e^{-\eta_q} + k_{\gamma,\perp} e^{-\eta_{\gamma}} }{2 E_p} \, , 
\end{align}
where $E_p$ is the energy of the proton. This relation follows from energy-momentum conservation\footnote{We assumed that the typical transverse momentum of the integrated antiquark is small and thus the antiquark does not carry a significant fraction of the longitudinal momentum.}. We will perform our study in the frame where the virtual photon and the proton have zero transverse momentum and the momentum of the proton is that of the laboratory frame. 

\subsection{Numerical results for azimuthal angle correlations in $e+p$ and $e+Au$ collisions}
\label{sec:azimuthal-correlations-numerical-results}

\begin{figure}[!htbp]
    \centering
    \includegraphics[scale=0.45]{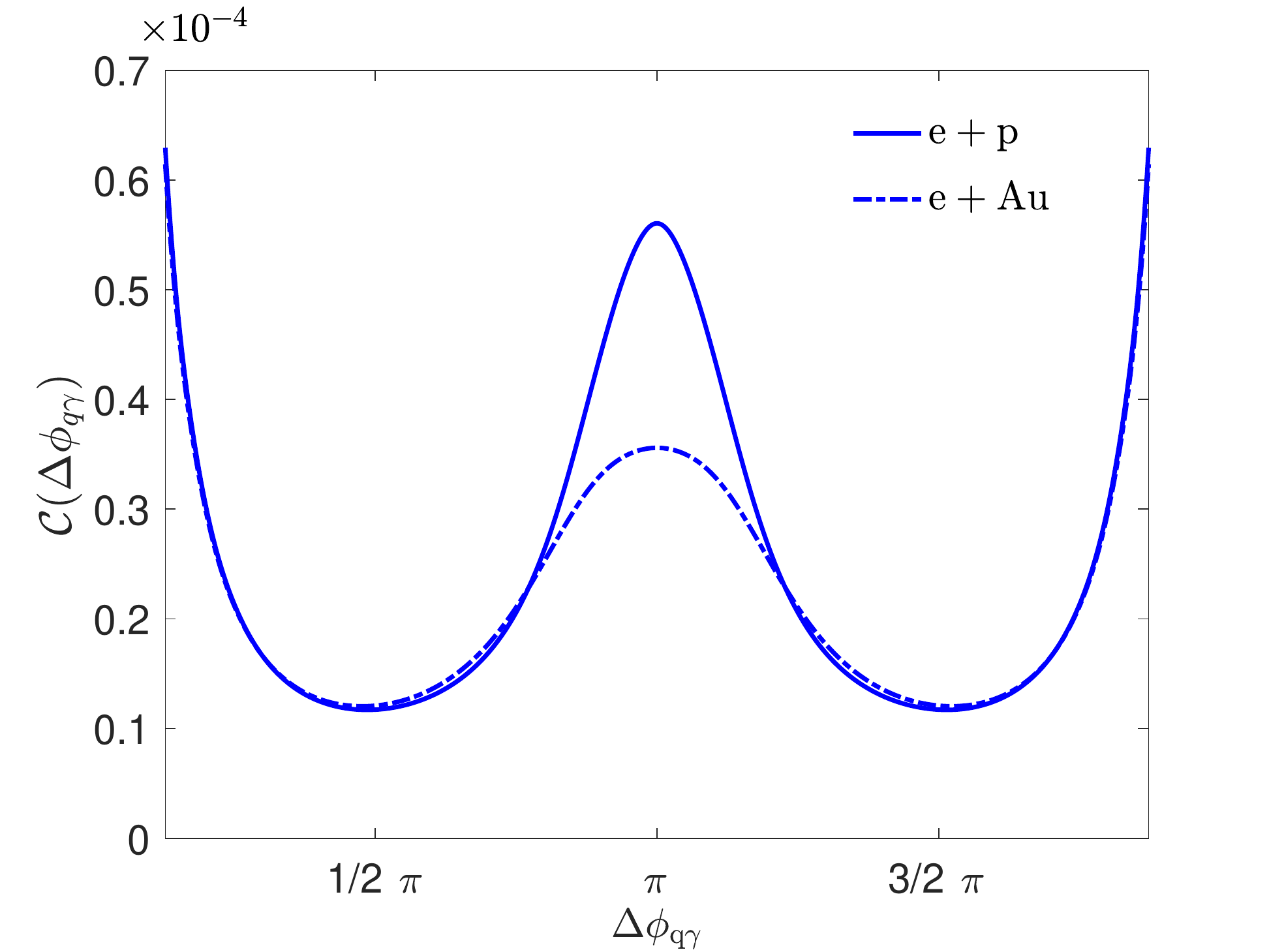}
    \includegraphics[scale=0.45]{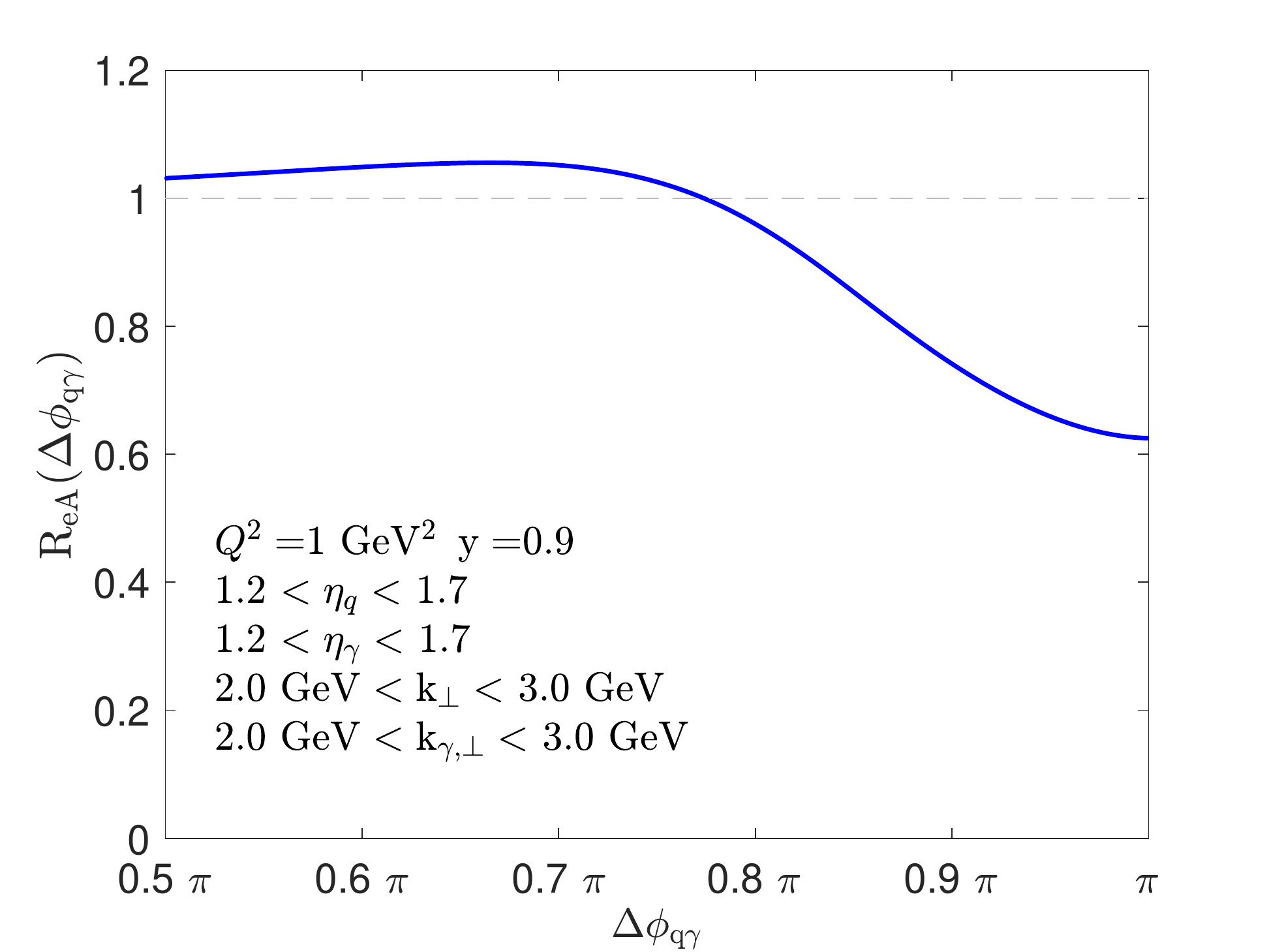}
    \caption{Left: The azimuthal correlator (defined in Eq.~(\ref{eq:coincidence-probability.})) as a function of the relative azimuthal angle $\Delta \phi_{q\gamma}$ between the direct photon and the quark in $e+p$ and $e+Au$ collisions. Right: The ratio $R_{eA}$ (see Eq.~(\ref{eq:R-eA})) of azimuthal angle correlations in $e+Au$ relative those in $e+p$ collisions as a function of $\Delta \phi_{q\gamma}$.}
    \label{fig:integrated_azimuthal_correlations}
\end{figure}

We present in Fig.\,\ref{fig:integrated_azimuthal_correlations} our numerical results for the direct photon-jet azimuthal angle correlations in $e+p$ and minimum bias $e+Au$  collisions at the center-of-mass energy $\sqrt{s} = 90 \,\mathrm{GeV}$ which is close to the anticipated EIC center-of-mass energy per nucleon. The kinematical range in transverse momenta and rapidities of the photon and quark have been choosen so that $x_g$ remains smaller than $10^{-2}$. Further, we select an inelasticity $y$ close to $1$ so that we maximize the center of mass energy squared $W^2$ of the photon-nucleon system for a fixed $s$.

The left panel of Fig.\,\ref{fig:integrated_azimuthal_correlations},
shows the yield of direct photons associated to DIS events with at least one quark/jet as a function of the relative azimuthal angle between the direct photon and the jet. We observe a back-to-back peak in the azimuthal angle correlation ($\Delta \phi_{q\gamma} \sim \pi$), which is suppressed and broadened for collisions with the denser nuclear target compared to those in proton. The origin of the supression in the back-to-back peak can be traced to the broadening of the dipole correlator in momentum space as the saturation scale is increased, as shown in Fig.\,\ref{fig:dipole_correlator}. Physically, the broadening of $C_Y(\vect{l})$ corresponds to an increase in the momentum transfer imparted to $q\gamma$ system as it multiple scatters from the nuclear target, resulting in the decorrelation of the back-to-back peak.  An additional imbalance of the transverse momenta of the $\gamma q$ is produced by the transverse momentum carried by the integrated antiquark $\bar{q}$; however, the integration over its phase-space is dominated by small transverse momenta.

We also observe an enhacement in the correlation as the photon is emitted collinearly from the quark ($\Delta \phi_{q\gamma} \sim 0, 2\pi$). The collinear enhancement can be easily identified in Eqs.~\eqref{eq:K-tilde-L-1} and \eqref{eq:K-tilde-T-1}. It occurs in the contribution where the photon is emitted from the quark after it scatters from the nuclear target. Our numerical result shows that the collinear region is less sensitive to nuclear effects, as little difference is observed when the nuclear species is changed from proton to gold.

To more efficiently characterize the supression in the back-to-back peak, we study the ratio
\begin{align}
    R_{eA} = \frac{\mathcal{C}(\Delta \phi_{q\gamma}) [\mathrm{eAu}]}{\mathcal{C}(\Delta \phi_{q\gamma}) [\mathrm{ep} ]} \,,
    \label{eq:R-eA}
\end{align}
shown in the right panel of Fig.\,\ref{fig:integrated_azimuthal_correlations}. Near the back-to-back peak we observe a significant supression for the yield of direct photons when comparing proton DIS to nuclear DIS in the same transverse momenta and rapidity bins. 

We can also study the dependence of $R_{eA}$ on the virtuality $Q^2$ of the virtual photon. We find that the ratio $R_{eA}$ converges towards unity as the $Q^2$ of the DIS probe increases, signaling the weakening of the suppression of back-to-back correlations. This behavior is characteristic of gluon saturation: as the virtuality $Q^2$ is increased, the system becomes more dilute and the multiple scattering and shadowing of gluons becomes less dominant. In the dilute limit of $Q^2 \gg Q_s^2$, one can approximate the Wilson line correlators in  Eq.\,\eqref{eq:coincidence-color-structure-trans-invariance} by
\begin{align}
    \Theta(\bm{r}_{\perp},\bm{r'}_{\perp} \vert Y=0) \approx \frac{1}{2} Q^2_s \, \left[ \Gamma(\vect{r}) + \Gamma(\vect{r'}) - \Gamma(\vect{r}-\vect{r}') \right]\,,
\end{align}
where $S^{(2)}_{Y=0}(r_{\perp}) = \exp(-\frac{1}{4}Q_s^2 \Gamma(\vect{r}))$.
In this approximation, the saturation scale appears as a multiplicative factor in the differential cross-sections in Eqs.\,\eqref{eq:diff-cs-direct-photon-jet-limit-cspace},\,\eqref{eq:diff-cs-direct-photon-jet-limit-cspace-trans-pol},\,\eqref{eq-semi_inclusive_cs_long}\,and \eqref{eq-semi_inclusive_cs_trans}. It therefore cancels in the correlator $\mathcal{C}(\Delta \phi_{q\gamma})$, resulting in the ratio $R_{eA}$ becoming identical to unity. Thus deviations from unity in  $R_{eA}$ provide a good measure of the sensitivity of the azimuthal correlations to the many-body multiple scattering and color screening (``shadowing") effects due the dense gluon system comprising nuclear matter at high energies. The study of the $Q^2$ dependence of correlation measurements provides an additional handle in $e+A$ collions relative to $p+A$ collisions for systematic studies of gluon saturation phenomena.

\begin{figure}[!htbp]
    \centering
    \includegraphics[scale=0.52]{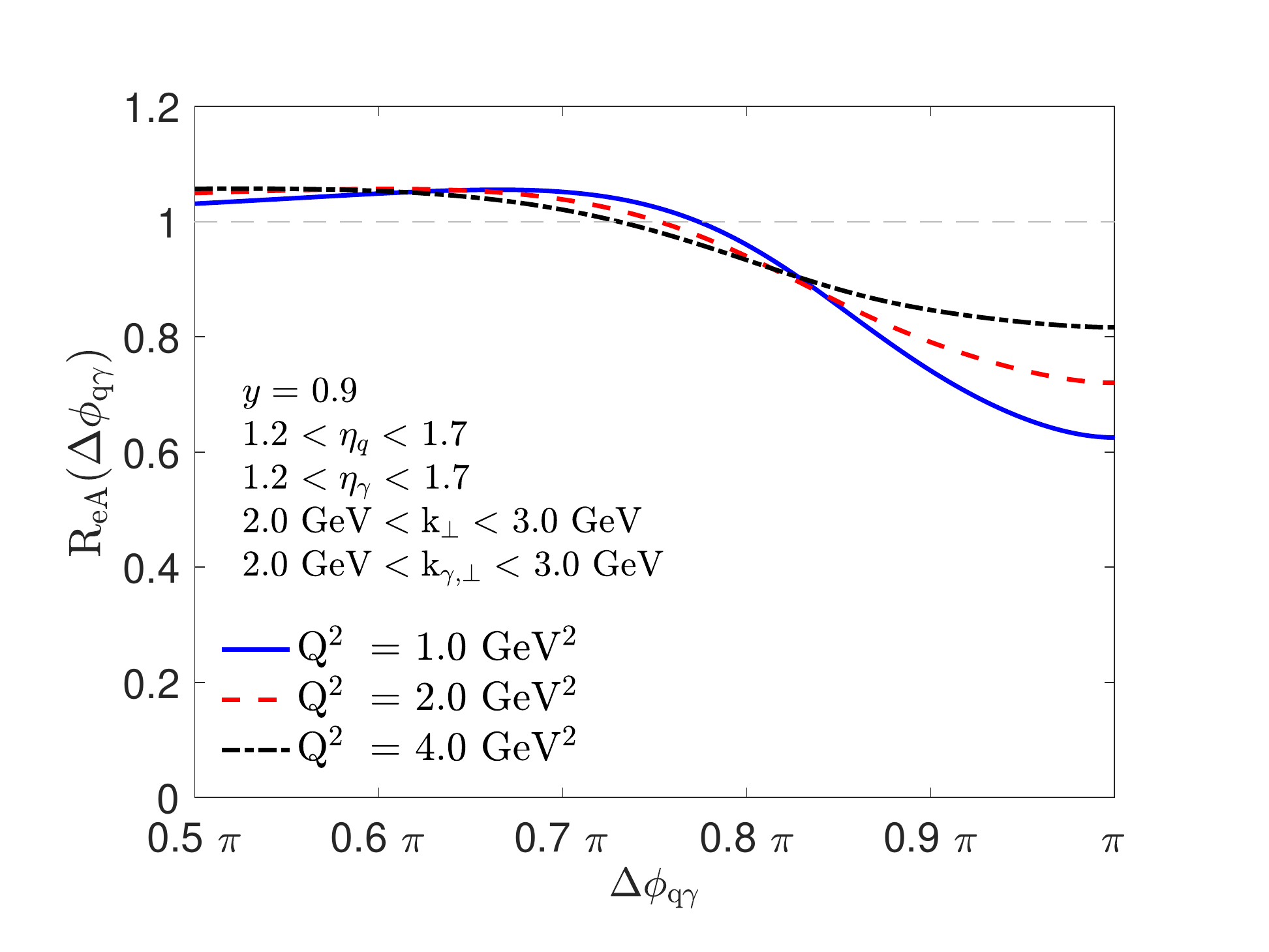}
    \caption{$R_{eA}$ dependence of photon+quark correlations as a function of $Q^2$.}
    \label{fig:Q2_dependence}
\end{figure}

\begin{figure}[!htbp]
    \centering
    \includegraphics[scale=0.45]{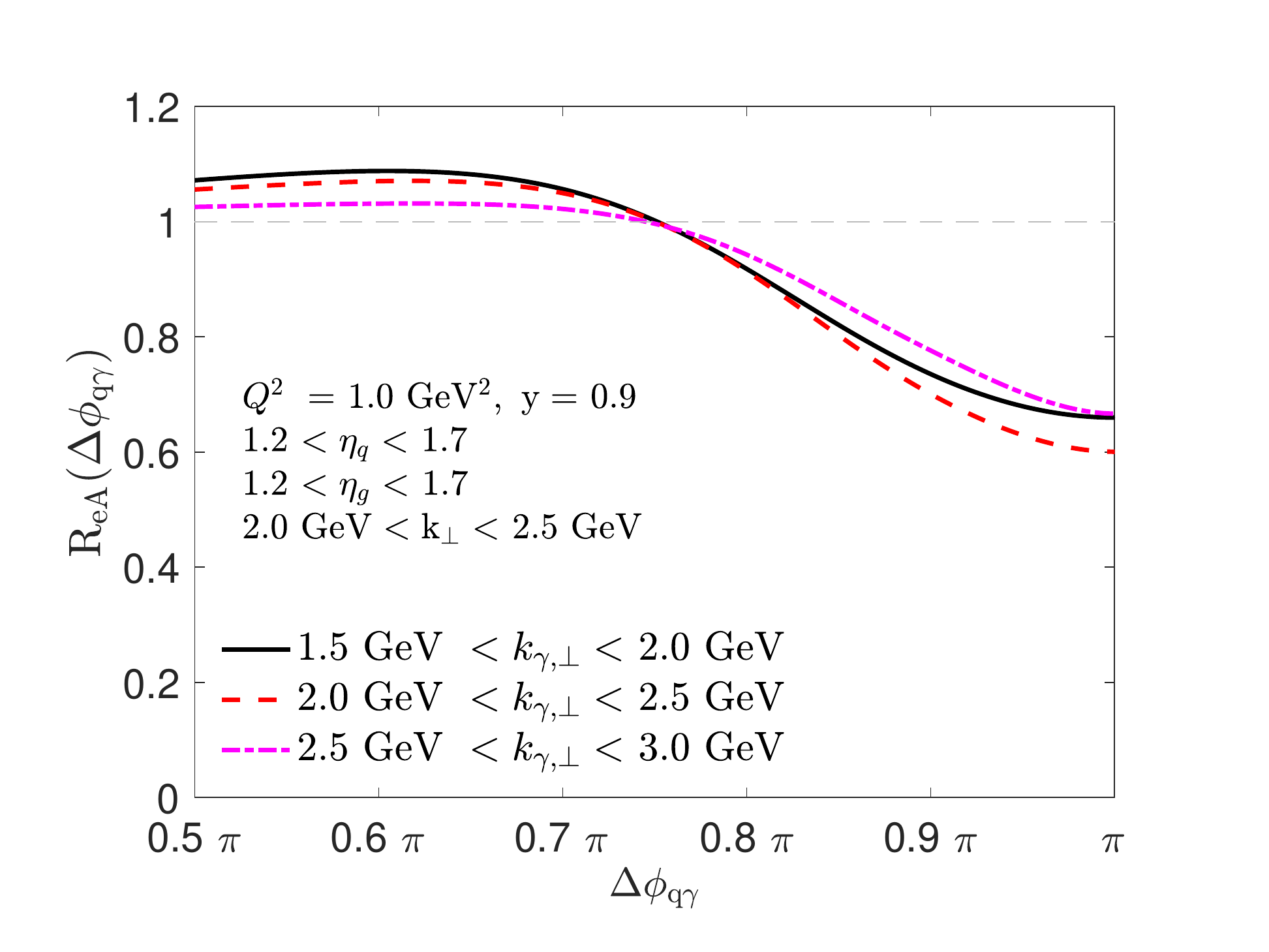}
    \includegraphics[scale=0.45]{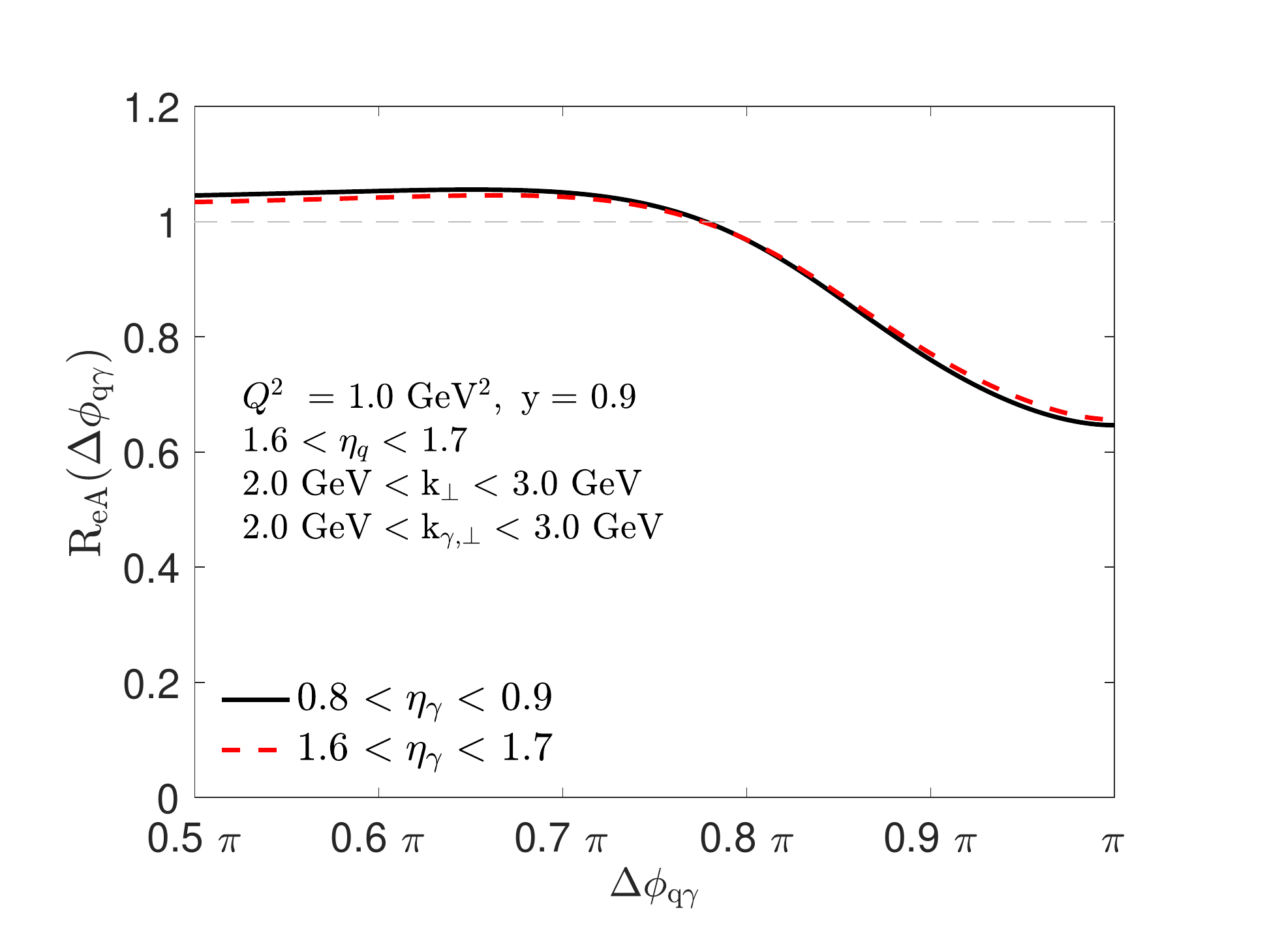}
    \caption{Left: $R_{eA}$ computed for different ranges in the photon transverse momentum. Right: $R_{eA}$ for different ranges in the photon rapidity.}
    \label{fig:ktg_etag_photon_dependence}
\end{figure}

More detailed information can be obtained by studying the azimuthal correlations at different transverse momenta and rapidities of the photon. The panel on the left in Fig.~\ref{fig:ktg_etag_photon_dependence} shows the $R_{eA}$ parameter defined in Eq.\,\eqref{eq:R-eA} as a function of the azimuthal angular separation $\Delta \phi_{q\gamma}$ for different ranges of the photon transverse momentum in the same rapidity ranges for quark and photon. We find that the suppression is maximal when the transverse momenta of the quark and photon are closer in magnitude or equivalently, when the back-to-back imbalance in transverse momenta is close to zero. As the difference in the magnitudes of the transverse momenta of the observed final state particles is increased, the suppression factor between that for the proton and for the gold nucleus is reduced. Thus precise reconstruction and identification of the jet and photon transverse momenta respectively are needed to carefully measure the back-to-back suppression.

In the case of dihadron production in $p+A$ collisions, it has been argued \cite{Albacete:2018ruq} that the rapidity dependence on the suppression provides an additional handle to probe gluon saturation. It is anticipated that because of small $x$ evolution, particle production at forward rapidities will result in more visible effects of saturation. However, we find very little dependence of the suppression on the rapidity of the photon (right panel in Fig.\,\ref{fig:ktg_etag_photon_dependence}). This might be because of the limited phase-space allowed by the kinematics at EIC energies, which results in limited small $x$ evolution. It may also be due in part to the integration over the antiquark phase space which results in a more complicated dependence of the impact factor on the rapidity of the jet and the photon.

\section{Summary and Outlook} \label{sec:summary}

We derived in this paper analytic expressions for inclusive prompt photon+quark production in $e+A$ DIS at small $x$ and at LO in the CGC EFT power counting. We presented a complete calculation accounting for both fragmentation and direct photon contributions. We find that the former is a convolution of the differential cross-section for inclusive dijet production and the antiquark-to-photon fragmentation function; it is sensitive to both dipole and quadrupole Wilson line correlators. In contrast, at large transverse momenta (which effectively impose an isolation cut on the photon), the direct photon contribution only depends on the dipole Wilson line correlator. 

We expressed our results for direct photon+jet production fully in momentum space, and computed the dipole correlator using the solution of the rcBK small $x$ evolution equation. This allowed us to study for the first time the inclusive direct photon-jet azimuthal angle correlations in the kinematic range of the Electron-Ion Collider. We observed a significant suppression and broadening of the back-to-back peak as the saturation scale is increased by comparing $e+p$ collisions to $e+Au$ min-bias collisions. Our results suggest that the back-to-back suppression will be  further amplified if it were possible to identify more central $e+A$ collisions~\cite{Lappi:2014foa} corresponding to larger values of the saturation scale. While our results  suggest that photon-jet correlations are a promising channel to probe gluon saturation, this will require jet reconstruction for  $k_\perp$ values in the range of $2-3$ GeV at EIC energies. 

Since such jet measurements are extremely challenging, photon-hadron correlations (whereby the final state quark fragments into a hadron), may provide an alternative channel that is sensitive to the process we have computed at the parton level. A useful study, outside the scope of the present work, would be to estimate the relative uncertainties in low $k_t$ jet reconstruction and those arising from our anticipated knowledge of fragmentation functions in the EIC era.

In future work, we plan to extend our numerical results to NLO in the CGC power counting. This will require the numerical evalution of the impact factor \cite{Roy:2019cux,Roy:2019hwr} and the numerical implementation of the small $x$ evolution of Wilson line correlators employing renormalization group evolution via the NLO BK \cite{Balitsky:2008zza,Iancu:2015vea,Lappi:2015fma,Lappi:2016fmu} or NLO JIMWLK \cite{Balitsky:2013fea,Kovner:2013ona} equations. 

An important subset of  NLO contributions are those of Sudakov type that have been studied in the context of two particle correlations at small $x$ \cite{Mueller:2012uf,Mueller:2013wwa}. They  provide an additional suppression of the back-to-back peak in the production of di-hadrons \cite{Zheng:2014vka} which may complicate the extraction of a clean signal for gluon saturation. We expect that due to the integration over the phase space of the antiquark in our process, Sudakov effects will be realized differently for photon+quark correlations. If so, a comparative study of the dijet and photon+jet channels can help isolate the relative contributions of Sudakov and gluon saturation effects.

An interesting extension of the computation discussed here would be to integrate over the phase space of the quark and extract the inclusive prompt photon cross-section in $e+A$ DIS at small $x$. This has been computed at small $x$ for $p+A$ collisions  at LO~\cite{Gelis:2002ki} and NLO~\cite{Benic:2016yqt,Benic:2016uku,Benic:2018hvb} in the CGC power counting. Because one has clean initial and final states, essentially free from  hadronization uncertainties, the measurement of prompt photons at the EIC provides another important channel to probe the gluon saturation regime. We will explore this possibility in the future.

\section*{Acknowledgements}
We are grateful to Renaud Boussarie, Oscar Garcia-Montero, Yacine Mehtar-Tani and Alba Soto-Ontoso for useful discussions. Special thanks to Heikki Mäntysaari for kindly providing the data for the dipole solution to rcBK. I.K. is supported by the U.S. Department of Energy under grant No. DE-FG02-00ER41132 as well as the Multifarious Minds grant provided by the Simons Foundation. F.S., B.P.S., and R.V. are supported by the U.S. Department of Energy under Contract No. DE-SC0012704. R.V's  work is supported by the US DOE within the framework of the TMD Theory Topical Collaboration. R.V and I.K's work was also supported in part by an LDRD grant from Brookhaven Science Associates. K.R is supported by the Joint BNL-Stony Brook Center for Frontiers in Nuclear Science (CFNS).  I.K. would also like to thank the Nuclear Theory Group at Brookhaven National Laboratory for their hospitality during the early stages of this work.

\appendix

\section{Conventions} \label{sec:conventions}
We work in light-cone coordinates suitable for computations in the eikonal limit. These are defined as
\begin{align}
    x^{+} = \frac{1}{\sqrt{2}}\left( x^0 + x^3\right), \ \ \ x^{-} = \frac{1}{\sqrt{2}}\left( x^0 - x^3\right)  \, ,
\end{align}
with transverse coordinates remaining the same as in Minkowski space. We write four vectors as $a^\mu = (a^+,a^-,\vect{a})$, where $\vect{a}$ is the two-dimensional vector of transverse components of $a$. The magnitude of the two dimensional vector will always be denoted by $a_{\perp}$. The metric  has the following non-zero entries: $g^{+-}=g^{-+}=1$ and $g^{ij} = -\delta^{ij}\  (i,j=1,2)$. The same definition holds for $\gamma^+$ and $\gamma^-$, with the Clifford algebra defined by the anticommutation relation
\begin{align}
    \left \{ \gamma^\mu,\gamma^\nu \right\} = 2 g^{\mu\nu} \ \mathbb{1}_{4x4} \, .
    \label{eq:dirac-algebra}
\end{align}
The scalar product is given by $a\cdot b = a^+ b^- + a^- b^+ - \vect{a}\cdot\vect{b}$. For example, $\slashed{a} = a^+ \gamma^- + a^- \gamma^+ - \vect{a} \cdot \vect{\gamma}$.

We will use the following shorthand notations for the momentum integrals,
\begin{equation}
 \int_{l}=\int \frac{\mathrm{d}^4 l}{(2\pi)^4} \quad \, , \quad \int_{\vect{l}}=\int \frac{\mathrm{d}^2 \vect{l}} {(2\pi)^2} \quad \, , \quad \int_{l^{\pm}}=\int \frac{\mathrm{d}l^{\pm}}{2\pi} \quad \, , \quad \int_{l}=\int_{\vect{l}} \int_{l^{+}} \int_{l^{-}} \, .
 \label{eq:mom-integrals-shorthand}
\end{equation}

\section{CGC Feynman rules} \label{sec:Feynman rules}
\begin{figure}[!htbp]
\centering
\includegraphics[scale=1]{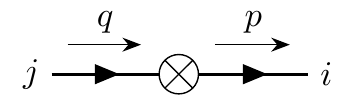}
\caption{Feynman diagram for the dressed quark propagator in the classical background field of the nucleus. $i$ and $j$ denote the color indices in the fundamental representation of $SU(N_{\mathrm{c}})$. \label{fig:effective-quark-propagator}}
\end{figure}

We work in the light cone gauge $A^{-}=0$. In addition to the regular Feynman rules that follow from the QED and QCD Lagrangians (see for example, the book by Schwartz~\cite{Schwartz:2013pla}), we also need an expression for the dressed quark propagator in the background of the classical field of small $x$ gluons inside the dense nucleus. In our computation, the Feynman diagram for the dressed quark propagator is shown in Fig.~\ref{fig:effective-quark-propagator}. The momentum space expression for the dressed propagator can be written as~\cite{McLerran:1994vd,McLerran:1998nk,Balitsky:2001mr},
\begin{equation}
S_{ij}(p,q) =  S^{0}_{ik}(p)\,\mathcal{T}_{km}(p,q)\,S^0_{mj} (q)\, ,
\label{eq:dressed-quark-mom-prop}
\end{equation}
where we have considered the two fermion lines on either side of the effective interaction vertex, shown by a crossed circle in Fig.~\ref{fig:effective-quark-propagator}, to be internal (off-shell). $i,j,k,m$ represent color indices in the fundamental representation of $SU(N_{\mathrm{c}})$. For an outgoing on-shell quark line with momentum $p$ in Fig.~\ref{fig:effective-quark-propagator}, we will replace $S^{0}(p)$ by the Dirac spinor $\bar{u}(p)$ in Eq.\,\ref{eq:dressed-quark-mom-prop}.

In Eq.\,\eqref{eq:dressed-quark-mom-prop}, the free massless quark propagator with momentum $p$ is given by 
\begin{equation}
    S^0_{ij} (p) = \frac{i \slashed{p}}{p^{2}+i \varepsilon} \, \delta_{ij} \, . 
\end{equation}
The effective vertex appearing in the dressed quark propagator has the following expression in $A^{-}=0$ gauge~\cite{McLerran:1998nk,Balitsky:2001mr}
\begin{align}
 \mathcal{T}_{km}(p,q) &= 2 \pi \, \delta(p^{-}-q^{-}) \gamma^{-} {\text{sign} (p^-)}  \int_{\vect{z}}  e^{-i(\bm{p}_{\perp} - \bm{q}_{\perp})\cdot \bm{z}_{\perp}} \tilde{U}_{km}^{\text{sign}(p^-)}(\bm{z}_{\perp}) \enskip .
\label{eq:effective-quark-vertex-2}
\end{align}
Here $\tilde{U}$ is a Wilson line in the fundamental representation of $SU(N_{\mathrm{c}})$ which can be written as 
\begin{equation}
\tilde{U}(\bm{x}_{\perp})=P_{-} \Bigg( \text{exp} \Bigg\{ -ig \int_{-\infty}^{+\infty} \mathrm{d}z^{-} A_{\rm cl}^{+,a} (z^{-},\bm{x}_{\perp}) \,t^a\Bigg\} \Bigg) \, ,  
\label{eq:Wilson-line-fund}
\end{equation}
where $P_{-}$ means path ordering along the `$-$' LC direction, $t^a$ ($a=1,\ldots,N_{\mathrm{c}})$ represents the generators of the fundamental representation of $SU(N_{\mathrm{c}})$ and $A^{+}_{\rm cl}$ is a solution (see Eq.\,\eqref{eq:A+}) of the classical Yang-Mills equations in the $A^{-}=0$ gauge. The latter can be written as
\begin{equation}
[D_{\mu},F^{\mu \nu}](x)=g \,  \delta^{\nu +}  \rho_{A}(x^{-},\bm{x}_{\perp}) \, ,
\label{eq:Yang-Mills}
\end{equation}
where the covariant derivative $D_\mu = \partial_\mu - ig A_\mu$, $F^{\mu \nu}$ is the field strength tensor, $g$ represents the QCD gauge coupling, and $\rho_{A}(x^{-},\bm{x}_{\perp})\approx \rho_{A}(\bm{x}_{\perp})\delta(x^-)$ represents the color charge density of large $x$ static sources for the small $x$ dynamical fields $A^\mu$. The delta function in the color charge density denotes that we are working in a frame where the nucleus is right moving at nearly the speed of light. The solutions of the YM equations in the light cone (LC) $A^-=0$ gauge are given by
\begin{align}
& A_{\rm cl}^+ = \frac{1}{4\pi }\int_{\vect{z}}  \ln\frac{1}{(\bm{x}_\perp-\bm{z}_\perp)^2\Lambda^2}\, \rho_A(x^-,\bm{z}_\perp) \,,\nonumber\\
& A_{\rm cl}^{-}=0\,\,;\, \, A_{{\rm cl},\perp} =0\,,
\label{eq:A+}
\end{align}
where $\Lambda$ is an infrared cutoff  necessary to invert the Laplace equation $-\nabla_\perp^2 A_{\rm cl}^+ = g\rho_A$. 

\begin{figure}[!htbp]
    \centering
    \includegraphics[scale=1]{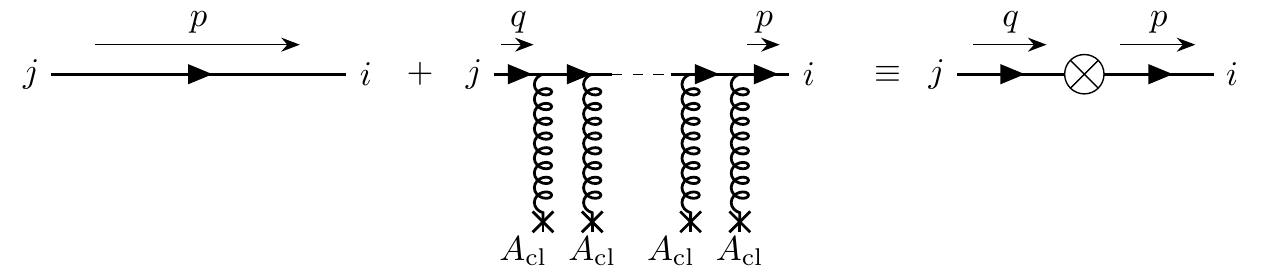}
    \caption{Diagrammatic representation of the dressed quark propagator in Eq.\,\eqref{eq:dressed-quark-mom-prop} as a sum of the ``no-scattering" contribution as shown by the free propagator on the left and   all possible multiple scattering with the background classical field $A_{\rm cl}$ of small $x$ gluons. $i$ and $j$ denote the color indices in the fundamental representation of $SU(N_{\mathrm{c}})$. }
    \label{fig:quark-shockwave-propagator}
\end{figure}

In Eq.\,\eqref{eq:dressed-quark-mom-prop}, we include the possibility of ``no scattering'' (given by $\tilde{U}=\mathds{1}$) within the definition of the effective vertex . So  the dressed propagator also contains a free part given by $(2\pi)^{4} \delta^{(4)}(p-q) S^{0}_{ij}(p)$ and an interacting part which contains all possible scattering with the nuclear shock wave. This is pictorially depicted by Fig.~\ref{fig:quark-shockwave-propagator}. From this equivalent representation and from the expressions for the fundamental Wilson line in Eq.\,\eqref{eq:Wilson-line-fund} and the small $x$ classical gluon field given by Eq.\,\eqref{eq:A+}, we observe that the dressed quark (and antiquark) propagators in the CGC EFT efficiently resum all higher twist contributions $\frac{\rho_{A}}{\nabla_\perp^2}\rightarrow \frac{Q_S^2}{Q^2}$ from the multiple scattering of the $q\bar{q}$ dipole off the color field of the nucleus. 

\section{Derivation of  $\mathcal{R}_{L}^{q\bar{q} \gamma}$ and $\mathcal{R}_{T}^{q\bar{q}\gamma}$} \label{sec:hadron-amplitude-squared}

In obtaining the differential cross-section for $\gamma+q$ production in $e+A$ DIS at small $x$, two essential elements are the ``hard" coefficient functions $R^{q\bar{q}\gamma}_{L}$ and  $R^{q\bar{q}\gamma}_{T}$ which enter the $\gamma+q\bar{q}$ cross-section in Eqs.~\eqref{eq:diff-cs-long-pol} and \eqref{eq:diff-cs-trans-pol} respectively. These functions are extracted from the modified hadron tensor given by Eq.\,\eqref{eq:modified-hadron-tensor-gamma-dijet}. Using this definition for the hadron tensor and comparing it with Eqs.~\eqref{eq:diff-cs-gamma-dijet-general} and \eqref{eq:diff-cs-long-pol}, \eqref{eq:diff-cs-trans-pol}, we can write 
\begin{align}
    R^{q\bar{q}\gamma}_{L}&=\frac{1}{64\,   z_{q}^{3} z^{3}_{\bar{q}} (q^{-})^{2}} \sum_{\text{spins} \, , \, \text{pols.}} \sum_{m,n=1}^{4} \mathcal{A}^{L \, \dagger}_{m}(\vect{x}',\vect{y}') \, \mathcal{A}^{L}_{n}(\bm{x}_{\perp}, \bm{y}_{\perp}) \, , \label{eq:R-qqbargamma-L-definition} \\
  R^{q\bar{q}\gamma}_{T}&=\frac{1}{32\,   z_{q} z_{\bar{q}} (q^{-})^{2}} \sum_{T=\pm 1} \, \,  \sum_{\text{spins} \, , \, \text{pols.}} \sum_{m,n=1}^{4} \mathcal{A}^{T \, \dagger}_{m}(\vect{x}',\vect{y}') \, \mathcal{A}^{T}_{n}(\bm{x}_{\perp}, \bm{y}_{\perp}) \, , \label{eq:R-qqbargamma-T-definition}   
\end{align}
where $\mathcal{A}^{\lambda}_{m}$ ($m=1,\ldots,4$) are the reduced amplitudes for the four subprocesses. Their expressions for the processes labeled $\mathcal{M}_1$ and $\mathcal{M}_2$ in Fig.~\ref{fig:photon_jet_LO_diagrams} are given by Eqs.~\eqref{eq:A1full} and \eqref{eq:A2full}. For their $q\leftrightarrow \bar{q}$ interchanged counterparts, we simply replace $k \leftrightarrow p$ and $\vect{x} \leftrightarrow \vect{y} $ in these expressions and put an overall minus sign. Also, we use the prime symbol to label transverse coordinates and momenta in the complex conjugate of these amplitudes. 

Firstly, we note that we do not need to evaluate all the combinations of the squared amplitudes because most of them are related by charge and complex conjugation. We can write the  reduced amplitudes in Eqs.~\eqref{eq:A1full} and \eqref{eq:A2full} schematically as
\begin{align}
    \mathcal{A}_{1}^{\lambda}(\vect{x},\vect{y}) &= e^{-i \vect{k_{\gamma}} \cdot \vect{x}} \int_{\vect{l}} e^{i\vect{l} \cdot \vect{r}} \frac{A_{1}^{\lambda} (\vect{l})}{D_{1}} \, , \nonumber \\
  \mathcal{A}_{2}^{\lambda}(\vect{x},\vect{y}) &= e^{-i \vect{k_{\gamma}} \cdot \vect{x}} \int_{\vect{l}} e^{i\vect{l} \cdot \vect{r}} \Big( \frac{-A_{2,\text{reg}}^{\lambda} (\vect{l})}{D_{1} \, D_{2}} + \frac{-A^{\lambda}_{2,\text{ins}}(\vect{l})}{D_{2}}  \Big) \, ,
\end{align}
where the denominators $D_1$ and $D_2$ (and their quark-antiquark interchanged counterparts) are not important for the current discussion. The terms of interest are the $A^{\lambda}(\vect{l})$'s which contain the Dirac gamma matrix structures specific to the hadronic subprocesses. When we sum over the spins of the outgoing quark and antiquark at the level of amplitude squared, this results in traces over these gamma matrices. In addition, we have to sum over the two transverse polarizations of the outgoing photon which is denoted by the index $\bar{\lambda}$ in our amplitudes--see Eqs.~\eqref{eq:A1}, \eqref{eq:A2reg} and \eqref{eq:A2ins}. For the case of the longitudinally polarized virtual photon ($\lambda=L=0$), the computation of the trace can be simplified using the identity in Eq.\,\eqref{eq:gamma-structure-A1} and also by noting that there is no ``instantaneous" contribution from the amplitude $\mathcal{A}_{2}^{L}$. We will provide  below the final expressions obtained for the six independent terms in the sum appearing in Eq.\,\eqref{eq:R-qqbargamma-L-definition}.
\subsubsection{Longitudinally polarized virtual photon} 
\begingroup
\allowdisplaybreaks
\begin{align}
  &  \sum_{\text{spins} \, , \,  \text{pols.} (\bar{\lambda})} A^{L \, \dagger} _1(\vect{l'}) A^L_1(\vect{l})  =64 \,  (q^-)^2 z_q \, z^3_{\bar{q}} (1-z_{\bar{q}})^2 Q^2  \left[ (1-z_{\bar{q}})^2 + z^2_q \right] \mathcal{I}^{i}_1 \,  \mathcal{I}^{i}_1 \, , \label{eq:A1L-A1L}\\
  &  \sum_{\text{spins} \, , \,  \text{pols.} (\bar{\lambda})} A^{L \, \dagger}_3(\vect{l'}) A^L_1(\vect{l})  =-64 \,  (q^-)^2 z^2_q \, z^2_{\bar{q}} (1-z_{\bar{q}})(1-z_q) Q^2  \left[z_q(1-z_q) +z_{\bar{q}}(1- z_{\bar{q}}) \right] \mathcal{I}^{i}_1 \, \mathcal{I}^{i }_3  \, , \label{eq:A3L-A1L}\\
  &  \sum_{\text{spins} \, , \,  \text{pols.} (\bar{\lambda})} A^{L\, \dagger}_{2, \rm{reg}}(\vect{l'}) A^L_{2,\rm{reg}}(\vect{l})  =\frac{64 \, (q^-)^2 z^3_{\bar{q}} Q^2 }{z_q z^2_\gamma } \left[ (1-z_{\bar{q}})^2 + z^2_q \right] \mathcal{I}^{i}_2(\vect{l}) \,  \mathcal{I}^{i }_2(\vect{l'}) \, , \label{eq:A2L-A2L}\\
  &  \sum_{\text{spins} \, , \,  \text{pols.} (\bar{\lambda})} A^{L \, \dagger}_{4,\rm{reg}}(\vect{l'}) A^L_{2, \rm{reg}}(\vect{l})  =-\frac{64 \, (q^-)^2 z_{\bar{q}} \,  z_q Q^2 }{z^2_\gamma } \left[ z_q(1-z_{q}) + z_{\bar{q}}(1-z_{\bar{q}}) \right] \mathcal{I}^{i}_2(\vect{l}) \, \mathcal{I}^{i }_4 (\vect{l'}) \, ,\label{eq:A4L-A2L} \\
   & \sum_{\text{spins} \, , \,  \text{pols.} (\bar{\lambda})} A^{L \, \dagger}_{2,\rm{reg}}(\vect{l'}) A^L_1(\vect{l})  =  \frac{64 \, (q^-)^2 z^3_{\bar{q}} (1-z_{\bar{q}}) Q^2 }{z_\gamma} \left[ (1-z_{\bar{q}})^2 + z^2_q \right] \mathcal{I}^{i}_1 \,  \mathcal{I}^{i }_2(\vect{l'}) \, , \label{eq:A2L-A1L} \\
   &  \sum_{\text{spins} \, , \,  \text{pols.} (\bar{\lambda})} A^{L \, \dagger}_{4,\rm{reg}}(\vect{l'}) A^L_1(\vect{l})  =-\frac{64 \, (q^-)^2 z^2_q z_{\bar{q}} (1-z_{\bar{q}}) Q^2 }{z_\gamma} \left[ z_q (1-z_q) + z_{\bar{q}}(1-z_{\bar{q}}) \right] \mathcal{I}^{i}_1 \,  \mathcal{I}^{i }_4(\vect{l'}) \, , \label{eq:A4L-A1L}
\end{align}
\endgroup
where the vector functions $\mathcal{I}^{i}_{1,2}$ are shorthand for 
\begin{align}
    \mathcal{I}_{1}^{i}& = \frac{z_{q} \, \bm{k}_{\gamma \perp}^{i}-z_{\gamma} \, \bm{k}_{\perp}^i}{ \vert z_{q} \, \bm{k}_{\gamma \perp}-z_{\gamma} \, \bm{k}_{\perp} \vert^2} \, ,  \nonumber \\
    \mathcal{I}_{2}^{i}(\vect{l})&= (1-z_{\bar{q}}) \, \bm{k}_{\gamma \perp}^{i} -z_{\gamma} \, \bm{l}_{\perp}^{i}
 \, .
 \label{eq:I-1-I-2}
 \end{align}
 The expressions for $\mathcal{I}^{i}_{3}$ and $\mathcal{I}^{i}_{4}$ are obtained respectively from $\mathcal{I}_{1}^i$ and $\mathcal{I}_{2}^i$ by the replacements $z_{q} \leftrightarrow z_{\bar{q}}$ and $\bm{k}_{\perp} \rightarrow \bm{p}_{\perp}$. It is easy to check that the expressions for the remaining combinations of squared amplitudes can be evaluated from the quantities in Eqs.~\eqref{eq:A1L-A1L}-\eqref{eq:A4L-A1L}.
 
 To obtain the desired form for $R^{q\bar{q}\gamma}_{L}$ as it appears in Eq.\eqref{eq:R-long-pol-gamma-dijet}, we have combined the vector functions in Eqs.~\eqref{eq:I-1-I-2} (and their $q\leftrightarrow \bar{q}$ counterparts) with the denominators $D_1$ and $D_2$ in the transverse integration over $\bm{l}_{\perp}$ to define the new vector functions which we call $\mathcal{J}_{L,p}^{i}$ ($p=1,\ldots,4$). The coefficients $\xi$'s are formed by the functions of $z_{q}$ and $z_{\bar{q}}$ that survive after cancellation with the prefactors appearing in Eq.\,\eqref{eq:R-qqbargamma-L-definition}.
 
\subsubsection{Transversely polarized virtual photon}

In the case of the transversely polarized virtual photon, we have additional contributions due to the non-vanishing of the instantaneous term in Eq.\,\eqref{eq:A2full}. In addition to the sums over the four allowed processes, outgoing quark-antiquark spins and polarizations of the produced photon, we have to add the contributions from the two transverse polarization states of $\gamma^{*}$ to obtain the desired expression for $R^{q\bar{q}\gamma}_{T}$. We will represent this by the shorthand notation,
\begin{equation}
  \sum_{T,S,P}= \sum_{T=\pm 1} \, \,  \sum_{\text{spins} \, , \, \text{pols.}(\bar{\lambda})}   \, .
\end{equation}
As in the previous section, we present here the final results for the various independent terms in the squared amplitude. These are categorized into the following combinations:
\begin{enumerate}
    \item Regular-Regular:
    \begingroup
    \allowdisplaybreaks
\begin{align}
 & \sum_{T,S,P} A^{T \, \dagger}_1(\vect{l'}) A^{T}_1(\vect{l}) = 32 \, (q^-)^2 z_q z_{\bar{q}} \left[(1-z_{\bar{q}})^2 + z^2_{\bar{q}} \right] \left[(1-z_{\bar{q}})^2 + z^2_{q} \right]  \left[ \mathcal{I}^{i}_1 \vect{l}^j \right] \left[\mathcal{I}^{i}_1  \vect{l'}^j \right] \, ,  \\
 &  \sum_{T,S,P} A^{T \, \dagger }_3(\vect{l'})  A^{T}_1(\vect{l}) = 32 \, (q^-)^2  z_q z_{\bar{q}}\left[(1-z_{\bar{q}})z_q + z_{\bar{q}} (1-z_q) \right] \left[z_q (1-z_{q}) + z_{\bar{q}} (1-z_{\bar{q}}) \right] \left[ \mathcal{I}^{i}_1 \vect{l}^j \left]\right[ \mathcal{I}^{i}_3 \vect{l'}^j \right] \nonumber  \\
   &  - 32 \, (q^-)^2 z_q z_{\bar{q}} \left[(1-z_{\bar{q}})z_q - z_{\bar{q}} (1-z_q) \right] \left[z_q (1-z_{q}) - z_{\bar{q}} (1-z_{\bar{q}}) \right]  \epsilon^{im} \epsilon^{jn} \left[  \mathcal{I}^{i}_1 \vect{l}^j \right]\left[ \mathcal{I}^{m}_3 \vect{l'}^n \right] \, , \\
  & \sum_{T,S,P} A^{T \, \dagger}_{2, \rm{reg}} (\vect{l'}) A^{T}_{2, \rm{reg}} (\vect{l})  = \frac{32 \,  (q^-)^2 z_{\bar{q}} \left[ (1-z_{\bar{q}})^2 + z^2_{\bar{q}} \right] \left[ (1-z_{\bar{q}})^2 + z^2_{q} \right]\left[ \mathcal{I}^i_2(\vect{l}) \vect{l}^j   \right] \left[ \mathcal{I}^{i }_2(\vect{l'}) \vect{l'}^j \right]}{z_q z^2_\gamma (1-z_{\bar{q}})^2}  \nonumber \\
    &+ \frac{32 \, (q^-)^2 z_{\bar{q}}\left[ (1-z_{\bar{q}})^2 - z^2_{\bar{q}} \right] \left[ (1-z_{\bar{q}})^2 - z^2_{q} \right] \epsilon^{im} \epsilon^{jn} \left[ \mathcal{I}^i_2(\vect{l}) \vect{l}^j \right] \left[   \mathcal{I}^{m }_2(\vect{l'})  \vect{l'}^n \right] }{z_q z^2_\gamma (1-z_{\bar{q}})^2}  \, , \\
   & \sum_{T,S,P} A^{T \, \dagger}_{4,\rm{reg}}(\vect{l'})  A^{T}_{2,\rm{reg}} (\vect{l})   = \frac{32 \, (q^-)^2 \left[(1-z_{\bar{q}}) z_q + z_{\bar{q}}(1-z_q) \right]  \left[ z_q(1-z_q) + z_{\bar{q}}(1-z_{\bar{q}})\right]\left[ \mathcal{I}^i_2 (\vect{l}) \vect{l}^j \right] \left[ \mathcal{I}^{i }_4(\vect{l'}) \vect{l'}^j \right] }{z^2_\gamma (1-z_{\bar{q}})(1-z_q)} \nonumber \\
    &  - \frac{32 \, (q^-)^2   \left[(1-z_{\bar{q}}) z_q - z_{\bar{q}}(1-z_q) \right] \left[z_q (1-z_{q})  - z_{\bar{q}}(1-z_{\bar{q}})\right] \epsilon^{im} \epsilon^{jn} \left[ \mathcal{I}^i_2 (\vect{l}) \vect{l}^j  \right] \left[ \mathcal{I}^{m }_4(\vect{l'})  \vect{l'}^n \right]}{z^2_\gamma (1-z_{\bar{q}})(1-z_q)} \, , \\
 &  \sum_{T,S,P} A^{T \, \dagger}_{2,\rm{reg}} (\vect{l'})  A^{T}_1 (\vect{l})  = \frac{32 \,  (q^-)^2  z_{\bar{q}} \left[(1-z_{\bar{q}})^2 + z^2_{\bar{q}} \right] \left[(1-z_{\bar{q}})^2 + z^2_{q} \right]  \left[ \mathcal{I}^{i}_1  \vect{l}^j \right] \left[ \mathcal{I}^{i}_2(\vect{l'}) \vect{l'}^j \right] }{z_\gamma (1-z_{\bar{q}})}  \nonumber \\
    &+ \frac{32 \, (q^-)^2  z_{\bar{q}} \left[(1-z_{\bar{q}})^2 - z^2_{\bar{q}} \right] \left[(1-z_{\bar{q}})^2 - z^2_{q} \right] \epsilon^{im} \epsilon^{jn} \left[ \mathcal{I}^{i}_1  \vect{l}^j \right] \left[ \mathcal{I}^{m}_2(\vect{l'}) \vect{l'}^n \right]}{z_\gamma (1-z_{\bar{q}})}  \, ,  \\
 &  \sum_{T,S,P} A^{T \, \dagger}_{4,\rm{reg}} (\vect{l'})  A^{T}_1 (\vect{l}) = \frac{32 \, (q^-)^2 z_q \left[(1-z_{\bar{q}}) z_q + z_{\bar{q}}(1-z_q) \right]  \left[ z_q(1-z_q) + z_{\bar{q}}(1-z_{\bar{q}})\right] \left[ \mathcal{I}^{i}_1  \vect{l}^j \right] \left[ \mathcal{I}^{i}_4(\vect{l'}) \vect{l'}^j \right] }{z_\gamma (1-z_q)}   \nonumber \\
    &  - \frac{32 \, (q^-)^2 z_q \left[(1-z_{\bar{q}}) z_q - z_{\bar{q}}(1-z_q) \right]    \left[z_q (1-z_{q})  - z_{\bar{q}}(1-z_{\bar{q}})\right] \epsilon^{im} \epsilon^{jn} \left[ \mathcal{I}^{i}_1  \vect{l}^j \right] \left[ \mathcal{I}^{m}_4(\vect{l'}) \vect{l'}^n \right] }{z_\gamma (1-z_q)} \, . 
\end{align}
\endgroup
\item Regular-Instantaneous:
\begin{align}
  & \sum_{T,S,P}  A^{T \, \dagger}_{2,\rm{ins}}(\vect{l'}) A^{T }_{2,\rm{reg}}(\vect{l}) = \frac{32 \, (q^-)^2 z_q z^2_{\bar{q}} \,  \mathcal{I}^i_2(\vect{l}) \, \vect{l}^i}{(1-z_{\bar{q}})^2 z_\gamma} \, , \\
   &  \sum_{T,S,P} A^{T \, \dagger}_{4,\rm{ins}}(\vect{l'}) A^{T }_{2,\rm{reg}}(\vect{l})  =  \frac{32 \,  (q^-)^2  z_{\bar{q} } (1-z_{\bar{q}}) \, \mathcal{I}^i_2 (\vect{l}) \, \vect{l}^i }{z_\gamma (1-z_q)} \, , \\
   &  \sum_{T,S,P} A^{T \, \dagger}_{2,\rm{ins}} (\vect{l'}) A^{T }_1 (\vect{l})  =  \frac{32 \, (q^-)^2 z^2_q z^2_{\bar{q}} \, \mathcal{I}^i_1 \, \vect{l}^i}{(1-z_{\bar{q}})} \, ,  \\
  &  \sum_{T,S,P}  A^{T \, \dagger}_{4,\rm{ins}} (\vect{l'}) A^{T}_1 (\vect{l}) =  \frac{32 \, (q^-)^2 z_q z_{\bar{q}} \, (1-z_{\bar{q}})^2 \, \mathcal{I}^i_1 \, \vect{l}^i }{(1-z_q)}  \, .
\end{align}
\item Instantaneous-Instantaneous:
\begin{align}
    & \sum_{T,S,P} A^{T \, \dagger}_{2,\rm{ins}}(\vect{l'}) A^{T}_{2,\rm{ins}}(\vect{l})  = \frac{32 \,  (q^-)^2 z_q \,  z_{\bar{q}}}{(1-z_{\bar{q}})^2} \, , \\
  &   \sum_{T,S,P} A^{T \, \dagger}_{4,\rm{ins}}(\vect{l'})A^{T}_{2,\rm{ins}}(\vect{l})  =0 \, .
\end{align}
\end{enumerate}
We see that in all these expressions, the same vector functions defined in  Eqs.~\eqref{eq:I-1-I-2} appear, similar to the case of longitudinally polarized virtual photon, albeit with an additional factor of $\bm{l}_{\perp}^{i}$. Following a similar procedure, we can combine these functions with the relevant denominators that appear under the integration over $\vect{l}$ in Eqs.~\eqref{eq:A1full} and \eqref{eq:A2full} to define new tensor functions $\mathcal{J}_{T,p}^{ij}$ ($p=1,\ldots,4$) which constitute the quantity $R^{q\bar{q}\gamma}_{T}$ given by Eq.\,\eqref{eq:R-trans-pol-gamma-dijet}. The functions of $z_q$ and $z_{\bar{q}}$ that appear as prefactors in the above equations can be suitably moulded to give rise to the various coefficients $\zeta$, $\chi$, $\kappa$ defined in Appendix~\ref{sec:coefficients-J-functions}.

\section{Identities with polarization vector structures and transverse gamma matrices} \label{sec:gamma-matrix-identities}

In the previous section, we provided explicit results for the independent terms in the squared amplitude for both longitudinal and transverse polarizations of the virtual photon. While the computation for the longitudinal case is very straightforward, the same cannot be said for transversely polarized virtual photon. We will present here a detailed list of the traces over various non-trivial structures involving transverse gamma matrices that appear in this computation. We have explicitly indicated the sum over the two polarization states $\bar{ \lambda}=\pm 1$ of the outgoing photon.

In this and the following discussion, $c_1$ and $c_2$ denote multiplicative factors.  We will make extensive use of the following useful identities involving gamma matrices in four dimensions to derive the desired expressions.
\begin{align}
    \gamma^{j} \gamma^{k} \gamma^{j} &=0 \, , \, \, \, \, \, \, \, 
    \gamma^{j} \gamma^{r} \gamma^{l} \gamma^{j}= -2 \, \gamma^{l} \gamma^{r} \, , \, \, \, \, \, \, \,
\gamma^{j} \gamma^{r} \gamma^{t} \gamma^{k} \gamma^{j} = 2 \, \gamma^{r} \gamma^{k} \gamma^{t} - 2 \, \gamma^{t} \gamma^{k} \gamma^{r}   \, , \nonumber \\
\text{Tr} ( \gamma^{i} \gamma^{j} )&=-4 \, \delta^{ij} \, , \, \, \, \, \, \,
\text{Tr}( \gamma^{i} \gamma^{j} \gamma^{k} \gamma^{l} ) =4 \, \delta^{ij} \delta^{kl}-4\, \epsilon^{ij} \epsilon^{kl} \, , \nonumber \\
\Tr[\slashed{a}\gamma^{-} \gamma^{i_{1}} \, \gamma^{i_{2}} \ldots \gamma^{i_{n}} ] &=a^{-} \Tr[\gamma^{i_{1}} \, \gamma^{i_{2}} \ldots \gamma^{i_{n}}] \, . 
\end{align}

\subsubsection*{Identities I: Two polarization vector structures only}
\begin{align}
   \sum_{\bar{\lambda}} \Tr[\left( \vect{\epsilon}^{\bar{\lambda}*,i} + c_1 \gamma^i \gamma^j \vect{\epsilon}^{\bar{\lambda}*,j} \right)  \left( \vect{\epsilon}^{\bar{\lambda},m} +c_2 \gamma^n \gamma^m \vect{\epsilon}^{\bar{\lambda},n} \right) ] &= 4 (1-c_1-c_2+2c_1 c_2)\, \delta^{im} \label{epsilon_tensor_identity_1} \, , \\
   \sum_{\bar{\lambda}} \Tr[\left( \vect{\epsilon}^{\bar{\lambda}*,i} + c_1 \gamma^i \gamma^j \vect{\epsilon}^{\bar{\lambda}*,j} \right)  \left( \vect{\epsilon}^{\bar{\lambda},m} +c_2 \gamma^m \gamma^n \vect{\epsilon}^{\bar{\lambda},n} \right) ] &= 4 (1-c_1-c_2)\, \delta^{im} \label{epsilon_tensor_identity_2} \, , \\
   \sum_{\bar{\lambda}} \Tr[\left( \vect{\epsilon}^{\bar{\lambda}*,i} + c_1 \gamma^j \gamma^i \vect{\epsilon}^{\bar{\lambda}*,j} \right)  \left( \vect{\epsilon}^{\bar{\lambda},m} +c_2 \gamma^m \gamma^n \vect{\epsilon}^{\bar{\lambda},n} \right) ] &= 4 (1-c_1-c_2+2c_1 c_2)\, \delta^{im} \label{epsilon_tensor_identity_3} \, , \\
   \sum_{\bar{\lambda}} \Tr[\left( \vect{\epsilon}^{\bar{\lambda}*,i} + c_1 \gamma^j \gamma^i \vect{\epsilon}^{\bar{\lambda}*,j} \right)  \left( \vect{\epsilon}^{\bar{\lambda},m} +c_2 \gamma^n \gamma^m \vect{\epsilon}^{\bar{\lambda},n} \right) ] &= 4 (1-c_1-c_2)\, \delta^{im} \, . \label{epsilon_tensor_identity_4}
\end{align}

\subsubsection*{Identities II: Two polarization vector structures and 4 transverse gamma matrices.}
Four gamma matrices between polarization vector structures:
\begin{align}
    \sum_{\bar{\lambda}} \Tr[\left( \vect{\epsilon}^{\bar{\lambda}*,i} +c_1 \gamma^i \gamma^j \vect{\epsilon}^{\bar{\lambda}*,j} \right) \gamma^k \gamma^r \gamma^t \gamma^k \left( \vect{\epsilon}^{\bar{\lambda},m} +c_2 \gamma^n \gamma^m \vect{\epsilon}^{\bar{\lambda},n} \right)]&=8(1-c_1-c_2+2c_1 c_2) \delta^{im} \delta^{rt} \nonumber \\ & - 8(c_1 +c_2 - 2c_1 c_2) \epsilon^{im} \epsilon^{rt} \, ,  \label{epsilon_tensor_identity_5} \\
    \sum_{\bar{\lambda}} \Tr[\left( \vect{\epsilon}^{\bar{\lambda}*,i} +c_1 \gamma^i \gamma^j \vect{\epsilon}^{\bar{\lambda}*,j} \right) \gamma^r \gamma^k  \gamma^k \gamma^t \left( \vect{\epsilon}^{\bar{\lambda},m} +c_2 \gamma^n \gamma^m \vect{\epsilon}^{\bar{\lambda},n} \right)]&=8(1-c_1-c_2+2c_1 c_2) \delta^{im} \delta^{rt}  \nonumber \\
    &+ 8(c_1 + c_2 - 2c_1 c_2) \epsilon^{im} \epsilon^{rt} \label{epsilon_tensor_identity_6} \, ,  \\
    \sum_{\bar{\lambda}} \Tr[\left( \vect{\epsilon}^{\bar{\lambda}*,i} +c_1 \gamma^i \gamma^j \vect{\epsilon}^{\bar{\lambda}*,j} \right) \gamma^k \gamma^r  \gamma^k \gamma^t \left( \vect{\epsilon}^{\bar{\lambda},m} +c_2 \gamma^n \gamma^m \vect{\epsilon}^{\bar{\lambda},n} \right)]&=0 \, , \label{epsilon_tensor_identity_7} \\
    \sum_{\bar{\lambda}} \Tr[\left( \vect{\epsilon}^{\bar{\lambda}*,i} +c_1 \gamma^i \gamma^j \vect{\epsilon}^{\bar{\lambda}*,j} \right) \gamma^r \gamma^k  \gamma^t \gamma^k \left( \vect{\epsilon}^{\bar{\lambda},m} +c_2 \gamma^n \gamma^m \vect{\epsilon}^{\bar{\lambda},n} \right)]&= 0 \, . \label{epsilon_tensor_identity_8}
\end{align}
Four gamma matrices (two and two) between polarization vector structures:
\begin{align}
    \sum_{\bar{\lambda}} \Tr[ \left( \vect{\epsilon}^{\bar{\lambda}*,i} +c_1 \gamma^i \gamma^j \vect{\epsilon}^{\bar{\lambda}*,j} \right) \gamma^k \gamma^r  \left( \vect{\epsilon}^{\bar{\lambda},m} +c_2 \gamma^m \gamma^n \vect{\epsilon}^{\bar{\lambda},n} \right)  \gamma^t \gamma^k] &= 8\left(1-c_1-c_2 \right)  \delta^{im} \delta^{rt} + 8(c_2-c_1)\epsilon^{im} \epsilon^{rt} \label{epsilon_tensor_identity_9} 
    \, , \\
    \sum_{\bar{\lambda}} \Tr[ \left( \vect{\epsilon}^{\bar{\lambda}*,i} +c_1 \gamma^i \gamma^j \vect{\epsilon}^{\bar{\lambda}*,j} \right) \gamma^r \gamma^k  \left( \vect{\epsilon}^{\bar{\lambda},m} +c_2 \gamma^m \gamma^n \vect{\epsilon}^{\bar{\lambda},n} \right)  \gamma^k \gamma^t] &= 8\left(1-c_1-c_2 \right)  \delta^{im} \delta^{rt} - 8(c_2-c_1)\epsilon^{im} \epsilon^{rt} \, , \label{epsilon_tensor_identity_10} \\
    \sum_{\bar{\lambda}} \Tr[ \left( \vect{\epsilon}^{\bar{\lambda}*,i} +c_1 \gamma^i \gamma^j \vect{\epsilon}^{\bar{\lambda}*,j} \right) \gamma^k \gamma^r  \left( \vect{\epsilon}^{\bar{\lambda},m} +c_2 \gamma^m \gamma^n \vect{\epsilon}^{\bar{\lambda},n} \right)  \gamma^k \gamma^t] &= 0 \, ,  \label{epsilon_tensor_identity_11}\\
    \sum_{\bar{\lambda}} \Tr[ \left( \vect{\epsilon}^{\bar{\lambda}*,i} +c_1 \gamma^i \gamma^j \vect{\epsilon}^{\bar{\lambda}*,j} \right) \gamma^r \gamma^k  \left( \vect{\epsilon}^{\bar{\lambda},m} +c_2 \gamma^m \gamma^n \vect{\epsilon}^{\bar{\lambda},n} \right)  \gamma^t \gamma^k] &= 0 \,  . \label{epsilon_tensor_identity_12}
\end{align}
Four gamma matrices between polarization vector structures:
\begin{align}
    \sum_{\bar{\lambda}} \Tr[\left( \vect{\epsilon}^{\bar{\lambda}*,i} +c_1 \gamma^j \gamma^i \vect{\epsilon}^{\bar{\lambda}*,j} \right) \gamma^k \gamma^r  \gamma^t \gamma^k \left( \vect{\epsilon}^{\bar{\lambda},m} +c_2 \gamma^m \gamma^n \vect{\epsilon}^{\bar{\lambda},n} \right)]&=8(1-c_1-c_2+2c_1 c_2) \delta^{im} \delta^{rt} \nonumber \\
    &+ 8(c_1 +c_2 - 2c_1 c_2) \epsilon^{im} \epsilon^{rt} \, ,  \label{epsilon_tensor_identity_13} \\
    \sum_{\bar{\lambda}} \Tr[\left( \vect{\epsilon}^{\bar{\lambda}*,i} +c_1 \gamma^j \gamma^i \vect{\epsilon}^{\bar{\lambda}*,j} \right) \gamma^r \gamma^k  \gamma^k \gamma^t \left( \vect{\epsilon}^{\bar{\lambda},m} +c_2 \gamma^m \gamma^n \vect{\epsilon}^{\bar{\lambda},n} \right)]&=8(1-c_1-c_2+2c_1 c_2) \delta^{im} \delta^{rt} \nonumber \\
    &- 8(c_1 + c_2 - 2c_1 c_2) \epsilon^{im} \epsilon^{rt} \, ,  \label{epsilon_tensor_identity_14} \\
    \sum_{\bar{\lambda}} \Tr[\left( \vect{\epsilon}^{\bar{\lambda}*,i} +c_1 \gamma^j \gamma^i \vect{\epsilon}^{\bar{\lambda}*,j} \right) \gamma^k \gamma^r  \gamma^k \gamma^t \left( \vect{\epsilon}^{\bar{\lambda},m} +c_2 \gamma^m \gamma^n \vect{\epsilon}^{\bar{\lambda},n} \right)]&=0 \, ,  \label{epsilon_tensor_identity_15} \\
    \sum_{\bar{\lambda}} \Tr[\left( \vect{\epsilon}^{\bar{\lambda}*,i} +c_1 \gamma^j \gamma^i \vect{\epsilon}^{\bar{\lambda}*,j} \right) \gamma^r \gamma^k  \gamma^t \gamma^k \left( \vect{\epsilon}^{\bar{\lambda},m} +c_2 \gamma^m \gamma^n \vect{\epsilon}^{\bar{\lambda},n} \right)]&= 0 \, . \label{epsilon_tensor_identity_16}
\end{align}
Four gamma matrices (two and two) between polarization vector structures:
\begin{align}
    \sum_{\bar{\lambda}} \Tr[ \left( \vect{\epsilon}^{\bar{\lambda}*,i} +c_1 \gamma^j \gamma^i \vect{\epsilon}^{\bar{\lambda}*,j} \right) \gamma^k \gamma^r  \left( \vect{\epsilon}^{\bar{\lambda},m} +c_2 \gamma^n \gamma^m \vect{\epsilon}^{\bar{\lambda},n} \right)  \gamma^t \gamma^k] &= 8\left(1-c_1-c_2 \right)  \delta^{im} \delta^{rt} - 8(c_2-c_1)\epsilon^{im} \epsilon^{rt} \, , \label{epsilon_tensor_identity_17} \\
    \sum_{\bar{\lambda}} \Tr[ \left( \vect{\epsilon}^{\bar{\lambda}*,i} +c_1 \gamma^j \gamma^i \vect{\epsilon}^{\bar{\lambda}*,j} \right) \gamma^r \gamma^k  \left( \vect{\epsilon}^{\bar{\lambda},m} +c_2 \gamma^n \gamma^m \vect{\epsilon}^{\bar{\lambda},n} \right)  \gamma^k \gamma^t] &= 8\left(1-c_1-c_2 \right)  \delta^{im} \delta^{rt} + 8(c_2-c_1)\epsilon^{im} \epsilon^{rt} \, ,  \label{epsilon_tensor_identity_18} \\
    \sum_{\bar{\lambda}} \Tr[ \left( \vect{\epsilon}^{\bar{\lambda}*,i} +c_1 \gamma^j \gamma^i \vect{\epsilon}^{\bar{\lambda}*,j} \right) \gamma^k \gamma^r  \left( \vect{\epsilon}^{\bar{\lambda},m} +c_2 \gamma^n \gamma^m \vect{\epsilon}^{\bar{\lambda},n} \right)  \gamma^k \gamma^t] &= 0 \, ,  \label{epsilon_tensor_identity_19}\\
    \sum_{\bar{\lambda}} \Tr[ \left( \vect{\epsilon}^{\bar{\lambda}*,i} +c_1 \gamma^j \gamma^i \vect{\epsilon}^{\bar{\lambda}*,j} \right) \gamma^r \gamma^k  \left( \vect{\epsilon}^{\bar{\lambda},m} +c_2 \gamma^n \gamma^m \vect{\epsilon}^{\bar{\lambda},n} \right)  \gamma^t \gamma^k] &= 0 \, . \label{epsilon_tensor_identity_20}
\end{align}
Four gamma matrices between polarization vector structures:
\begin{align}
    \sum_{\bar{\lambda}} \Tr[\left( \vect{\epsilon}^{\bar{\lambda}*,i} +c_1 \gamma^i \gamma^j \vect{\epsilon}^{\bar{\lambda}*,j} \right) \gamma^k \gamma^r  \gamma^t \gamma^k \left( \vect{\epsilon}^{\bar{\lambda},m} +c_2 \gamma^m \gamma^n \vect{\epsilon}^{\bar{\lambda},n} \right)]&=8(1-c_1-c_2) \delta^{im} \delta^{rt} + 8(c_2 -c_1) \epsilon^{im} \epsilon^{rt} \, , \label{epsilon_tensor_identity_21} \\
    \sum_{\bar{\lambda}} \Tr[\left( \vect{\epsilon}^{\bar{\lambda}*,i} +c_1 \gamma^i \gamma^j \vect{\epsilon}^{\bar{\lambda}*,j} \right) \gamma^r \gamma^k  \gamma^k \gamma^t \left( \vect{\epsilon}^{\bar{\lambda},m} +c_2 \gamma^m \gamma^n \vect{\epsilon}^{\bar{\lambda},n} \right)]&=8(1-c_1-c_2) \delta^{im} \delta^{rt}  - 8(c_2 - c_1 ) \epsilon^{im} \epsilon^{rt} \, , \label{epsilon_tensor_identity_22} \\
    \sum_{\bar{\lambda}} \Tr[\left( \vect{\epsilon}^{\bar{\lambda}*,i} +c_1 \gamma^i \gamma^j \vect{\epsilon}^{\bar{\lambda}*,j} \right) \gamma^k \gamma^r  \gamma^k \gamma^t \left( \vect{\epsilon}^{\bar{\lambda},m} +c_2 \gamma^m \gamma^n \vect{\epsilon}^{\bar{\lambda},n} \right)]&=0 \, , \label{epsilon_tensor_identity_23} \\
    \sum_{\bar{\lambda}} \Tr[\left( \vect{\epsilon}^{\bar{\lambda}*,i} +c_1 \gamma^i \gamma^j \vect{\epsilon}^{\bar{\lambda}*,j} \right) \gamma^r \gamma^k  \gamma^t \gamma^k \left( \vect{\epsilon}^{\bar{\lambda},m} +c_2 \gamma^m \gamma^n \vect{\epsilon}^{\bar{\lambda},n} \right)]&= 0 \, . \label{epsilon_tensor_identity_24}
\end{align}
Four gamma matrices (two and two) between polarization vector structures:
\begin{align}
    \sum_{\bar{\lambda}} \Tr[ \left( \vect{\epsilon}^{\bar{\lambda}*,i} +c_1 \gamma^i \gamma^j \vect{\epsilon}^{\bar{\lambda}*,j} \right) \gamma^k \gamma^r  \left( \vect{\epsilon}^{\bar{\lambda},m} +c_2 \gamma^n \gamma^m \vect{\epsilon}^{\bar{\lambda},n} \right)  \gamma^t \gamma^k] &= 8\left(1-c_1-c_2 + 2 c_1 c_2 \right)  \delta^{im} \delta^{rt} \nonumber \\ &- 8(c_1+c_2-2c_1 c_2)\epsilon^{im} \epsilon^{rt} \, ,  \label{epsilon_tensor_identity_25} \\
    \sum_{\bar{\lambda}} \Tr[ \left( \vect{\epsilon}^{\bar{\lambda}*,i} +c_1 \gamma^i \gamma^j \vect{\epsilon}^{\bar{\lambda}*,j} \right) \gamma^r \gamma^k  \left( \vect{\epsilon}^{\bar{\lambda},m} +c_2 \gamma^n \gamma^m \vect{\epsilon}^{\bar{\lambda},n} \right)  \gamma^k \gamma^t] &= 8\left(1-c_1-c_2 + 2 c_1 c_2 \right)  \delta^{im} \delta^{rt} \nonumber \\ &+ 8(c_1+c_2-2c_1 c_2)\epsilon^{im} \epsilon^{rt} \, , \label{epsilon_tensor_identity_26} \\
    \sum_{\bar{\lambda}} \Tr[ \left( \vect{\epsilon}^{\bar{\lambda}*,i} +c_1 \gamma^i \gamma^j \vect{\epsilon}^{\bar{\lambda}*,j} \right) \gamma^k \gamma^r  \left( \vect{\epsilon}^{\bar{\lambda},m} +c_2 \gamma^n \gamma^m \vect{\epsilon}^{\bar{\lambda},n} \right)  \gamma^k \gamma^t] &= 0 \, , \label{epsilon_tensor_identity_27}\\
    \sum_{\bar{\lambda}} \Tr[ \left( \vect{\epsilon}^{\bar{\lambda}*,i} +c_1 \gamma^i \gamma^j \vect{\epsilon}^{\bar{\lambda}*,j} \right) \gamma^r \gamma^k  \left( \vect{\epsilon}^{\bar{\lambda},m} +c_2 \gamma^n \gamma^m \vect{\epsilon}^{\bar{\lambda},n} \right)  \gamma^t \gamma^k] &= 0 \, . \label{epsilon_tensor_identity_28}
\end{align}
\subsubsection*{Identities III: One polarization vector structure and 4 transverse gamma matrices}
Four gamma matrices and one polarization vector structure:
\begin{align}
    \sum_{\bar{\lambda}} \Tr[\left( \vect{\epsilon}^{\bar{\lambda}*,i} +c_1 \gamma^j \gamma^i \vect{\epsilon}^{\bar{\lambda}*,j} \right) \gamma^k \gamma^r \gamma^k \gamma^m] \vect{\epsilon}^{\bar{\lambda},m} &= 0 \, ,  \\
    \sum_{\bar{\lambda}} \Tr[\left( \vect{\epsilon}^{\bar{\lambda}*,i} +c_1 \gamma^j \gamma^i \vect{\epsilon}^{\bar{\lambda}*,j} \right) \gamma^r \gamma^k \gamma^k \gamma^m] \vect{\epsilon}^{\bar{\lambda},m} &= 8 (1-2c_1) \ \delta^{ir} \, .
\end{align}
Four gamma matrices and one polarization vector structure:
\begin{align}
    \sum_{\bar{\lambda}} \Tr[\left( \vect{\epsilon}^{\bar{\lambda}*,i} +c_1 \gamma^j \gamma^i \vect{\epsilon}^{\bar{\lambda}*,j} \right) \gamma^r \gamma^k \gamma^m \gamma^k] \vect{\epsilon}^{\bar{\lambda},m} &= 0 \, ,  \\
    \sum_{\bar{\lambda}} \Tr[\left( \vect{\epsilon}^{\bar{\lambda}*,i} +c_1 \gamma^j \gamma^i \vect{\epsilon}^{\bar{\lambda}*,j} \right) \gamma^k \gamma^r \gamma^m \gamma^k] \vect{\epsilon}^{\bar{\lambda},m} &= 8\ \delta^{ir} \, .
\end{align}
Four gamma matrices and one polarization vector structure:
\begin{align}
    \sum_{\bar{\lambda}} \Tr[\left( \vect{\epsilon}^{\bar{\lambda}*,i} +c_1 \gamma^i \gamma^j \vect{\epsilon}^{\bar{\lambda}*,j} \right) \gamma^k \gamma^r \gamma^k \gamma^m] \vect{\epsilon}^{\bar{\lambda},m} &= 0 \, ,  \\
    \sum_{\bar{\lambda}} \Tr[\left( \vect{\epsilon}^{\bar{\lambda}*,i} +c_1 \gamma^i \gamma^j \vect{\epsilon}^{\bar{\lambda}*,j} \right) \gamma^r \gamma^k \gamma^k \gamma^m] \vect{\epsilon}^{\bar{\lambda},m} &= 8 \ \delta^{ir} \, .
\end{align}
Four gamma matrices and one polarization vector structure:
\begin{align}
    \sum_{\bar{\lambda}} \Tr[\left( \vect{\epsilon}^{\bar{\lambda}*,i} +c_1 \gamma^i \gamma^j \vect{\epsilon}^{\bar{\lambda}*,j} \right) \gamma^r \gamma^k \gamma^m \gamma^k] \vect{\epsilon}^{\bar{\lambda},m} &= 0 \, ,  \\
    \sum_{\bar{\lambda}} \Tr[\left( \vect{\epsilon}^{\bar{\lambda}*,i} +c_1 \gamma^i \gamma^j \vect{\epsilon}^{\bar{\lambda}*,j} \right) \gamma^k \gamma^r \gamma^m \gamma^k] \vect{\epsilon}^{\bar{\lambda},m} &= 8 (1-2c_1) \delta^{ir} \, .
\end{align}

\section{Expressions for the coefficients $\zeta$, $\chi$, $\kappa$, $\sigma$ and functions  $\mathcal{J}_{T}$ in $R^{q\bar{q}\gamma}_{T}$} \label{sec:coefficients-J-functions}

In this Appendix, we will provide the expressions for the various coefficients and functions that constitute the coefficient function $R_{T}^{q\bar{q}\gamma}$ and its three components given by  Eqs.~\eqref{eq:R-trans-pol-gamma-dijet}-\eqref{eq:R-T-ins-ins}. 

\begin{align}
    \zeta_{qq} &= \left[\left(\frac{1-z_{\bar{q}} }{z_q} \right)^2 + 1\right] \left[\left( 1- z_{\bar{q}} \right)^2  + z^2_{\bar{q}} \right] \quad , \quad \zeta_{\bar{q}\bar{q}} = \left[\left(\frac{1-z_{q} }{z_{\bar{q}}} \right)^2 + 1\right] \left[\left( 1- z_{q} \right)^2  + z^2_{q} \right] \, , \label{eq:zeta-qq-zeta-qbar-qbar} \\
    \zeta_{q\bar{q}} &= \zeta_{\bar{q}q}=\left[\frac{1-z_q}{z_{\bar{q}}} + \frac{1-z_{\bar{q}}}{z_{q}} \right]\left[(1-z_{\bar{q}})z_q + z_{\bar{q}} (1-z_q) \right] \, , \label{eq:zeta-q-qbar} \\
    \chi_{qq} &= \left[\left(\frac{1-z_{\bar{q}} }{z_q} \right)^2 - 1\right] \left[\left( 1- z_{\bar{q}} \right)^2  - z^2_{\bar{q}} \right] \quad ,  \quad  \chi_{\bar{q}\bar{q}} = \left[\left(\frac{1-z_{q} }{z_{\bar{q}}} \right)^2 - 1\right] \left[\left( 1- z_{q} \right)^2  - z^2_{q} \right] \, , \label{eq:chi-qq-chi-qbar-qbar} \\
    \chi_{q\bar{q}} &= \chi_{\bar{q}q}=\left[\frac{1-z_q}{z_{\bar{q}}} - \frac{1-z_{\bar{q}}}{z_{q}} \right]\left[(1-z_{\bar{q}})z_q - z_{\bar{q}} (1-z_q) \right] \, , \label{eq:chi-q-qbar} \\
    \kappa_{qq} &=
    \frac{z^2_{\bar{q}}(1-z_{\bar{q}})}{(1-z_q)} \quad \, , \quad \kappa_{\bar{q}\bar{q}} = \frac{z^2_{q}(1-z_{q})}{(1-z_{\bar{q}})} \, ,  \label{eq:kappa-qq-qbar-qbar} \\
   \kappa_{q\bar{q}} &= \kappa_{\bar{q}q} = (1-z_q)(1-z_{\bar{q}}) \label{eq:kappa-q-qbar} \, , \\
   \sigma_{qq} &= \frac{z^2_{\bar{q}}(1-z_{\bar{q}})^2}{(1-z_q)^2} \quad \, , \quad \sigma_{\bar{q}\bar{q}} = \frac{z^2_{q}(1-z_{q})^2}{(1-z_{\bar{q}})^2} \, . \label{eq:sigma-qq-qbar-qbar}
\end{align}
For the case of $\gamma+q$ production, where we integrate over the antiquark phase space, we should replace $z_{\bar{q}}=1-z_{q}-z_{\gamma}$ to get the corresponding expressions for the above coefficients. 
\begin{align}
    \mathcal{J}^{i j}_{T,1}(\vect{r})  &=   z_q \frac{\left(z_q \vect{k_\gamma}^i -z_\gamma \vect{k}^i \right) }{ \left(z_q \vect{k_\gamma} -z_\gamma \vect{k} \right)^2} \int_{\vect{l}}  \frac{\vect{l}^j e^{i \vect{l} \cdot \vect{r}}}{\vect{l}^2 + \Delta_1} \, ,  \qquad \Delta_{1}=z_{\bar{q}}(1-z_{\bar{q}}) \, Q^{2} \, ,   \label{eq:J-T-1}\\
    \mathcal{J}^{i j}_{T,2}(\vect{r})  &=- \frac{1}{z_\gamma (1-z_{\bar{q}})} \int_{\vect{l}} \frac{\left[(1-z_{\bar{q}})\vect{k_\gamma}^i -z_\gamma \vect{l}^i \right] \vect{l}^j e^{i \vect{l} \cdot \vect{r} }}{ \left( \vect{l}^2 + \Delta_1  \right) \left( Q^2 +\frac{\vect{l}^2}{z_{\bar{q}}}+\frac{\vect{k_\gamma}^2}{z_\gamma} + \frac{(\vect{l} - \vect{k_\gamma})^2}{z_q}  \right)} \, , \label{eq:J-T-2} \\
    \mathcal{J}_{T\rm{ins},2}(\vect{r})  &=- \frac{(1-z_q)}{  z_{\bar{q}} (1-z_{\bar{q}})^2} \int_{\vect{l}} \frac{e^{i \vect{l} \cdot \vect{r} }}{ \left( Q^2 +\frac{\vect{l}^2}{z_{\bar{q}}}+\frac{\vect{k_\gamma}^2}{z_\gamma} + \frac{(\vect{l} - \vect{k_\gamma})^2}{z_q}  \right)} \, . \label{eq:J-T-2-ins}
\end{align}
The expressions for  $\mathcal{J}^{ij}_{T,3}$, $\mathcal{J}^{ij}_{T,4}$ and $\mathcal{J}_{T\rm{ins},4}$ can be obtained from the above results simply by interchange of quark and antiquark (i.e. $k \leftrightarrow p$ and $\vect{r} \leftrightarrow -\vect{r}$).

\section{Photon+quark production cross-section for transversely polarized virtual photon} \label{sec:photon-jet-trans-pol}

In this section, we will provide expressions for the direct and fragmentation contributions to the $\gamma+q$ production cross-section in Eq.\,\eqref{eq:cs-in-terms-of-d-and-f} for the case of the transversely polarized virtual photon. The procedure is similar to the case discussed in the main text for longitudinally polarized virtual photon albeit with more contributions in each case. 

\subsection{Fragmentation photon+quark production cross-section} \label{sec:fragmentation-contribution-trans-pol}

We begin with the expression for the cross-section for $\gamma+q\bar{q}$ production in $\gamma^{*}_{T}+A$ scattering given by Eqs.~\eqref{eq:diff-cs-trans-pol} and \eqref{eq:R-trans-pol-gamma-dijet} and then integrate over $\bm{p}_{\perp}$ and $\eta_{\bar{q}}$ (or $z_{\bar{q}}$). The integration over the latter is trivial using the Dirac delta function and sets $z_{\bar{q}}=1-z_{q}-z_{\gamma}$. For the integration over $\bm{p}_{\perp}$, we have mostly terms in $R^{q\bar{q}\gamma}_{T}$ which are independent of $\bm{p}_{\perp}$ and therefore yields the delta: $\delta^{(2)}(\vect{y}-\bm{y'}_{\perp})$. Integrating over $\bm{y'}_{\perp}$, we get the reduced color structure containing only dipoles defined by Eq.\,\eqref{eq:coincidence-color-structure} for such terms.

There are two kinds of terms that depend on $\bm{p}_{\perp}$: those proportional to the tensor function $\mathcal{J}_{T,3}^{ij \, *} \, $ (or its complex conjugate) which yield finite results upon the $\bm{p}_{\perp}$ integration, and one term\footnote{It is easy to check that the term proportional to $\epsilon^{im }\epsilon^{jn} \mathcal{J}_{T,3}^{ij} \, \mathcal{J}_{T,3}^{mn \, *} $ is zero because of the symmetric nature of $ \mathcal{J}_{T,3}^{ij} \, \mathcal{J}_{T,3}^{mn \, *} $ with respect to the indices $i$ and $m$.} proportional to $\zeta_{\bar{q} \bar{q}} \,  \delta^{im} \delta^{jn} \, \mathcal{J}_{T,3}^{ij} \, \mathcal{J}_{T,3}^{mn \, *} $ which will exhibit collinear singularity. Both these terms have the general LO color structure in Eq.\,\eqref{eq:color-structure-LO}. We use an IR regulator $\Lambda$ and introduce a scale, $\mu_F$ to isolate the contribution from such fragmentation photons. The contributions above the scale $\mu_{F}$ are absorbed into the remaining finite terms and together they constitute the direct photon+quark contribution.

In terms of the momentum fraction, $\lambda_{\gamma/\bar{q}}$ of the photon with respect to the parent antiquark,  defined in Eq.\,\eqref{eq:photon-mom-frac-q-bar} we obtain 
\begin{equation}
    \zeta_{\bar{q}\bar{q}}= \big[ z_{q}^2+(1-z_q)^2 \big] \, \frac{\lambda_{\gamma/\bar{q}}}{(1-\lambda_{\gamma/\bar{q}})^2} \, P_{\gamma \bar{q}}(\lambda_{\gamma/\bar{q}}) \, ,
\end{equation}
where the photon splitting function is defined by Eq.\,\eqref{eq:gamma-splitting-function}. Also, using the identity
\begin{align}
    \int_{\vect{l}} e^{-i \vect{l} \cdot \vect{r}} \frac{\bm{l}_{\perp}^{j}}{\bm{l}_{\perp}^{2}+\Delta} = -\frac{i \bm{r}_{\perp}^{j}}{4 \pi} \, \Big( \frac{2 \, \Delta_{3}^{1/2}}{r_{\perp}} \Big) \, K_{1}(\Delta_{3}^{1/2}r_{\perp} ) \, ,
    \label{eq:Bessel-K-1}
\end{align}
where $K_{1}$ is a modified Bessel of the second kind, we can express the fragmentation contribution to $\gamma+q$ production in $\gamma^{*}_{T}+A$ scattering as
\begin{align}
    \sum_{T=\pm 1}   \frac{d \sigma^{\gamma^{*}_{T} A \rightarrow q \gamma X}_{F}}{\mathrm{d}^{2} \vect{k}  \mathrm{d}^{2} \vect{k_{\gamma}}    \mathrm{d} \eta_{q} \mathrm{d} \eta_{\gamma}} &= \frac{4 \, N_{\mathrm{c}} \alpha_{\rm em} q_{f}^2 z_{q}}{(2\pi)^6} \int_{\vect{x},\vect{y},\vect{x'},\vect{y'}} \!\!\!\!\!\!\!\!\!\!\!\!\!\!\!\!\!\!\!\!\!\!\!\!\!\!\! e^{-i\vect{k}.(\vect{x}-\vect{x'})} e^{-i \frac{\vect{k_{\gamma}}}{\lambda_{\gamma/\bar{q}}}.(\vect{y}-\vect{y'})} \, \Xi(\vect{x},\vect{y};\vect{y'},\vect{x'} \vert Y) \nonumber \\
    & \times \frac{D_{\bar{q}\rightarrow \gamma}(\lambda_{\gamma/\bar{q}},\mu_{F})}{\lambda_{\gamma/\bar{q}}} \, \big[z_{q}^2+(1-z_{q})^2 \big] \, \frac{\bm{r}_{\perp}.\bm{r'}_{\perp}}{r_{\perp} \, r'_{\perp}} \, \Delta_{3} \, K_{1}(\Delta_{3}^{1/2}r_{\perp} ) \, K_{1}(\Delta_{3}^{1/2}r'_{\perp} ) \, ,
    \label{eq:fragmentation-cs-trans-pol}
\end{align}
where the (anti)quark-to-photon fragmentation function at the scale $\mu_{F}$ is defined by Eq.\,\eqref{eq:photon-fragmentation function}. This can equivalently be written in terms of the sum of the squared lightcone wavefunctions for the two transverse polarization states of the virtual photon as
\begin{align}
   \sum_{T=\pm 1}  \frac{d \sigma^{\gamma^{*}_{T} A \rightarrow q \gamma X}_{F}}{\mathrm{d}^{2} \vect{k}  \mathrm{d}^{2} \vect{k_{\gamma}}    \mathrm{d} \eta_{q} \mathrm{d} \eta_{\gamma}} & =\frac{\sqrt{2} \, N_{c}\alpha_{\rm em}q^{2}_{f} \, k_{\perp} e^{\eta_{q}}}{ (2\pi)^8}  \int_{\vect{x},\vect{y},\vect{x'},\vect{y'}} \!\!\!\!\!\!\!\!\!\!\!\!\!\!\!\!\!\!\!\!\!\!\!\!\!\!\! e^{-i\vect{k}.(\vect{x}-\vect{x'})} e^{-i \frac{\vect{k_{\gamma}}}{\lambda_{\gamma/\bar{q}}}.(\vect{y}-\vect{y'})} \, \Xi(\vect{x},\vect{y};\vect{y'},\vect{x'} \vert Y) \nonumber \\
     &\times \frac{D_{\bar{q}\rightarrow \gamma}(\lambda_{\gamma/\bar{q}},\mu_{F})}{\lambda_{\gamma/\bar{q}}} \sum_{T=\pm 1, \,  \alpha,\beta} \Psi^{T \, *}_{\alpha \, \beta} (q^{-},z_{q},r'_{\perp}) \, \Psi^{T }_{\alpha \, \beta} (q^{-},z_{q},r_{\perp}) \, , 
     \label{eq:fragmentation-cs-trans-pol-wavefunctions}
\end{align}
where the wavefunctions are defined as~\cite{Dominguez:2011wm}
\begin{align}
  \Psi^{T }_{\alpha \, \beta} (q^{-},z_{q},r_{\perp})= 2\pi \, \sqrt{\frac{2}{q^-}} \begin{cases}
  i \, \Delta^{1/2} \, K_1(\Delta^{1/2} \, r_{\perp}) \, \frac{\bm{r}_{\perp}.\bm{\epsilon}_{\perp}^{T=+1}}{r_{\perp}} \, \big[ \delta_{\alpha +} \, \delta_{\beta +} \, (1-z) +\delta_{\alpha -} \, \delta_{\beta -} \, z \big] \quad , \quad T=+1 \, ,  \\[0.2cm]
   i \, \Delta^{1/2} \, K_1(\Delta^{1/2} \, r_{\perp}) \, \frac{\bm{r}_{\perp}.\bm{\epsilon}_{\perp}^{T=-1}}{r_{\perp}} \, \big[ \delta_{\alpha -} \, \delta_{\beta -} \, (1-z) +\delta_{\alpha +} \, \delta_{\beta +} \, z \big] \quad , \quad T=-1  \, .
  \end{cases}
\end{align}
The similarity with the results obtained in Eqs.~\eqref{eq:fragmentation-cs-long-pol} and \eqref{eq:fragmentation-cs-long-pol-wavefunctions} for the longitudinally polarized virtual photon can be clearly seen. We will now discuss the computation of the direct photon+quark production cross-section.

\subsection{Direct photon+quark production cross-section} \label{sec:direct-photon-trans-pol}

Following the discussion in
Sec.~\ref{sec:direct-contribution-gamma-jet-long}, we can write the direct photon+quark production cross-section as 
\begin{align}
 \sum_{T=\pm 1}     \frac{d \sigma^{\gamma^{*}_{T} A \rightarrow q \gamma X}_{D}}{\mathrm{d}^{2} \vect{k}  \mathrm{d}^{2} \vect{k_{\gamma}}    \mathrm{d} \eta_{q} \mathrm{d} \eta_{\gamma}} &= \mathcal{A}_{T}(\eta_{q},\eta_{\gamma},\vect{k},\vect{k_{\gamma}})+\mathcal{B}_{T} (\eta_{q},\eta_{\gamma},\vect{k},\vect{k_{\gamma}};\mu_{F}) \, ,
      \label{eq:direct-cs-trans-pol}
 \end{align}
where $\mathcal{A}_{T}$ contains only dipoles and can be obtained as
\begin{align}
\mathcal{A}_{T}(\eta_{q},\eta_{\gamma},\vect{k},\vect{k_{\gamma}})&= \frac{8 N_{\mathrm{c}} \alpha_{\rm em}^2 q^4_{f} z_{q} }{(2\pi)^4} \, S_{\perp} \int_{\vect{r},\bm{r'}_{\perp}} e^{-i \bm{k}_{\perp}.(\bm{r}_{\perp}-\bm{r'}_{\perp})} \, \Theta(\bm{r}_{\perp},\bm{r'}_{\perp} \vert Y ) \, \big( A_{T;\rm{reg-reg}}  + A_{T;\rm{reg-ins}}+A_{T;\rm{ins-ins}} \big) \, , 
\label{eq:direct-cs-trans-pol-A}
\end{align}
where the color structure is defined in Eq.\,\eqref{eq:coincidence-color-structure-trans-invariance} and the three contributions constituting $\mathcal{A}_{T}$ are written below
\begin{align}
  A_{T;\rm{reg-reg}}  (\eta_{q},\eta_{\gamma},\vect{k},\vect{k_{\gamma}}) & =  [\zeta_{qq} \, \delta^{im} \, \delta^{jn}+\chi_{qq} \, \epsilon^{im} \, \epsilon^{jn}  ] \, \Big[ \mathcal{K}_{T,1}^{ij}(\bm{r}_{\perp})+\mathcal{K}_{T,2}^{ij}(\bm{r}_{\perp}) \Big] \, \Big[ \mathcal{K}_{T,1}^{mn \, *}(\bm{r'}_{\perp})+\mathcal{K}_{T,2}^{mn \, *}(\bm{r'}_{\perp}) \Big] \nonumber \\
  &+ [\zeta_{q\bar{q}} \, \delta^{im} \, \delta^{jn}- \chi_{q\bar{q}} \, \epsilon^{im} \, \epsilon^{jn}  ] \,  \big[ \mathcal{K}_{T,1}^{ij}(\bm{r}_{\perp})+\mathcal{K}_{T,2}^{ij}(\bm{r}_{\perp}) \big] \, \mathcal{K}_{T,4}^{mn \, *}(\bm{r'}_{\perp})
  \nonumber \\
  & + [\zeta_{\bar{q}q} \, \delta^{im} \, \delta^{jn}- \chi_{\bar{q}q} \, \epsilon^{im} \, \epsilon^{jn}  ] \, \mathcal{K}_{T,4}^{ij }(\bm{r}_{\perp}) \, \big[ \mathcal{K}_{T,1}^{mn \, *}(\bm{r'}_{\perp})+\mathcal{K}_{T,2}^{mn \, *}(\bm{r'}_{\perp}) \big] 
  \nonumber \\
  & +[\zeta_{\bar{q}\bar{q}} \, \delta^{im} \, \delta^{jn}+\chi_{\bar{q}\bar{q}} \, \epsilon^{im} \, \epsilon^{jn}  ] \, \mathcal{K}_{T,4}^{ij}(\bm{r}_{\perp}) \, \mathcal{K}_{T,4}^{mn \, *} (\bm{r'}_{\perp}) \, ,
  \label{eq:A-T-reg-reg}
\end{align}
\begin{align}
    A_{T;\rm{reg-ins}}  (\eta_{q},\eta_{\gamma},\vect{k},\vect{k_{\gamma}}) & =  \kappa_{qq} \Big[  \left(\mathcal{K}^{ii}_{T,1} (\vect{r})  + \mathcal{K}^{ii}_{T,2} (\vect{r}) \right) \mathcal{K}^{*}_{T\rm{ins},2}(\vect{r}') + \mathcal{K}_{T\rm{ins},2}(\vect{r}) \left(\mathcal{K}^{ii*}_{T,1} (\vect{r}')  + \mathcal{K}^{ii*}_{T,2} (\vect{r}') \right) \Big] \nonumber \\
    &+ \kappa_{q\bar{q}} \Big[ \left(\mathcal{K}^{ii}_{T,1} (\vect{r})  + \mathcal{K}^{ii}_{T,2} (\vect{r}) \right) \mathcal{K}^{*}_{T\rm{ins},4}(\vect{r}') + \mathcal{K}_{T\rm{ins},2}(\vect{r}) \,  \mathcal{K}^{ii*}_{T,4} (\vect{r}')   \Big] \nonumber \\
    &+ \kappa_{\bar{q}q} \Big[  \mathcal{K}^{ii}_{T,4} (\vect{r}) \, \mathcal{K}^{*}_{T\rm{ins},2}(\vect{r}') + \mathcal{K}_{T\rm{ins},4}(\vect{r}) \left(\mathcal{K}^{ii*}_{T,1} (\vect{r}')  + \mathcal{K}^{ii*}_{T,2} (\vect{r}') \right)   \Big]  \nonumber
    \\
    &+ \kappa_{\bar{q}\bar{q}}  \Big[   \mathcal{K}^{ii}_{T4} (\vect{r}) \,  \mathcal{K}^{*}_{T\rm{ins},4}(\vect{r}') + \mathcal{K}_{T\rm{ins},4}(\vect{r}) \, \mathcal{K}^{ii*}_{T,4} (\vect{r}')  \Big] \, ,
    \label{eq:A-T-reg-ins}
\end{align}
\begin{align}
    A_{T;\rm{ins-ins}}  (\eta_{q},\eta_{\gamma},\vect{k_{\gamma}})= \sigma_{qq} \, \mathcal{K}_{T \rm{ins},2}(\bm{r}_{\perp}) \, \mathcal{K}^{*}_{T \rm{ins},2}(\bm{r'}_{\perp}) + \sigma_{\bar{q}\bar{q}} \,  \mathcal{K}_{T \rm{ins},4}(\bm{r}_{\perp}) \, \mathcal{K}^{*}_{T \rm{ins},4}(\bm{r'}_{\perp}) \, .
    \label{eq:A-T-ins-ins}
\end{align}
The functions $\mathcal{K}_{T}$ appearing in the above equations are defined in terms of the functions $\mathcal{J}_{T}$ given in Appendix~\ref{sec:coefficients-J-functions} as
\begin{align}
    \mathcal{K}_{T,1}^{ij}(\bm{r}_{\perp})&= \mathcal{J}_{T,1}^{ij}(\bm{r}_{\perp}) \, e^{-i \bm{k}_{\gamma \perp} \cdot \bm{r}_{\perp} } \quad , \quad  \mathcal{K}_{T,2}^{ij}(\bm{r}_{\perp})= \mathcal{J}_{T,2}^{ij}(\bm{r}_{\perp}) \, e^{-i \bm{k}_{\gamma \perp} \cdot \bm{r}_{\perp} } \, , \nonumber \\
    \mathcal{K}_{T \rm{ins},2}(\bm{r}_{\perp})&= \mathcal{J}_{T \rm{ins},2}(\bm{r}_{\perp}) \, e^{-i \bm{k}_{\gamma \perp} \cdot \bm{r}_{\perp} }  \quad , \quad  \mathcal{K}_{T \rm{ins},4}(\bm{r}_{\perp})= \mathcal{J}_{T \rm{ins},4}(\bm{r}_{\perp}) \, .
    \label{eq:K-T-definitions}
    \end{align}
To compute the second term, $\mathcal{B}_{T}$ in the direct contribution which contains both dipoles and a quadrupole, we need the following results
\begin{align}
    \int \mathrm{d}^2 \bm{p}_{\perp} e^{-i \vect{p} \cdot (\bm{y}_{\perp}-\bm{y'}_{\perp})} \begin{cases} \mathcal{J}^{ij}_{T,3}(\bm{r}_{\perp}) \\
    \mathcal{J}^{ij \, *}_{T,3}(\bm{r'}_{\perp})
    \end{cases}
    =2\pi i \, \frac{\bm{r}^{i}_{yy' \perp}}{\bm{r}^{2}_{yy' \perp}} e^{-i  \frac{(1-z_{q}) }{z_{\gamma}} \,  \vect{k_{\gamma}}.\bm{r}_{yy' \perp}} \, e^{i\vect{k_{\gamma}}.\bm{r}_{yy' \perp}} \,  \begin{cases} \hat{\mathcal{J}}^{j}_{T,3} (\bm{r}_{\perp}) \\
   \hat{\mathcal{J}}^{j \, *}_{T,3} (\bm{r'}_{\perp})
   \end{cases} \, ,
    \label{eq:integral-over-J3-T} 
\end{align}
where $\bm{r}_{yy' \perp}=\bm{y}_{\perp}-\bm{y'}_{\perp}$ and 
\begin{equation}
   \hat{\mathcal{J}}^{j}_{T,3} (\bm{r}_{\perp})=\frac{1-z_{q}-z_{\gamma}}{z_{\gamma}} \int_{\vect{l}} e^{-i \vect{l}.\vect{r}}\frac{\bm{l}_{\perp}^{j}}{\bm{l}_{\perp}^2+\Delta_{3}} = - \frac{i \bm{r}_{\perp}^{j}}{2\pi \, r_{\perp} } \frac{(1-z_{q}-z_{\gamma} ) \, \Delta_{3}^{1/2}}{z_{\gamma}} \, K_{1}(\Delta_{3}^{1/2} \, r_{\perp}) \, .
   \label{eq:J-hat-T-3}
\end{equation}
 We have used Eq.\,\eqref{eq:Bessel-K-1} in getting to the second equality in the above equation. The corresponding expression for $ \hat{\mathcal{J}}^{j \, *}_{T,3} (\bm{r'}_{\perp})$ is obtained taking the complex conjugate of the r.h.s of Eq.\,\eqref{eq:J-hat-T-3} and replacing $\bm{r}_{\perp} \rightarrow \bm{r'}_{\perp}$.

Finally we can write the second term in Eq.\,\eqref{eq:direct-cs-trans-pol} as
\begin{align}
   \mathcal{B}_{T} (\eta_{q},\eta_{\gamma},\vect{k},\vect{k_{\gamma}};\mu_{F})&=  \frac{8 N_{\mathrm{c}} \alpha_{\rm em}^2 q^4_{f} z_{q} }{(2\pi)^6} \int_{\vect{x},\vect{y},\vect{x'},\vect{y'}} \!\!\!\!\!\!\!\!\!\!\!\!\!\!\!\!\!\!\!\!\!\!\!\!\!\!\! e^{-i\vect{k}.(\vect{x}-\vect{x'})} e^{-i \frac{1-z_{q}}{z_{\gamma}} \, \vect{k_{\gamma}} . \bm{r}_{yy' \perp}} \, \Xi(\vect{x},\vect{y};\vect{y'},\vect{x'} \vert Y) \nonumber \\
   & \times \big( B_{T;\rm{reg-reg}}(\eta_{q},\eta_{\gamma},\bm{k}_{\perp},\bm{k}_{\gamma \perp}; \mu_{F})   + B_{T;\rm{reg-ins}} (\eta_{q},\eta_{\gamma},\bm{k}_{\gamma \perp}) \big) \, , 
   \label{eq:direct-cs-trans-pol-B}
\end{align}
where 
\begin{align}
  B_{T;\rm{reg-reg}} &=  \zeta_{\bar{q}\bar{q}} \,  \pi \ln \Big( \frac{1}{\bm{r}_{yy' \perp}^2 \, \mu_{F}^2} \Big) \, \hat{\mathcal{J}}^{j}_{T,3} (\bm{r}_{\perp}) \, \hat{\mathcal{J}}^{j \, *}_{T,3} (\bm{r'}_{\perp}) \nonumber \\ &+ 2\pi i \, \frac{\bm{r}_{yy' \perp}^{m}}{\bm{r}_{yy' \perp}^2} \, [\zeta_{q\bar{q}} \delta^{im} \delta^{jn}-\chi_{q\bar{q}} \epsilon^{im} \epsilon^{jn}] \, \big[ \mathcal{K}^{ij}_{T,1}(\bm{r}_{\perp})  +\mathcal{K}^{ij}_{T,2}(\bm{r}_{\perp}) \big] \hat{\mathcal{J}}^{n \, *}_{T,3} (\bm{r'}_{\perp})
  \nonumber \\
  &+ 2\pi i \, \frac{\bm{r}_{yy' \perp}^{m}}{\bm{r}_{yy' \perp}^2} \, [\zeta_{\bar{q}q} \delta^{im} \delta^{jn}-\chi_{\bar{q}q} \epsilon^{im} \epsilon^{jn}] \, \hat{\mathcal{J}}^{n }_{T,3} (\bm{r}_{\perp}) \,  \big[ \mathcal{K}^{ij \, *}_{T,1}(\bm{r'}_{\perp}) +\mathcal{K}^{ij \, *}_{T,2}(\bm{r'}_{\perp} ) \big]  \nonumber \\ &  + 2\pi i \, \frac{\bm{r}_{yy' \perp}^{m}}{\bm{r}_{yy' \perp}^2} \, [\zeta_{\bar{q}\bar{q}} \delta^{im} \delta^{jn}+ \chi_{\bar{q}\bar{q}} \epsilon^{im} \epsilon^{jn}] \, \big[ \mathcal{K}^{ij}_{T,4}(\bm{r}_{\perp}) \hat{\mathcal{J}}^{n \, *}_{T,3} (\bm{r'}_{\perp})+ \hat{\mathcal{J}}^{n }_{T,3} (\bm{r}_{\perp}) \mathcal{K}^{ij \, *}_{T,4}(\bm{r'}_{\perp}) \big] \Big\} \, , 
 \label{eq:B-T-reg-reg}
\end{align}
and 
\begin{align}
    B_{T;\rm{reg-ins}}= 2\pi i \, \frac{\bm{r}_{yy' \perp}^{i}}{\bm{r}_{yy' \perp}^2} &\, \Big\{ \kappa_{q\bar{q}} \big[ \mathcal{K}_{T \rm{ins},2}(\vect{r}) \, \hat{\mathcal{J}}^{i \, *}_{T,3} (\bm{r'}_{\perp}) + \hat{\mathcal{J}}^{i}_{T,3} (\bm{r}_{\perp}) \, \mathcal{K}^{*}_{T \rm{ins},2}(\vect{r}') \big] \nonumber \\
     &+ \kappa_{\bar{q}\bar{q}} \big[ \mathcal{K}_{T \rm{ins},4}(\vect{r}) \, \hat{\mathcal{J}}^{i \, *}_{T,3} (\bm{r'}_{\perp})  + \hat{\mathcal{J}}^{i}_{T,3} (\bm{r}_{\perp}) \, \mathcal{K}^{*}_{T \rm{ins},4}(\vect{r}') \big] \Big\} \, . 
    \label{eq:B-T-reg-ins}
\end{align}
The functions appearing in the above expressions are defined in Eqs.~\eqref{eq:K-T-definitions} and \eqref{eq:J-hat-T-3}. The first term on the r.h.s of Eq.\,\eqref{eq:B-T-reg-reg} is the contribution in $$\int \mathrm{d}^2  \vect{p} e^{- i \vect{p} \cdot \bm{r}_{yy' \perp}} \mathcal{J}^{ij}_{T,3} (\vect{r}) \mathcal{J}^{ij \, *}_{T,3} (\bm{r'}_{\perp})  $$ above the fragmentation scale $\mu_{F}$. Combining the expressions for $\mathcal{A}_{T}$ and $\mathcal{B}_{T}$ given by Eqs.~\eqref{eq:direct-cs-trans-pol-A} and \eqref{eq:direct-cs-trans-pol-B} we get the total cross-section for the direct photon+quark production of transversely polarized virtual photons.

\subsection{Simplifying the direct photon+quark cross-section} \label{sec:direct-photon-limit-trans}

In this section, we will follow the procedure discussed for the longitudinal case in Sec.~\ref{sec:direct-photon-limit-long} where we simplify the direct photon+quark production cross-section into an expression which contains only the dipole Wilson line correlators. This leads to expressions for the cross-section that are fully in momentum space. This  makes the subsequent numerical analyses easier relative to the general result that also contains the quadrupole. 

We remind the reader that this simplification occurs in the limit $\bm{y}_{\perp} \rightarrow \bm{y'}_{\perp}$ or, equivalently, in terms of the momenta of the problem, for $$k_{\gamma \perp} \gg \frac{z_{\gamma}}{1-z_{q}} \, Q_{s} \, ,  $$ where $Q_{s}$ is the saturation scale that sets the scale of gradients in the correlators embedded in the LO color structure $\Xi (\vect{x},\vect{y}; \bm{y'}_{\perp}, \bm{x'}_{\perp})$. By Taylor expanding this color structure around $\bm{y}_{\perp}=\bm{y'}_{\perp}$ and using known identities (see for example Eq.\,\eqref{eq:identity-leading-term-first-integral}), we have shown that the higher order terms in the expansion of quantities like those in Eqs.~\eqref{eq:integral-type-1-B} and \eqref{eq:integral-type-2-B} are suppressed by powers of $(Q_{s}/v_{\perp})$, where $v_{\perp}=(1-z_{q}) /z_{\gamma } \, k_{\gamma \perp}$. Therefore, as long as the limit defined in Eq.\,\eqref{eq:limit} is satisfied for the choice of kinematics, our direct photon+quark production cross-section will be dominated by the leading term in such an expansion which contains only dipoles. 

We will use the following expressions for the leading terms in the expansion around $\bm{y}_{\perp} \rightarrow \bm{y'}_{\perp}$ for quantities that appear in the quadrupole dependent contribution, $\mathcal{B}_{T}$ defined by Eqs.~\eqref{eq:direct-cs-trans-pol-B} and \eqref{eq:B-T-reg-reg} and \eqref{eq:B-T-reg-ins}:
\begin{align}
\int_{\bm{r}_{yy' \perp}} \ln \Big( \frac{1}{\bm{r}^2_{yy' \perp} \, \mu_{F}^2} \Big) e^{- i \frac{1-z_{q}}{z_{\gamma}} \, \bm{k}_{\gamma \perp} . \bm{r}_{yy' \perp}}    \, \Xi(\vect{x},\vect{y}; \bm{y'}_{\perp}, \bm{x'}_{\perp} \vert Y) \Big \vert_{\rm{lead.}}&= \Big( \frac{z_{\gamma}}{1-z_{q}} \Big)^2 \,\frac{4\pi}{k_{\gamma \perp}^2} \,  \Theta(\vect{x},\vect{x}', \vect{y} \vert Y) \, , \\
i \int_{\bm{r}_{yy' \perp}}  \frac{\bm{r}_{yy' \perp}^{i}}{\bm{r}_{yy' \perp}^{2}} e^{- i \frac{1-z_{q}}{z_{\gamma}} \, \bm{k}_{\gamma \perp} . \bm{r}_{yy' \perp}}    \, \Xi(\vect{x},\vect{y}; \bm{y'}_{\perp}, \bm{x'}_{\perp} \vert Y) \Big \vert_{\rm{lead.}}&= 2\pi \frac{z_{\gamma}}{1-z_{q}} \, \frac{\bm{k}_{\gamma \perp}^{i}}{\bm{k}_{\gamma \perp}^2} \, \Theta(\vect{x},\vect{x}', \vect{y} \vert Y) \, ,
\end{align}
where `lead.' indicates the leading term in the limit defined by Eq.\,\eqref{eq:limit} and the limiting color structure is defined in Eq.\,\eqref{eq:coincidence-color-structure}. Finally we will assume translational invariance of the Wilson line correlators and make the redefinition
\begin{align}
    \mathcal{K}_{T,3}^{ij}(\bm{r}_{\perp}) = \frac{z_{\gamma}}{1-z_{q}} \, \frac{\bm{k}_{\gamma \perp}^{i}}{\bm{k}_{\gamma \perp}^2} \, \hat{\mathcal{J}}^{j}_{T,3} (\bm{r}_{\perp}) \, ,
    \label{eq:K-T-3}
\end{align}
to obtain the leading term in the direct photon+quark cross-section, in the limit of Eq.\,\eqref{eq:limit}, for the transversely polarized case as 
\begin{align}
   \sum_{T=\pm 1}  \frac{d \sigma^{\gamma^{*}_{T} A \rightarrow q \gamma X}_{D}}{\mathrm{d}^{2} \vect{k}  \mathrm{d}^{2} \vect{k_{\gamma}}    \mathrm{d} \eta_{q} \mathrm{d} \eta_{\gamma}} \Bigg \vert_{\rm lead.}&= \frac{8N_{c} \alpha_{\rm em}^{2}q^{4}_{f} z_{q}}{(2\pi)^{4}} \, S_{\perp}  \int_{\bm{r}_{\perp}, \bm{r'}_{\perp}} \!\!\!\!\!\!\!\!\! e^{-i\bm{k}_{\perp}.(\bm{r}_{\perp}-\bm{r'}_{\perp})} \Theta(\bm{r}_{\perp},\bm{r'}_{\perp} \vert Y) \, \mathcal{R}^{q\gamma}_{T}(\bm{r}_{\perp},\bm{r'}_{\perp};k,k_{\gamma}) \, ,
     \label{eq:diff-cs-direct-photon-jet-limit-cspace-trans-pol}
\end{align}
where $S_{\perp}$ is the transverse size of the nucleus and the color structure is defined in Eq.\,\eqref{eq:coincidence-color-structure-trans-invariance}. The perturbative piece is given by
\begin{align}
    \mathcal{R}^{q\gamma}_{T}= \mathcal{R}^{q\gamma}_{T;\rm{reg-reg}} + \mathcal{R}^{q\gamma}_{T;\rm{reg-ins}}+ \mathcal{R}^{q\gamma}_{T;\rm{ins-ins}} \, ,
     \label{eq:R-trans-pol-gamma-jet}
\end{align}
where the three components can be expressed compactly as
\begin{align}
    \mathcal{R}^{q \gamma}_{T;\rm{reg-reg}}&=\left[\zeta_{qq} \delta^{im} \delta^{jn} + \chi_{qq} \epsilon^{im} \epsilon^{jn}  \right]\left[ \mathcal{K}^{ij}_{T,1}(\vect{r})  + \mathcal{K}^{ij}_{T,2}(\vect{r})  \right] \left[ \mathcal{K}^{mn*}_{T,1}(\vect{r}')  + \mathcal{K}^{mn*}_{T,2}(\vect{r}')  \right]  \nonumber \\
    &+\left[\zeta_{q\bar{q}} \delta^{im} \delta^{jn} - \chi_{q\bar{q}} \epsilon^{im} \epsilon^{jn}  \right]\left[ \mathcal{K}^{ij}_{T,1}(\vect{r})  + \mathcal{K}^{ij}_{T,2}(\vect{r})  \right] \left[ \mathcal{K}^{mn*}_{T,3}(\vect{r}')  + \mathcal{K}^{mn*}_{T,4}(\vect{r}')  \right] \nonumber \\
    & + \left[\zeta_{\bar{q}q} \delta^{im} \delta^{jn} - \chi_{\bar{q}q} \epsilon^{im} \epsilon^{jn}  \right]\left[ \mathcal{K}^{ij}_{T,3}(\vect{r})  + \mathcal{K}^{ij}_{T,4}(\vect{r})  \right] \left[ \mathcal{K}^{mn*}_{T,1}(\vect{r}')  + \mathcal{K}^{mn*}_{T,2}(\vect{r}')  \right] \nonumber \\
    &+\left[\zeta_{\bar{q}\bar{q}} \delta^{im} \delta^{jn} + \chi_{\bar{q}\bar{q}} \epsilon^{im} \epsilon^{jn}  \right]\left[ \mathcal{K}^{ij}_{T,3}(\vect{r})  + \mathcal{K}^{ij}_{T,4}(\vect{r})  \right] \left[ \mathcal{K}^{mn*}_{T,3}(\vect{r}')  + \mathcal{K}^{mn*}_{T,4}(\vect{r}')  \right]  \, ,
    \label{R-T-gammajet-reg-reg}
\end{align}
\begin{align}
    \mathcal{R}^{q\gamma}_{T;\rm{reg-ins}} &=\kappa_{qq} \Big[  \left(\mathcal{K}^{ii}_{T,1} (\vect{r})  + \mathcal{K}^{ii}_{T,2} (\vect{r}) \right) \mathcal{K}^{*}_{T\rm{ins},2}(\vect{r}') + \mathcal{K}_{T\rm{ins},2}(\vect{r}) \left(\mathcal{K}^{ii*}_{T,1} (\vect{r}')  + \mathcal{K}^{ii*}_{T,2} (\vect{r}') \right) \Big] \nonumber \\
    &+ \kappa_{q\bar{q}} \Big[ \left(\mathcal{K}^{ii}_{T,1} (\vect{r})  + \mathcal{K}^{ii}_{T,2} (\vect{r}) \right) \mathcal{K}^{*}_{T\rm{ins},4}(\vect{r}') + \mathcal{K}_{T\rm{ins},2}(\vect{r}) \left(\mathcal{K}^{ii*}_{T,3} (\vect{r}')  + \mathcal{K}^{ii*}_{T,4} (\vect{r}') \right)  \Big] \nonumber \\
    &+ \kappa_{\bar{q}q} \Big[ \left(\mathcal{K}^{ii}_{T,3} (\vect{r})  + \mathcal{K}^{ii}_{T,4} (\vect{r}) \right) \mathcal{K}^{*}_{T\rm{ins},2}(\vect{r}') + \mathcal{K}_{T\rm{ins},4}(\vect{r}) \left(\mathcal{K}^{ii*}_{T,1} (\vect{r}')  + \mathcal{K}^{ii*}_{T,2} (\vect{r}') \right)   \Big]  \nonumber
    \\
    &+ \kappa_{\bar{q}\bar{q}}  \Big[  \left(\mathcal{K}^{ii}_{T,3} (\vect{r})  + \mathcal{K}^{ii}_{T4} (\vect{r}) \right) \mathcal{K}^{*}_{T\rm{ins},4}(\vect{r}') + \mathcal{K}_{T\rm{ins},4}(\vect{r}) \left(\mathcal{K}^{ii*}_{T,3} (\vect{r}')  + \mathcal{K}^{ii*}_{T,4} (\vect{r}') \right) \Big]  \, ,
    \label{eq:R-T-gammajet-reg-ins}
\end{align}
and
\begin{align}
    \mathcal{R}^{q\bar{q}\gamma}_{T;\rm{ins-ins}} &= \sigma_{qq} \,  \mathcal{K}_{T\rm{ins},2}(\vect{r})\mathcal{K}^{*}_{T\rm{ins},2}(\vect{r'}) \,  + \sigma_{\bar{q}\bar{q}} \, \mathcal{K}_{T\rm{ins},4}(\vect{r})\mathcal{K}^{*}_{T\rm{ins},4}(\vect{r'})
    \label{eq:R-T-gammajet-ins-ins} \, .
\end{align}
 Expressions for all the coefficients $\zeta$, $\chi$, $\kappa$ and $\sigma$ are provided in Appendix~\ref{sec:coefficients-J-functions}. The functions $\mathcal{K}_{T}$ are all defined in terms of the functions $\mathcal{J}_{T}$ that are used to write the $\gamma+q\bar{q}$ cross-section. The relations connecting these are given in Eqs.~\eqref{eq:K-T-definitions}, \eqref{eq:K-T-3} and \eqref{eq:J-hat-T-3} and the expressions for the different $\mathcal{J}_{T}$'s are given in  Appendix~\ref{sec:coefficients-J-functions}.

We can now express the cross-section entirely in momentum space by defining the Fourier conjugates of the various $\mathcal{K}_T$ functions as
\begin{align}
    \begin{bmatrix}
    \mathcal{K}_{T,1}^{ij}(\vect{r}) \\
    \mathcal{K}_{T,2}^{ij}(\vect{r}) \\
    \mathcal{K}_{T \rm{ins},2}(\vect{r}) 
    \end{bmatrix} = \int e^{i \vect{l}.\vect{r}} \begin{bmatrix}
   \widetilde{ \mathcal{K}}_{T,1}^{ij}(\vect{l}+\vect{k_\gamma}) \\
     \widetilde{ \mathcal{K}}_{T,2}^{ij}(\vect{l}+\vect{k_\gamma}) \\
    \widetilde{ \mathcal{K}}_{T \rm{ins},2}(\vect{l}+\vect{k_\gamma}) 
    \end{bmatrix} \quad \, , \qquad   \begin{bmatrix}
    \mathcal{K}_{T,3}^{ij}(\vect{r}) \\
    \mathcal{K}_{T,4}^{ij}(\vect{r}) \\
    \mathcal{K}_{T \rm{ins},4}(\vect{r}) 
    \end{bmatrix} = \int e^{i \vect{l}.\vect{r}} \begin{bmatrix}
   \widetilde{ \mathcal{K}}_{T,3}^{ij}(\vect{l}) \\
     \widetilde{ \mathcal{K}}_{T,4}^{ij}(\vect{l}) \\
    \widetilde{ \mathcal{K}}_{T \rm{ins},4}(\vect{l}) 
    \end{bmatrix}  \, ,
\end{align}
and similarly for the dipole correlators as
\begin{equation}
   S^{(2)}_{Y}(\vect{r}) = \int_{\vect{l}} e^{i \vect{l}\cdot \vect{r}} \,  C_{Y}(\vect{l})    \, .
\end{equation}
The direct photon+quark cross-section for the transversely polarized $\gamma^*$ can now be written as
\begin{align}
   \sum_{T=\pm 1}  \frac{d \sigma^{\gamma^{*}_{T} A \rightarrow q \gamma X}_{D}}{\mathrm{d}^{2} \vect{k}  \mathrm{d}^{2} \vect{k_{\gamma}}    \mathrm{d} \eta_{q} \mathrm{d} \eta_{\gamma}} \Bigg \vert_{\rm lead.}&= \frac{8N_{c} \alpha_{\rm em}^{2}q^{4}_{f} z_{q}}{(2\pi)^{4}} \, S_{\perp}  \int_{\bm{l}_{\perp}} C_{Y}(\bm{l}_{\perp} ) \mathcal{H}^{q\gamma}_{T}(\bm{l}_{\perp};k,k_{\gamma}) \, ,
     \label{eq:diff-cs-direct-photon-jet-limit-mspace-trans-pol}
\end{align}
where 
\begin{align}
 \mathcal{H}^{q\gamma}_{T}(\bm{l}_{\perp};k,k_{\gamma})=\mathcal{H}^{q\gamma}_{T;\rm{reg-reg}}(\bm{l}_{\perp};k,k_{\gamma})+\mathcal{H}^{q\gamma}_{T;\rm{reg-ins}}(\bm{l}_{\perp};k,k_{\gamma})+\mathcal{H}^{q\gamma}_{T;\rm{ins-ins}}(\bm{l}_{\perp};k,k_{\gamma})   \, ,
 \label{eq:H-T-q-gamma}
\end{align}
and the three terms can be expressed compactly as
\begin{align}
   \mathcal{H}^{q\gamma}_{T;\rm{reg-reg}}&= \mathfrak{C}^{im;jn}_{qq} \sum_{p,r=1}^{2} \Big[  \widetilde{\mathcal{K}}^{ij}_{T,p}(\vect{k}+\vect{k_{\gamma}})  - \widetilde{\mathcal{K}}^{ij}_{T,p}(\vect{k}+\vect{k_{\gamma}}-\vect{l})  \Big] \,  \Big[ \widetilde{\mathcal{K}}^{mn}_{T,r}(\vect{k}+\vect{k_{\gamma}})  - \widetilde{\mathcal{K}}^{mn}_{T,r}(\vect{k}+\vect{k_{\gamma}}-\vect{l}) \Big] \nonumber \\
    &+ \mathfrak{C}^{im;jn}_{q\bar{q}} \sum_{p,r=1}^{2} \Big[ \widetilde{\mathcal{K}}^{ij}_{T,p}(\vect{k}+\vect{k_{\gamma}})  - \widetilde{\mathcal{K}}^{ij}_{T,p}(\vect{k}+\vect{k_{\gamma}}-\vect{l})  \Big] \,  \Big[ \widetilde{\mathcal{K}}^{mn}_{T,r+2}(\vect{k})  - \widetilde{\mathcal{K}}^{mn}_{T,r+2}(\vect{k}-\vect{l}) \Big] \nonumber \\
    & + \mathfrak{C}^{im;jn}_{\bar{q}q} \sum_{p,r=1}^{2} \Big[  \widetilde{\mathcal{K}}^{ij}_{T,p+2}(\vect{k})  - \widetilde{\mathcal{K}}^{ij}_{T,p+2}(\vect{k}-\vect{l})  \Big] \,  \Big[ \widetilde{\mathcal{K}}^{mn}_{T,r}(\vect{k}+\vect{k_{\gamma}})  - \widetilde{\mathcal{K}}^{mn}_{T,r}(\vect{k}+\vect{k_{\gamma}}-\vect{l}) \Big] \nonumber \\
    &+ \mathfrak{C}^{im;jn}_{\bar{q}\bar{q}} \sum_{p,r=1}^{2} \Big[  \widetilde{\mathcal{K}}^{ij}_{T,p+2}(\vect{k})  - \widetilde{\mathcal{K}}^{ij}_{T,p+2}(\vect{k}-\vect{l})  \Big] \,  \Big[ \widetilde{\mathcal{K}}^{mn}_{T,r+2}(\vect{k})  - \widetilde{\mathcal{K}}^{mn}_{T,r+2}(\vect{k}-\vect{l}) \Big] \, ,
    \label{eq:H-T-gammajet-reg-reg}
\end{align}
\begin{align}
   \mathcal{H}^{q\gamma}_{T;\rm{reg-ins}} & = 2 \kappa_{qq} \sum_{p=1}^{2} \Big[  \widetilde{\mathcal{K}}^{ii}_{T,p}(\vect{k}+\vect{k_{\gamma}})  - \widetilde{\mathcal{K}}^{ii}_{T,p}(\vect{k}+\vect{k_{\gamma}}-\vect{l})  \Big] \,  \Big[ \widetilde{\mathcal{K}}_{T\rm{ins},2}(\vect{k}+\vect{k_{\gamma}})  - \widetilde{\mathcal{K}}_{T\rm{ins},2}(\vect{k}+\vect{k_{\gamma}}-\vect{l}) \Big] \nonumber \\
    &+ 2 \kappa_{q\bar{q}} \sum_{p=1}^{2} \Big[  \widetilde{\mathcal{K}}^{ii}_{T,p}(\vect{k}+\vect{k_{\gamma}})  - \widetilde{\mathcal{K}}^{ii}_{T,p}(\vect{k}+\vect{k_{\gamma}}-\vect{l})  \Big] \,  \Big[ \widetilde{\mathcal{K}}_{T\rm{ins},4}(\vect{k})  - \widetilde{\mathcal{K}}_{T\rm{ins},4}(\vect{k}-\vect{l}) \Big] \nonumber \\
    &+ 2  \kappa_{\bar{q}q} \sum_{p=1}^{2}  \Big[ \widetilde{\mathcal{K}}_{T\rm{ins},2}(\vect{k}+\vect{k_{\gamma}})  - \widetilde{\mathcal{K}}_{T\rm{ins},2}(\vect{k}+\vect{k_{\gamma}}-\vect{l}) \Big] \, \Big[  \widetilde{\mathcal{K}}^{ii}_{T,p+2}(\vect{k})  - \widetilde{\mathcal{K}}^{ii}_{T,p+2}(\vect{k}-\vect{l})  \Big]  \nonumber \\
    &+ 2 \kappa_{\bar{q}\bar{q}} \sum_{p=1}^{2} \Big[  \widetilde{\mathcal{K}}^{ii}_{T,p+2}(\vect{k})  - \widetilde{\mathcal{K}}^{ii}_{T,p+2}(\vect{k}-\vect{l})  \Big] \,  \Big[ \widetilde{\mathcal{K}}_{T\rm{ins},4}(\vect{k})  - \widetilde{\mathcal{K}}_{T\rm{ins},4}(\vect{k}-\vect{l}) \Big] \, ,
    \label{eq:H-T-gammajet-reg-ins}
\end{align}
and 
\begin{align}
     \mathcal{H}^{q\gamma}_{T;\rm{ins-ins}}& = \sigma_{qq} \Big[ \widetilde{\mathcal{K}}_{T\rm{ins},2}(\vect{k}+\vect{k_{\gamma}})  - \widetilde{\mathcal{K}}_{T\rm{ins},2}(\vect{k}+\vect{k_{\gamma}}-\vect{l})  \Big]^2 + \sigma_{\bar{q}\bar{q}} \Big[ \widetilde{\mathcal{K}}_{T\rm{ins},4}(\vect{k})  - \widetilde{\mathcal{K}}_{T\rm{ins},4}(\vect{k}-\vect{l}) \Big]^2
   \label{eq:H-T-gammajet-ins-ins}  
\end{align} 
In Eq.\,\eqref{eq:H-T-gammajet-reg-reg}, the prefactors are defined as
\begin{align}
    \mathfrak{C}_{qq}^{im;jn}&= \zeta_{qq} \, \delta^{im} \, \delta^{jn} + \chi_{qq} \, \epsilon^{im} \, \epsilon^{jn}  \quad \, \quad  \mathfrak{C}_{\bar{q}\bar{q}}^{im;jn}= \zeta_{\bar{q}\bar{q}} \, \delta^{im} \, \delta^{jn} + \chi_{\bar{q}\bar{q}} \, \epsilon^{im} \, \epsilon^{jn} \, , \\
      \mathfrak{C}_{q\bar{q}}^{im;jn}&= \zeta_{q\bar{q}} \, \delta^{im} \, \delta^{jn} - \chi_{q\bar{q}} \, \epsilon^{im} \, \epsilon^{jn} \quad , \quad \mathfrak{C}_{\bar{q}q}^{im;jn}=\zeta_{\bar{q}q} \, \delta^{im} \, \delta^{jn} - \chi_{\bar{q}q} \, \epsilon^{im} \, \epsilon^{jn} \, .
\end{align}
The final expressions for the momentum space  $\widetilde{\mathcal{K}}_{T}$ terms, that constitute the cross-section in Eq.\,\eqref{eq:diff-cs-direct-photon-jet-limit-mspace-trans-pol} above, are as follows:
\begin{align}
  \widetilde{ \mathcal{K}}^{ij}_{T,1}(\vect{l})&= \frac{\bm{k}_{\gamma \perp}^{i} - \frac{z_{\gamma}}{z_{q}} \, \bm{k}_{\perp}^{i}}{(\bm{k}_{\gamma \perp}- \frac{z_{\gamma}}{z_{q}} \, \bm{k}_{\perp})^2} \, \frac{\bm{l}_{\perp}^{j}}{\bm{l}_{\perp}^2+\Delta_{1}}   \, , \label{eq:K-tilde-T-1} \\
 \widetilde{ \mathcal{K}}^{ij}_{T,2}(\vect{l}) &= -\frac{1}{z_{\gamma} \, (z_{q}+z_{\gamma})} \, \frac{\big[(z_q+z_{\gamma}) \, \bm{k}_{\gamma \perp}^{i}-z_{\gamma} \, \bm{l}_{\perp}^{i} \big] \, \bm{l}_{\perp}^{j}}{(\bm{l}_{\perp}^2+\Delta_{1}) \, \Big( Q^{2}+\frac{\bm{l}^{2}_{\perp}}{1-z_{q}-z_{\gamma}} +\frac{\bm{k}_{\gamma \perp}^{2}}{z_{\gamma}}+\frac{(\bm{l}_{\perp}-\bm{k}_{\gamma \perp})^{2}}{z_{q}} \Big) } \, , \label{eq:K-tilde-T-2} \\
  \widetilde{ \mathcal{K}}^{ij}_{T,3}(\vect{l}) &= -\frac{1-z_{q}-z_{\gamma}}{1-z_{q}} \, \frac{\bm{k}_{\gamma \perp}^{i}}{\bm{k}_{\gamma \perp}^2} \, \frac{\bm{l}_{\perp}^{j}}{\bm{l}_{\perp}^2+\Delta_{3}} \, , \label{eq:K-tilde-T-3} \\
  \widetilde{ \mathcal{K}}^{ij}_{T,4}(\vect{l})  &= \frac{1}{z_{\gamma} \, (1-z_{q}) } \, \frac{\big[ (1-z_{q}) \, \bm{k}_{\gamma \perp}^{i}+z_{\gamma} \, \bm{l}_{\perp}^{i} \big] \, \bm{l}_{\perp}^{j}}{(\bm{l}^{2}_{\perp}+\Delta_{3}) \, \Big( Q^{2}+\frac{\bm{l}^{2}_{\perp}}{z_{q}} +\frac{\bm{k}_{\gamma \perp}^{2}}{z_{\gamma}}+\frac{(\bm{l}_{\perp}+\bm{k}_{\gamma \perp})^{2}}{1-z_{q}-z_{\gamma}} \Big)} \, , \label{eq:K-tilde-T-4} \\
 \widetilde{ \mathcal{K}}_{T\rm{ins},2}(\vect{l})  &= -\frac{1-z_q}{(1-z_q-z_\gamma)\, (z_{q}+z_{\gamma})^2 } \, \frac{1}{\Big( Q^{2}+\frac{\bm{l}^{2}_{\perp}}{1-z_{q}-z_{\gamma}} +\frac{\bm{k}_{\gamma \perp}^{2}}{z_{\gamma}}+\frac{(\bm{l}_{\perp}-\bm{k}_{\gamma \perp})^{2}}{z_{q}} \Big)} \, , \label{eq:K-tilde-T-ins-2} \\
  \widetilde{ \mathcal{K}}_{T\rm{ins},4}(\vect{l})  &= -\frac{z_q+z_{\gamma}}{z_q \, (1-z_{q})^2 } \, \frac{1}{\Big( Q^{2}+\frac{\bm{l}^{2}_{\perp}}{z_{q}} +\frac{\bm{k}_{\gamma \perp}^{2}}{z_{\gamma}}+\frac{(\bm{l}_{\perp}+\bm{k}_{\gamma \perp})^{2}}{1-z_{q}-z_{\gamma}} \Big)} \, . \label{eq:K-tilde-T-ins-4} 
\end{align}

\section{Differential cross-section for inclusive quark production from $q\bar{q}$ production in DIS} 
\label{sec:single-inclusive-parton-cs}

The unpolarized differential cross-section for $q\bar{q}$ production in $\gamma^* + A$ scattering is given by \cite{Dominguez:2011wm}
\begin{align}
    \frac{\mathrm{d} \sigma^{\gamma^{*}_{L} A \rightarrow q \bar{q} X}}{\mathrm{d}^{2} \vect{k}  \mathrm{d}^{2} \vect{p}    \mathrm{d} \eta_{q} \mathrm{d} \eta_{\bar{q}}}  =  \frac{8N_{c}\alpha_{\rm em}q^{2}_{f}}{(2\pi)^{6}} \delta(1-z_q-z_{\bar{q}}) \int_{\vect{x},\vect{y},\vect{x'},\vect{y'}} \!\!\!\!\!\!\!\!\!\!\!\!\!\!\!\!\!\!\!\!\!\!\!\!\!\!\! e^{-i\vect{k}.(\vect{x}-\vect{x'})} &e^{-i \vect{p}.(\vect{y}-\vect{y'})} \, \Xi(\vect{x},\vect{y},\vect{y'},\vect{x'} \vert Y) \nonumber \\  &\times z_{q}^{3}(1-z_{q})^{3} Q^{2} K_{0}(\Delta^{1/2} r_{\perp}) \,  K_{0}(\Delta^{1/2} r'_{\perp})  \, , 
\end{align}
\begin{align}
    \sum_{\lambda=T=\pm 1} \, \frac{\mathrm{d} \sigma^{\gamma^{*}_{\lambda} A \rightarrow q \bar{q} X}}{\mathrm{d}^{2} \vect{k}  \mathrm{d}^{2} \vect{p}    \mathrm{d} \eta_{q} \mathrm{d} \eta_{\bar{q}}}  = \frac{4N_{c}\alpha_{\rm em}q^{2}_{f}}{(2\pi)^{6}} \delta(1-z_q-z_{\bar{q}}) & \int_{\vect{x},\vect{y},\vect{x'},\vect{y'}} \!\!\!\!\!\!\!\!\!\!\!\!\!\!\!\!\!\!\!\!\!\!\!\!\!\!\! e^{-i\vect{k}.(\vect{x}-\vect{x'})} e^{-i \vect{p}.(\vect{y}-\vect{y'})} \, \Xi(\vect{x},\vect{y},\vect{y'},\vect{x'} \vert Y) \nonumber \\ & \!\!\!\!\!\!\!\!\!\! \times z_{q} (1-z_{q}) \big[z_{q}^2+(1-z_{q})^2 \big] \, \frac{\bm{r}_{\perp}.\bm{r'}_{\perp}}{r_{\perp} \, r'_{\perp}} \, \Delta \, K_{1}(\Delta^{1/2}r_{\perp} ) \, K_{1}(\Delta^{1/2}r'_{\perp} )  \, .
\end{align}
where the color structure $\Xi(\vect{x},\vect{y};\vect{y'},\vect{x'} \vert Y)$ is given by Eq.\,\eqref{eq:color-structure-LO} contains dipoles and a quadrupole, and $\Delta = z_q(1-z_q) Q^2$.

The expression for the unpolarized differential cross-section for inclusive quark production can be obtained by integrating the antiquark phase space:
\begin{align}
    \frac{\mathrm{d} \sigma ^{\gamma^{*}_{\lambda} A \rightarrow q X}}{\mathrm{d}^{2} \vect{k}     \mathrm{d} \eta_{q}} = \int \mathrm{d}^{2} \vect{p} \mathrm{d}\eta_{\bar{q}} \, \frac{d \sigma ^{\gamma^{*}_{\lambda} A \rightarrow q \bar{q} X}}{\mathrm{d}^{2} \vect{k} \mathrm{d}^{2} \vect{p} \mathrm{d} \eta_{q} \, \mathrm{d} \eta_{\bar{q}}} \, .
\end{align}    
The integration over $\vect{p}$ results in a delta function $\delta^{(2)}(\vect{y}-\vect{y'})$ which reduces one of the transverse integration. Furthermore, assuming translation invariance of correlators we obtain
\begin{align}
    \frac{\mathrm{d} \sigma^{\gamma^{*}_{L} A \rightarrow q X} }{\mathrm{d}^{2} \vect{k}     \mathrm{d} \eta_{q}}  &=  \frac{8N_{c}\alpha_{\rm em}q^{2}_{f}z_{q}^{3}(1-z_{q})^{2}}{(2\pi)^{4}} S_\perp \int_{\vect{r},\vect{r'}} \!\!\!\!\!\!\!\!\!\! e^{-i\vect{k}.(\vect{r}-\vect{r'})} \, \Theta(\vect{r},\vect{r'} \vert Y)   Q^{2} K_{0}(\Delta^{1/2} r_{\perp}) \,  K_{0}(\Delta^{1/2} r'_{\perp}) \, , \label{eq-semi_inclusive_cs_long} \\
    \sum_{\lambda=T=\pm 1} \!\!\!\!\! \frac{\mathrm{d} \sigma^{\gamma^{*}_{\lambda} A \rightarrow q X} }{\mathrm{d}^{2} \vect{k}     \mathrm{d} \eta_{q}}  &=  \frac{4N_{c}\alpha_{\rm em}q^{2}_{f}z_{q}}{(2\pi)^{4}} S_\perp \!\! \int_{\vect{r},\vect{r'}} \!\!\!\!\!\!\!\!\!\!\!\!\! e^{-i\vect{k}.(\vect{r}-\vect{r'})} \, \Theta(\vect{r},\vect{r'} \vert Y)  \big[z_{q}^2+(1-z_{q})^2 \big] \, \frac{\bm{r}_{\perp} \cdot \bm{r'}_{\perp}}{r_{\perp} \, r'_{\perp}} \, \Delta \, K_{1}(\Delta^{1/2} r_{\perp}) \,  K_{1}(\Delta^{1/2} r'_{\perp})  \, . \label{eq-semi_inclusive_cs_trans}
\end{align} 
where the color structure $\Theta(\vect{r},\vect{r'} \vert Y)$ is given by Eq.\,\eqref{eq:coincidence-color-structure-trans-invariance} and only involves dipoles.

In momentum space these expressions take the form:
\begin{align}
    \frac{\mathrm{d} \sigma^{\gamma^{*}_{L} A \rightarrow q X} }{\mathrm{d}^{2} \vect{k}     \mathrm{d} \eta_{q}}  &=  \frac{8N_{c}\alpha_{\rm em}q^{2}_{f}z_{q}^{3}(1-z_{q})^{2}}{(2\pi)^{2}} S_\perp \int_{\vect{l}} C_{Y}(\vect{l}) \left | \frac{ Q}{\vect{k}^2 + \Delta} - \frac{ Q}{(\vect{k}-\vect{l})^2 + \Delta} \right |^2  \, , \\
    \sum_{\lambda=T=\pm 1} \, \frac{\mathrm{d} \sigma^{\gamma^{*}_{\lambda} A \rightarrow q X} }{\mathrm{d}^{2} \vect{k}     \mathrm{d} \eta_{q}}  &=  \frac{4N_{c}\alpha_{\rm em}q^{2}_{f}z_{q} \big[z_{q}^2+(1-z_{q})^2 \big] }{(2\pi)^{2}} S_\perp \int_{\vect{l}} C_{Y}(\vect{l}) \left | \frac{ \vect{k}}{\vect{k}^2 + \Delta} - \frac{(\vect{k}-\vect{l})}{(\vect{k}-\vect{l})^2 + \Delta} \right |^2 \, .
\end{align}
We can now write the general expression for the differential cross-section for inclusive quark production in $e+A$ DIS at small $x$ as
\begin{align}
    \frac{d \sigma ^{eA \rightarrow e'q X}}{\mathrm{d}W^{2} \mathrm{d}Q^{2} \mathrm{d}^{2} \vect{k} \mathrm{d} \eta_{q}  }&=f_{L}(Q^{2},W^{2}) \, \frac{d \sigma ^{\gamma^{*}_{L} A \rightarrow q  X}}{\mathrm{d}^{2} \vect{k}      \mathrm{d} \eta_{q}}  +f_{T}(Q^{2},W^{2}) \sum_{\lambda=T=\pm 1} \,  \frac{d \sigma ^{\gamma^{*}_{\lambda} A \rightarrow q  X}}{\mathrm{d}^{2} \vect{k} \mathrm{d} \eta_{q}} \, .
\end{align}
It is also interesting to observe that upon integration over the quark phase space ($\vect{k},\eta_{q}$), one recovers the LO cross-section for  fully inclusive DIS at small $x$.

\bibliography{bibliography.bib}

\end{document}